\documentclass[journal, draftcls,onecolumn]{IEEEtran}
\usepackage{enumerate}
\usepackage{cite}
\usepackage{amscd}
\usepackage{dsfont}
\usepackage{latexsym}
\usepackage{array}

\usepackage{tikz}\usetikzlibrary{calc}
\usepackage{tabularx}
\usepackage{epsfig}
\usepackage{graphicx,subfigure}
\usepackage{graphicx}
\usepackage{amsmath,mathrsfs}
\usepackage{amsthm}
\usepackage{amssymb}  %used for the font style \mathbb{}
\usepackage{amsbsy}
\usepackage{color}
\usepackage{stfloats}
\usepackage{algorithm}
\usepackage{algpseudocode}
\usepackage{bm}
\usepackage{pifont}
\usepackage{accents}
\usepackage{multirow}
\usepackage{etoolbox}
\makeatletter
\patchcmd{\@makecaption}
  {\scshape}
  {}
  {}
  {}
\makeatother
\newtheoremstyle{note}
{3pt}
{1pt}
{}
{\parindent}
{\itshape}
{:}
{.5em}
{\thmname{#1}\thmnumber{ #2}\thmnote{\thmnote{ (#3)}}}% <Theorem head spec (can be left empty, meaning `normal')>
\theoremstyle{note}

\newtheorem{lem}{Lemma}
\newtheorem{ther}{Theorem}
\newtheorem{deft}{Definition}
\newtheorem{prop}{Proposition}

\theoremstyle{definition}
\newtheorem{rem}{Remark}
\newtheoremstyle{dotless}{}{}{\itshape}{}{\bfseries}{}{ }{}
\theoremstyle{dotless}

\newcommand {\aplt} {\ {\raise-.5ex\hbox{$\buildrel<\over{\mbox{\scriptsize $\sim$}}$}}\ }

\DeclareMathOperator{\SNR}{\mathsf{SNR}}
\newcommand{\argmax}{\operatornamewithlimits{argmax}}
\newcommand{\ubar}[1]{\underaccent{\bar}{#1}}

\begin{document}
%
% paper title
% can use linebreaks \\ within to get better formatting as desired
\title{Polar Lattices for Lossy Compression}
\author{Ling~Liu, Jinwen Shi, \emph{Student Member, IEEE} and Cong Ling, \emph{Member, IEEE}
\thanks{This work was presented in part at the IEEE Information Theory Workshop (ITW) 2015, Jeju Island, Korea, October 2015 and at the IEEE International Conference on Wireless Communications and Signal Processing 2016,
Yangzhou, China, October 2016. This work was supported in part by the China Scholarship Council and in part by the Engineering and Physical Sciences Research Council (EPSRC).}
\thanks{Ling Liu is with the College of Computer Science and Software Engineering, Shenzhen University, Shenzhen, China (e-mails: liulingcs@szu.edu.cn).}
\thanks{Jingwen Shi and Cong Ling are with the Department of Electrical and Electronic Engineering, Imperial College London, London, UK (e-mails: jinwen.shi12@imperial.ac.uk, cling@ieee.org).}}

\maketitle
\begin{abstract}
%\boldmath
In this work, we propose a new construction of polar lattices to achieve the rate-distortion bound of a memoryless Gaussian source. The structure of the proposed polar lattices allows to integrate entropy coding into the lattice quantizer, which greatly simplifies the compression process. The overall complexity of encoding and decoding is $O(N \log^2 N)$ for any target distortion and fixed rate larger than the rate-distortion bound. Moreover, the nesting structure of polar lattices provides solutions to various multi-terminal coding problems. The Wyner-Ziv coding problem for a Gaussian source can be solved by using a capacity-achieving polar lattice for the Gaussian channel, nested with a rate-distortion bound achieving lattice, while the Gelfand-Pinsker problem can be solved in a reversed manner. The polar lattice quantizer is further extended to extract Wyner's common information of a pair of Gaussian sources or multiple Gaussian sources.
\end{abstract}

\IEEEpeerreviewmaketitle

\section{Introduction}
% no \IEEEPARstart
% no \IEEEPARstart
Vector quantization (VQ) \cite{VecQz} has been widely used for source coding of image and speech data since the 1980s. Compared with scalar quantization, the advantage of VQ, guaranteed by Shannon's rate-distortion theory, is that better performance can always be achieved by encoding vectors instead of scalars, even in the case of memoryless sources. However, the Shannon theory does not provide us any constructive VQ design scheme. During the past several decades, many practical VQ techniques with relatively low complexity have been proposed, such as lattice VQ \cite{yellowbook}, multistage VQ \cite{MultiStageQZ}, tree-structured VQ \cite{TreeQZ}, gain-shape VQ \cite{GainShapQZ}, etc. Among them, lattice VQ is of particular interest because its highly regular structure makes compact storage and fast quantization possible.

In this work, we present an explicit construction of polar lattices for quantization, which achieves the rate-distortion bound of the continuous Gaussian source. It is well known that the optimal output alphabet size is infinite for continuous-amplitude sources. Particularly, the rate-distortion function for the i.i.d. Gaussian source of variance $\sigma_s^2$ under the squared-error distortion measure $d(x,y)=\|x-y\|^2$ is given by \cite{gallager1968information}
\begin{eqnarray}\label{eqn:Eq1}
R(\Delta)=\max\Big\{\frac{1}{2}\log\Big(\frac{\sigma_s^2}{\Delta}\Big),0\Big\},
\end{eqnarray}
where $\Delta$ and $R$ denote the average distortion and rate per symbol, respectively. However, in practice, the size of the reconstruction alphabet needs to be finite. Using the argument of random coding ensembles, the author proved the existence of a block code with a finite number of output letters that achieves performance arbitrarily close to the rate-distortion bound in \cite[Theorem 9.6.2]{gallager1968information}. Then, the rate-distortion function $R_M(\Delta)$ for a size-constrained output alphabet was defined in \cite{pearlman} with $M$ denoting the size of output alphabet. The well-known trellis coded quantization (TCQ) \cite{marcellin} was motivated by this alphabet constrained rate-distortion theory. It was shown that for a given encoding rate of $R$ bits per symbol, the rate-distortion function $R(\Delta)$ can be approached by using a TCQ encoder with rate $R+1$ after an initial Lloyd-Max quantization. It is equivalent to the trellis coded modulation (TCM) in the sense that $m$ information bits are transmitted using $2^{m+1}$ constellation points. A near-optimum lattice quantization scheme based on tailbiting convolutional codes was introduced in \cite{NSMboundLattice}. Despite good practical performance, a theoretical proof of the rate-distortion bound achieving TCQ with low complexity is still missing. More recently, a scheme based on low density Construction-A (LDA) lattices \cite{GoodnessLDA} was proved to be quantization-good (defined in Sect. \ref{sec:qzgood}) using the minimum-distance lattice decoder. However, the ideal performance cannot be realized by the suboptimal belief-propagation decoding algorithm in practice.

Polar lattices, which are multilevel lattices constructed from polar codes, have the potential to solve this problem with low complexity. A construction of polar lattices was given in \cite{polarlatticeJ} to achieve the capacity of the Gaussian channel with quasi-linear complexity. A salient feature is the discrete Gaussian distribution it employed, which shares many similar properties to the continuous Gaussian distribution when its associated flatness factor is negligible. We may use the discrete Gaussian distribution instead of the continuous one as the distribution of the reconstruction alphabet \cite{COngQz}. It is also shown in \cite{polarlatticeJ} that with binary lattice partition, the number of the levels $r$ does not need to be very large ($r=O(\log\log N)$) to achieve the capacity $\frac{1}{2}\log(1+\SNR)$ of the additive white Gaussian noise (AWGN) channel, where $\SNR$ denotes the signal noise ratio. By the duality between source coding and channel coding, the quantization lattices can be roughly viewed as a channel coding lattice constructed on the test channel. For a Gaussian source with variance $\sigma_{s}^{2}$ and an average distortion $\Delta$, the test channel is actually an AWGN channel with noise variance $\Delta$. In this case, the $``\SNR"$ of the test channel is $\frac{\sigma_{s}^{2}-\Delta}{\Delta}$, and its ``capacity\char`\"{} is exactly $\frac{1}{2}\log\left(\frac{\sigma_{s}^{2}}{\Delta}\right)$, which implies that the rate of the polar lattice quantizer can be made arbitrarily close to $\frac{1}{2}\log\left(\frac{\sigma_{s}^{2}}{\Delta}\right)$. Therefore, based on this idea, we propose the construction of polar lattices which are capable of achieving the rate-distortion bound of Gaussian sources in this work. We note that the difference between quantization polar lattices and AWGN-coding polar lattices not only lies in the construction of their component polar codes, but also in the role of their associate flatness factors. For AWGN-coding polar lattices, the flatness factor is required to be negligible to ensure a coding rate close to the channel capacity and it has no impact on the error correction performance. For quantization polar lattices, however, the flatness factor affects both the compression rate and the distortion performance. This is also the reason why the lattice Gaussian distribution can be optimal for both channel coding and quantization, and consequently be utilized for Gaussian Wyner-Ziv and Gelfand-Pinsker coding.

As another application, we extract Wyner's common information (CI) of correlated Gaussian sources using polar lattices. In the literature, there are different ways to characterize the amount of CI of correlated sources. Apart from Shannon's mutual information \cite{shannon2001mathematical} and G\'{a}cs-K\"{o}rner's CI \cite{gacs1973common}, Wyner proposed an alternative definition to quantify the CI of a pair of correlated sources $(X,Y)$ with finite alphabet \cite{WynerCI} as
\begin{eqnarray}
C(X,Y)=\inf_{X-W-Y}I(X,Y;W),\label{eq:WynerCI}
\end{eqnarray}
where the infimum is taken over all $W$, such that $X-W-Y$ forms a Markov chain.
Wyner and G\'{a}cs-K\"{o}rner's works on CI can be considered two different viewpoints of the lossless Gray-Wyner region. Their works were then extended by \cite{viswanatha2014lossy} to the lossy case, where the reconstructed sequences have certain distortions. Moreover, a generalized lossy source coding interpretation of Wyner's CI was given in \cite{GeXulossyCI} for multiple dependent random variables with arbitrary number of alphabets.

\subsection{Contributions}
The novel technical contribution of this paper is three-fold:
\begin{itemize}
  \item The construction of polar lattices for the Gaussian source and the proof of achieving the rate-distortion bound. This is a dual work of capacity-achieving polar lattices for the AWGN channel, and can also be considered as an extension of binary polar lossy coding to the multilevel coding scenario. Compared with traditional lattice quantization schemes \cite{ZamirQzNoise,BK:Zamir}, which generally require a separate entropy encoding process after obtaining the quantized lattice points, our scheme naturally integrates these two processes together. The analysis of these quantization polar lattices prepares us for further discussions of Gaussian Wyner-Ziv and Gelfand-Pinsker problems.
  \item The solutions of the Gaussian Wyner-Ziv and Gelfand-Pinsker problems, which consist of two nested polar lattices. One is AWGN-good and the other is Gaussian rate-distortion bound achieving. The two lattices are simultaneously shaped according to a proper lattice Gaussian distribution. Note that the Wyner-Ziv and Gelfand-Pinsker problems for the binary case have been solved by Korada and Urbanke \cite{KoradaSource} using nested polar codes. However, in the Gaussian case, the problems turn out to be more complicated as the Wyner-Ziv bound becomes lower (the Gelfand-Pinsker capacity becomes larger by duality). As mentioned in \cite{WynerZiv}, the severe conditions \cite[eq. (12)]{WynerZiv} and \cite[eq. (18)]{WynerZiv} for the bound, which corresponds to the scenario where both encoder and decoder know the side information, can be satisfied in the Gaussian case rather than the binary case because of infinite alphabet size, meaning that more effort should be made for the Gaussian case. Our polar lattices achieve the whole region of the Wyner-Ziv bound and have no requirement on the $\SNR$ for the Gelfand-Pinsker capacity.
  \item An explicit construction based on polar lattices to achieve the lossy Gray-Wyner region \cite{GeXulossyCI} for two Gaussian sources. Note that the lossy Gray-Wyner region not only contains the case where lossy CI equals lossless CI, but also the case where lossy CI equals the optimal rate for a certain distortion pair of the source. Finally, Wyner's CI of multiple Gaussian random variables with a specific covariance matrix is also achieved by polar lattices.
\end{itemize}

\subsection{Relation to Prior Works}
As mentioned above, although the TCQ technique performs well in practice, its theoretical limit is still unclear, to the best of our knowledge. Polar lattices, as we will see, can be proved to achieve the rate-distortion bound. Moreover, thanks to their low complexity, considerably high-dimensional polar lattices are available in practice. According to the simulation results in Section \ref{sec:simu}, the achieved performance is within a gap of $0.2$ dB to the theoretic bound when the lattice dimension $N=2^{18}$.

The sparse regression codes were also proved to achieve the the optimal rate-distortion bound of i.i.d Gaussian sources with polynomial complexity \cite{VenkaSparse1,VenkaSparse2}. In fact, there exists a trade-off between the distortion performance and encoding complexity. For a block length $N$, typical encoding complexity of this kind of codes is $O\left((N/\log N)^{2}\right)$ for an exponentially decaying excess distortion with exponent $O(N/\log N)$, and their designed random matrix incurs $O\left(N^{3}\right)$ storage complexity. In comparison, the construction of polar lattices is as explicit as that of polar codes themselves, and the complexity is quasi-linear $O\left(N\log^{2}N\right)$ for a sub-exponentially \footnote{By saying a sub-exponentially decaying excess distortion, we mean that the excess distortion vanishes as $2^{-N^\beta}$ for some $0< \beta < 1$. In fact, $\beta$ can be arbitrarily close to $\frac{1}{2}$ in our work.}
decaying excess distortion with exponent roughly $O(\sqrt{N})$. The sparse regression codes have also been used in multi-terminal source and channel coding \cite{SPCmulti12}.

The saliently nesting structure of polar lattices also gives us solutions to the Gaussian Wyner-Ziv and Gelfand-Pinsker problems. According to the prior work by Zamir, Shamai and Erez \cite{zamir3,ErezDirtyPaper1}, the two problems can be solved by nested quantization-good and AWGN-good lattices. However, due to the lack of explicit construction of such
good lattices, no explicit solution is known. A practical scheme based on multidimensional nested lattice codes for the Gaussian Wyner-Ziv problem was also proposed in \cite{CongGaoWZ}. A lattice-based Gelfand-Pinsker coding scheme
using repeat-accumulate codes, which were concatenated with trellis shaping, was presented in \cite{ErezDirtyPaper}. This scheme was shown to be able to obtain a very close-to-capacity performance. Unfortunately, the complexity grows exponentially to achieve the shaping gain and a theoretical proof for the Gelfand-Pinsker capacity-achieving is also missing.

Wyner's CI of two Gaussian random variables was presented in \cite{GeXulossyCI,viswanatha2014lossy}. A generalized formula of Wyner's CI of jointly Gaussian vectors was deduced in \cite{satpathy2015gaussian}. The dual problem was considered in \cite{yang2014wyner}, where the CI of the outputs of two additive Gaussian channels with a common input was computed. For general continuous sources, the upper bound on Wyner's CI of multiple continuous random variables has been established in terms of the dual total correlation in \cite{li2016distributed}. A lower bound on Wyner's common information for continuous random variables was given in \cite{SulaCIlowbnd}. Some interesting extensions of Wyner's common information can be also found in \cite{GastparRelaxCI} and \cite{GastparCCA}.

The use of polar codes for the CI was recently proposed in \cite{GoelaCommonInfor}, which discussed polarization from the perspective of the maximal correlation of two discrete sources. Furthermore, it proved that polar codes are
optimal to extract Wyner's CI of discrete sources.

\subsection{Outline of the Paper}
The paper is organized as follows: Section II presents the background of polar codes and polar lattices. The construction of rate-distortion bound achieving polar lattices is given in Section III, followed by simulation results in Section IV. In Section V and Section VI, we present the solutions of the Gaussian
Wyner-Ziv and Gelfand-Pinsker problems, respectively, by combining the AWGN polar lattices and the proposed quantization polar lattices. In Section VII, we construct polar lattices for a pair of Gaussian random variables for the lossy Gray-Wyner network; then we extend the method to multiple Guassian sources. The paper is concluded in Section VIII.

\subsection{Notations}
All random variables (RVs) will be denoted by capital letters. Let $P_{X}$ denote the probability distribution of a RV $X$ taking values $x$ in a set $\mathcal{X}$. For multilevel coding, we denote by $X_{\ell}$ a RV $X$ at level $\ell$. The $i$-th realization of $X_{\ell}$ is denoted by $x_{\ell}^{i}$. We also use the notation $x_{\ell}^{i:j}$ as a shorthand for a vector $(x_{\ell}^{i},...,x_{\ell}^{j})$, which is a realization of RVs $X_{\ell}^{i:j}=(X_{\ell}^{i},...,X_{\ell}^{j})$. Similarly, $x_{\ell:\jmath}^{i}$ will denote the realization of the
$i$-th RV from level $\ell$ to level $\jmath$, i.e., of $X_{\ell:\jmath}^{i}=(X_{\ell}^{i},...,X_{\jmath}^{i})$. For a set $\mathcal{I}$, $|\mathcal{I}|$ represents its cardinality. For an integer $N$, $[N]$ denotes the set of all integers from $1$ to $N$. $\mathds{1}(\cdot)$ denotes an indicator function. Let $I(X;Y)$ denote the mutual information between $X$ and $Y$. The notations $R\to I(X;Y)^{+}$ and $R\to I(X;Y)^{-}$ will be used to represent a rate approaching $I(X;Y)$ from the right side (equal or greater than $I(X;Y)$) and the left side (equal or less than $I(X;Y)$),
respectively. The variational distance between probability density functions $f(x)$ and $g(x)$ is defined by
$\mathbb{V}(f(x),g(x)) \triangleq \frac{1}{2}\int_{x}|f(x)-g(x)|dx$;
the Kullback-Leibler divergence is defined by $\mathbb{D}(f(x)\|g(x))=\int_{x}f(x)\log\frac{f(x)}{g(x)}dx$. These are defined for two probability mass functions similarly.
Throughout this paper, we use the binary logarithm and information is measured in bits.

\section{Background}
\subsection{Lattice codes and lattice Gaussian distribution}
A lattice is a discrete subgroup of $\mathbb{R}^{n}$ which can be described by
\begin{eqnarray}
\Lambda=\{\lambda=\mathbf{B}z:z\in\mathbb{Z}^{n}\},
\end{eqnarray}
where $\mathbf{B}=[\mathrm{b}_{1},\cdots,\mathrm{b}_{n}] \in \mathbb{R}^{n\times n}$ is the generator matrix. In this paper, the columns of $\mathbf{B}$ are assumed to be linearly independent.

For a vector $x\in\mathbb{R}^{n}$, the nearest-neighbor quantizer associated with $\Lambda$ is $Q_{\Lambda}(x)=\text{arg}\min_{\lambda\in\Lambda}\|\lambda-x\|$, where ties are resolved arbitrarily. We define the modulo lattice operation by $x\text{ mod }\Lambda\triangleq x-Q_{\Lambda}(x)$. The Voronoi region of $\Lambda$, defined by $\mathcal{V}(\Lambda)=\{x:Q_{\Lambda}(x)=0\}$, specifies the nearest-neighbor decoding region. The volume of a fundamental region is equal to that of the Voronoi region $\mathcal{V}(\Lambda)$, which is given by $V(\Lambda)=|\text{det}(\mathbf{B})|$.

Given an $n$-dimensional lattice $\Lambda$, the block error probability $P_{e}(\Lambda,\sigma^2)$ of lattice decoding is the probability $\mathbb{P}\{{x}\notin \mathcal{V}(\Lambda)\}$ that an $n$-dimensional independent and identically distributed (i.i.d.) Gaussian noise vector ${x}$ with zero mean and variance $\sigma^{2}$ per dimension falls outside the Voronoi region $\mathcal{V}(\Lambda)$. Define the volume-to-noise ratio (VNR) by \cite{forney6, LingBel13}
\begin{eqnarray}
\gamma_{\Lambda}(\sigma)\triangleq\frac{V(\Lambda)^\frac{2}{n}}{\sigma^2}. \notag\
%\label{eqn:VNR}
\end{eqnarray}
The VNR stands for the normalized volume of $\Lambda$ to the normalized volume of a noise sphere of squared radius $n\sigma^2$ for large $n$.

\begin{deft}[AWGN-good lattices]\label{deft:awgngood}
A sequence of lattices $\Lambda^{(N)}$ of increasing dimension $N$ is AWGN-good if, for any fixed VNR greater than $2\pi e$, \[
\lim_{N\rightarrow\infty} P_e\left(\Lambda^{(N)},\sigma^2\right) = 0.
\]
\end{deft}

For $\sigma>0$ and $c\in\mathbb{R}^{n}$, the Gaussian distribution of variance $\sigma^{2}$ centered at $c$ is defined as
\begin{eqnarray*}
f_{\sigma,c}(x)=\frac{1}{\left(\sqrt{2\pi}\sigma\right)^{n}}e^{-\frac{\|x-c\|^{2}}{2\sigma^{2}}},\:\:x\in\mathbb{R}^{n}.
\end{eqnarray*}
Let $f_{\sigma,0}(x)=f_{\sigma}(x)$ for short. We define a $\Lambda$-periodic function
\begin{eqnarray*}
f_{\sigma,\Lambda}(x)=\sum\limits _{\lambda\in\Lambda}f_{\sigma,\lambda}(x)=\frac{1}{\left(\sqrt{2\pi}\sigma\right)^{n}}\sum\limits _{\lambda\in\Lambda}e^{-\frac{\|x-\lambda\|^{2}}{2\sigma^{2}}}.\
\end{eqnarray*}
We note that $f_{\sigma,\Lambda}(x)$ is a probability density function (PDF) if $x$ is restricted to the fundamental region $\mathcal{R}(\Lambda)$. This distribution is actually the PDF of the $\Lambda$-aliased Gaussian
noise, i.e., the Gaussian noise after the mod-$\Lambda$ operation \cite{forney6}.

The flatness factor of a lattice $\Lambda$ is defined as \cite{cong2}
\begin{eqnarray}
\epsilon_{\Lambda}(\sigma)\triangleq\max\limits _{x\in\mathcal{R}(\Lambda)}\left|V(\Lambda)f_{\sigma,\Lambda}(x)-1\right|.\ \label{eq:flatnessFactor}
\end{eqnarray}

\begin{rem}
The flatness factor used in this work originated from \cite{LingBel13} and \cite{cong2}. It is a measure of the ``goodness" of the lattice Gaussian distribution in terms of its approximation capability to the capacity of a Gaussian test channel. Roughly speaking, the flatness factor represents how dense the underlying lattice is compared with the sampled continuous Gaussian distribution. For a given variance $\sigma^2$, the flatness factor is large when the underlying lattice is coarse, and one may scale it down to a desired level by using denser lattice.
\end{rem}

We define the discrete Gaussian distribution over $\Lambda$ centered at $c$ as the following discrete distribution taking values in $\lambda\in\Lambda$:
\begin{eqnarray}
D_{\Lambda,\sigma,c}(\lambda)=\frac{f_{\sigma,c}(\lambda)}{f_{\sigma,c}(\Lambda)},\;\forall\lambda\in\Lambda,
\end{eqnarray}
where $f_{\sigma,\mathrm{c}}(\Lambda)=\sum_{\lambda\in\Lambda}f_{\sigma,\mathrm{c}}(\lambda)$. For convenience, we write $D_{\Lambda,\sigma}=D_{\Lambda,\sigma,\mathrm{0}}$. Some recently developed algorithms to generate $D_{\Lambda,\sigma}$ for a general lattice can be found in \cite{ZhengWangTIT15} and \cite{ZhengWangTIT17}. In our work, we utilize the technique of source polarization to obtain $D_{\Lambda,\sigma}$ over polar lattices.

A sublattice $\Lambda'\subset\Lambda$ induces a partition (denoted by $\Lambda/\Lambda'$) of $\Lambda$ into equivalence groups modulo $\Lambda'$. The order of the partition is denoted by $|\Lambda/\Lambda'|$, which is equal to the number of the cosets. If $|\Lambda/\Lambda'|=2$, we call this a binary partition. Let $\Lambda(\Lambda_{0})/\Lambda_{1}/\cdots/\Lambda_{r-1}/\Lambda'(\Lambda_{r})$ for $r\geq1$ be an $n$-dimensional lattice partition chain. If only one level is applied ($r=1$), the construction is known as ``Construction
A\char`\"{}. If multiple levels are used, the construction is known as ``Construction D\char`\"{} \cite[p.232]{yellowbook}. For each partition $\Lambda_{\ell-1}/\Lambda_{\ell}$ ($1\leq\ell\leq r$) a code $C_{\ell}$ over $\Lambda_{\ell-1}/\Lambda_{\ell}$ selects a sequence of coset representatives $a_{\ell}$ in a set $A_{\ell}$
of representatives for the cosets of $\Lambda_{\ell}$. This construction requires a set of nested linear binary codes $C_{\ell}$ with block length $N$ and dimension of information bits $k_{\ell}$, which are represented as $[N,k_{\ell}]$ codes for $1\leq\ell\leq r$ and $C_{1}\subseteq C_{2}\cdot\cdot\cdot\subseteq C_{r}$. Let $\psi$ be the natural embedding of $\mathbb{F}_{2}^{N}$ into $\mathbb{Z}^{N}$, where $\mathbb{F}_{2}$ is the binary field. Consider
$\mathbf{g}_{1},\mathbf{g}_{2},\cdots,\mathbf{g}_{N}$ be a basis of $\mathbb{F}_{2}^{N}$ such that $\mathbf{g}_{1},\cdots\mathbf{g}_{k_{\ell}}$ span $C_{\ell}$. When $n=1$, the binary lattice $L$ of Construction D consists of
all vectors of the form
\begin{eqnarray}
\sum_{\ell=1}^{r}2^{\ell-1}\sum_{j=1}^{k_{\ell}}u_{\ell}^{j}\psi(\mathbf{g}_{j})+2^{r}z,\label{eqn:multilvl}
\end{eqnarray}
where $u_{\ell}^{j}\in\{0,1\}$ and $z\in\mathbb{Z}^{N}$.

%A well-known example of Construction D is the Barnes-Wall lattices constructed from Reed-Muller codes. Reed-Muller codes RM$(N,k,d)$ are a class of linear block codes over GF$(2)$, where $N$ is the length of the codeword, $k$ is the length of the information block and $d$ is the minimum Hamming distance. Conventionally, Reed-Muller codes are denoted by RM$(r',m)$ $(0\leq r'\leq m)$ with following relation among $N$, $k$ and $d$:
%\begin{eqnarray*}
%N=2^{m},k=1+\binom{m}{1}+\cdots+\binom{m}{r'},d=2^{m-r'}.\
%\end{eqnarray*}
%The $m$-th member of the family of Barnes-Wall lattices is a $2N$ dimensional real lattice. For example, the code formula of the $1024$-dimensional Barnes-Wall lattice is:
%\begin{eqnarray}
%BW_{1024}=\text{RM}(1,10)+2\text{RM}(3,10)+\cdot\cdot\cdot+2^{5}\mathbb{Z}^{1024}.\label{eqn:BW1024}
%\end{eqnarray}
%When Reed-Muller codes are replaced with polar codes in Construction D, we obtain polar lattices, which will be defined in the following section.

\subsection{Polar codes and polar lattices}
Polar codes are the first kind of codes that are proved to achieve the capacity of any binary memoryless symmetric (BMS) channel. Let $\tilde{W}(y|x)$ be a BMS channel with input alphabet $\mathcal{X}=\{0,1\}$ and output alphabet $\mathcal{Y}\subseteq\mathbb{R}$. Given the capacity $I(\tilde{W})$ of $\tilde{W}$ and any rate $R<I(\tilde{W})$, the information bits of a polar code with block length $N=2^{m}$ are indexed by a set of $RN$ rows of the generator matrix $G_{N}=\left[\begin{smallmatrix}1 & 0\\
1 & 1
\end{smallmatrix}\right]^{\otimes m}$, where $\otimes$ denotes the Kronecker product. Let $\tilde{W}^N: X^{1:N} \to Y^{1:N}$ denote the $N$ independent and identical copies of $\tilde{W}$. By the polarization transform $U^{1:N}=X^{1:N}G_N$, $\tilde{W}^N$ is converted to a so-called combined channel $\tilde{W}_N$ with input $U^{1:N}$ and output $Y^{1:N}$. Then, this combined channel can be successively split into $N$ BMS subchannels, denoted by $\tilde{W}_{N}^{(i)}$ with $1\leq i\leq N$. By channel polarization, the fraction of good (roughly error-free) subchannels is about $I(\tilde{W})$ as $m\rightarrow\infty$. Therefore, to achieve capacity, information bits should be sent over those good subchannels and the rest are fed with frozen bits which are known before transmission. The indices of good subchannels can be identified according to their
associated Bhattacharyya parameters.

\begin{deft}
{[}Bhattacharyya Parameter for Symmetric Channel \cite{arikan2009channel}{]}\label{deft:symZ}
Given a BMS channel $\tilde{W}$ with transition probability $P_{Y|X}$, the Bhattacharyya parameter $\tilde{Z}\in[0,1]$ is defined as
\begin{eqnarray*}
\tilde{Z}(\tilde{W}) & \triangleq\sum\limits _{y}\sqrt{P_{Y|X}(y|0)P_{Y|X}(y|1)}.\
\end{eqnarray*}
\end{deft}

Based on the Bhattacharyya parameter, the information set $\tilde{\mathcal{I}}$ is defined as $\{i:\tilde{Z}(\tilde{W}_{N}^{(i)})\leq2^{-N^{\beta}}\}$ for some constant $\beta<\frac{1}{2}$. The frozen set $\tilde{\mathcal{F}}$ is defined as the complement of $\tilde{\mathcal{I}}$. Let $P_{B}$ denote the block error probability of a polar code under the successive cancellation (SC) decoding. It can be upper-bounded as $P_{B}\leq\Sigma_{i\in\tilde{\mathcal{I}}}\tilde{Z}(\tilde{W}_{N}^{(i)})$ according to the analysis given in \cite{arikan2009channel}. The methods of \cite{Ido,polarconstruction} can be adapted to efficiently  evaluate the Bhattacharyya parameter of subchannels when
$\tilde{W}$ is a binary $\Lambda_{\ell-1}/\Lambda_{\ell}$ channel.

A polar lattice for the unconstrained Gaussian channel is constructed by using a set of nested polar codes as the component codes in \eqref{eqn:multilvl}. An explicit construction of AWGN-good polar lattices based on the multilevel approach of Forney \textit{et al.} \cite{forney6} has been presented in \cite{polarlatticeJ}. The key idea is to design a polar code to achieve the capacity for each level $\ell=1,2,...,r$ in Construction D.

To achieve the capacity of the power-constrained Gaussian channel, we need to apply Gaussian shaping over the AWGN-good polar lattice, which is difficult to implement directly. Motivated by \cite{LingBel13}, we apply Gaussian shaping to the top lattice $\Lambda$ instead. This shaping process generally leads to a nonuniform input distribution and an binary memoryless asymmetric (BMA) channel for each level. In this case, we need polar coding technique for asymmetric channels.

\begin{deft}
{[}Bhattacharyya Parameter for BMA Channel \cite{polarsource,aspolarcodes}{]}\label{deft:asymZ}
Let $W$ be a BMA channel with input $X\in\mathcal{X}=\{0,1\}$ and
output $Y\in\mathcal{Y}$, and let $P_{X}$ and $P_{Y|X}$ denote
the input distribution and channel transition probability, respectively.
The Bhattacharyya parameter $Z$ for channel $W$ is the defined as
\begin{eqnarray*}
Z(X|Y) & = & 2\sum\limits _{y}P_{Y}(y)\sqrt{P_{X|Y}(0|y)P_{X|Y}(1|y)}\ \\
 & = & 2\sum\limits _{y}\sqrt{P_{X,Y}(0,y)P_{X,Y}(1,y)}.
\end{eqnarray*}
Note that Definition \ref{deft:asymZ} is the same as Definition \ref{deft:symZ} when $P_{X}$ is uniform.
\end{deft}

Let $X^{1:N}$ and $Y^{1:N}$ be the input and output vector after $N$ independent uses of $W$. For a positive constant $\beta<\frac{1}{2}$, the following property of the polarized random variables $U^{1:N}=X^{1:N}G_{N}$ holds almost surely.
\begin{eqnarray}
\begin{cases}
\begin{aligned} & \left.\left.\lim_{N\rightarrow\infty}\frac{1}{N}\right\vert \left\{ i:Z(U^{i}|U^{1:i-1},Y^{1:N})\leq2^{-N^{\beta}}\text{ and }Z(U^{i}|U^{1:i-1})\geq1-2^{-N^{\beta}}\right\} \right\vert =I(X;Y),\\
 & \left.\left.\lim_{N\rightarrow\infty}\frac{1}{N}\right\vert \left\{ i:Z(U^{i}|U^{1:i-1},Y^{1:N})\geq2^{-N^{\beta}}\text{ or }Z(U^{i}|U^{1:i-1})\leq1-2^{-N^{\beta}}\right\} \right\vert =1-I(X;Y),
\end{aligned}
\end{cases}
\end{eqnarray}
which provides a method of achieving the capacity of a BMA channel. Moreover, the Bhattacharyya parameter of a BMA channel can be related to that of a BMS channel, and the decoding of a polar code for the BMA channel can also be converted to that for the BMS channel (see \cite{polarlatticeJ,aspolarcodes} for more details.)

\section{Polar Lattices for Quantization}\label{sec:qzgood}
Let $Y\sim N(0,\sigma_{s}^{2})$ denote a one-dimensional Gaussian source with zero mean and variance $\sigma_{s}^{2}$.
Let $Y^{1:N}$ ($\ubar{Y}$) be $N$ independent copies of $Y$ and $y^{1:N}$ ($\ubar{y}$) be a realization of $Y^{1:N}$. The PDF of $\ubar{Y}$ is given by $f_{\ubar{Y}}(\ubar{y})=f_{\sigma_{s}}(\ubar{y})$. For an $N$-dimensional polar lattice $L$ and its associated quantizer $Q_{L}(\cdot)$, the average distortion $\Delta$ per dimension after
quantization is given by
\begin{eqnarray}
\Delta=\frac{1}{N}\int_{\mathbb{R}^{N}}\parallel\ubar{y}-Q_{L}(\ubar{y})\parallel^{2}f_{\ubar{Y}}(\ubar{y})d\ubar{y}.
\end{eqnarray}

The normalized second moment (NSM) of a quantization lattice $L$ is defined as
\begin{eqnarray}
G(L)=\frac{\frac{1}{N}\int_{\mathcal{V}(L)}\|v\|^{2}dv}{V(L)^{1+2/N}},
\end{eqnarray}
where vector $v$ is uniformly distributed in $\mathcal{V}(L)$.
An $N$-dimensional lattice $L$ is called quantization-good \cite{ZamirQzNoise}, if
$\lim\limits _{N\rightarrow\infty}G(L)=\frac{1}{2\pi e}$.

In \cite{BK:Zamir}, an entropy-coded dithered quantization (ECDQ) scheme based on quantization-good lattices was proposed to achieve the rate-distortion bound \eqref{eqn:Eq1}. This scheme requires a pre-shared dither which is uniformly distributed on the Voronoi region of a quantization-good lattice and an entropy encoder after lattice
quantization. For our quantization scheme, we will show that the entropy encoder can be integrated in the lattice quantization process, which brings much convenience for practical application.

Our task is to construct a polar lattice which achieves the rate-distortion bound of the Gaussian source with reconstruction distribution $D_{\Lambda,\sqrt{\sigma_{s}^{2}-\Delta}}$. Firstly, we shall prove that the rate achieved by $D_{\Lambda,\sqrt{\sigma_{s}^{2}-\Delta}}$ can be arbitrarily close to $\frac{1}{2}\log\left(\frac{\sigma_{s}^{2}}{\Delta}\right)$. Note that the following theorem is essentially the same as \cite[Theorem
2]{LingBel13}. Here we just reexpress it in the source coding formulation.

\begin{figure}[ht]
\centering{} \includegraphics[width=10cm]{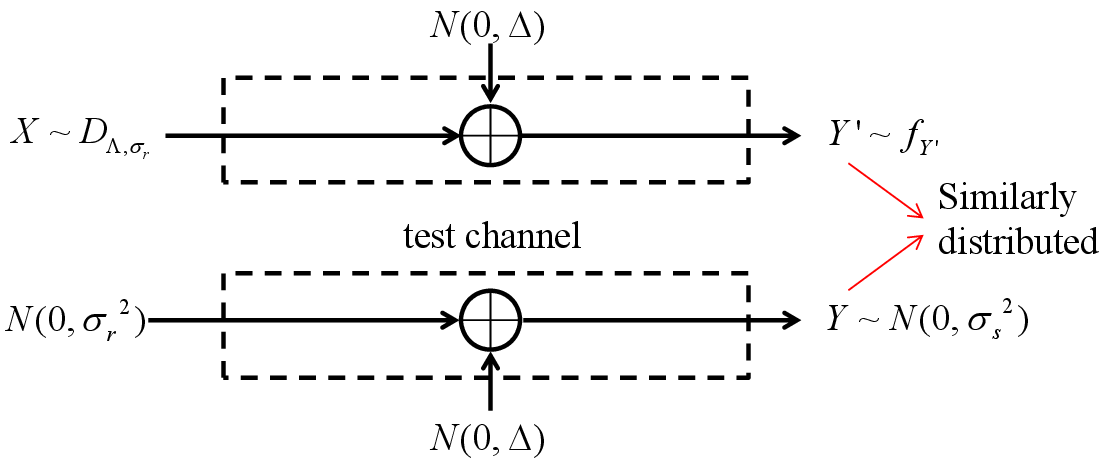} %\vspace{-1em}
 \caption{The relationship between the quantization of source $Y'$ and $Y$.}
\label{fig:testchannel}
\end{figure}

\begin{ther}
\label{theorem:lingDisGau}
Let $X$ denote a reconstruction variable which has a discrete Gaussian distribution $D_{\Lambda-\mathrm{c},\sigma_{r}}$ for arbitrary
$\mathrm{c}\in\mathbb{R}^{n}$, where $\sigma_{r}^{2}=\sigma_{s}^{2}-\Delta$. Consider an additive Gaussian test channel with input $X$ and output $Y'$ (see Fig. \ref{fig:testchannel}). The independent Gaussian noise has zero mean and variance $\Delta$. Let $\tilde{\sigma}_{\Delta}\triangleq\frac{\sigma_{r}\sqrt{\Delta}}{\sigma_{s}}$. Then, if $\epsilon=\epsilon_{\Lambda}(\tilde{\sigma}_{\Delta})<\frac{1}{2}$ and $\frac{\pi\epsilon_{t}}{1-\epsilon_{t}}<\epsilon$ where
\begin{equation}
\epsilon_{t}\triangleq\begin{cases}
\epsilon_{\Lambda}\left(\sigma_{r}/\sqrt{\frac{\pi}{\pi-t}}\right),\;\;\;\;\;\;\;\;\;\;\;\;\;\;\;\;\;\;\;\;\;\;t\geq1/e\\
(t^{-4}+1)\epsilon_{\Lambda}\left(\sigma_{r}/\sqrt{\frac{\pi}{\pi-t}}\right),\;\;\;\;\;\;\;\;\;0<t<1/e
\end{cases}
\end{equation}
the discrete Gaussian constellation results in mutual information $I(X;Y')\geq\frac{1}{2}\log\left(\frac{\sigma_{s}^{2}}{\Delta}\right)-\frac{6\epsilon}{n}\log(e)$
per channel use.
\end{ther}
The statement of Theorem \ref{theorem:lingDisGau} is non-asymptotic, i.e., it can hold even if $n=1$. Therefore, it is possible to construct a good polar lattice over one-dimensional lattice partition such as $\mathbb{Z}/2\mathbb{Z}/4\mathbb{Z}...$. The flatness factor $\epsilon$ of the top lattice $\Lambda$ can be made negligible by scaling this binary partition. This technique has already been used to construct AWGN capacity-achieving polar lattices \cite{polarlatticeJ,LiuL15w}.

In this paper, we often require the flatness factor $\epsilon_{\Lambda}(\tilde{\sigma})$ be exponentially small. The following proposition shows that this is possible with a partition of $r=O(\log N)$ levels.

\begin{prop}\label{prop:morelevel}
Given a one-dimensional binary partition chain $\Lambda/\Lambda_1/\cdots/\Lambda_{r-1}/\Lambda_r/\cdots$, let the quotient group $\Lambda_{\ell-1}/\Lambda_{\ell}$ be indexed by $X_{\ell}\in \{0,1\}$ for ${\ell}=1, 2, \ldots,r,\ldots$, such that the input $X \sim D_{\Lambda-\mathrm{c},\sigma_{r}}$ of the test channel are expressed uniquely by the binary sequence $X_1, X_2, \ldots, X_{r},\ldots$.  Then, for any $\tilde{\sigma}$, there exists $r=O(\log N)$ such that $\epsilon_{\Lambda}(\tilde{\sigma})=O\left(e^{-N}\right)$ and that using the first $r$ levels only incurs a capacity loss $\sum_{\ell>r}I(Y';X_{\ell}|X_{1:\ell-1})=O\left(e^{-N}\right)$ where $Y'$ denotes the output of a test channel. Then, the rate of convergence to the rate-distortion bound is $O\left(e^{-N}\right)$ by using $r$ partition levels.
\end{prop}

\begin{IEEEproof}
Since the partition is with dimension one, we can assume that $\Lambda=\eta \mathbb{Z}$ for some scaling $\eta$. Let $\Lambda^*=\frac{1}{\eta}\mathbb{Z}$ be the dual lattice of $\Lambda$. By \cite[Corollary 1]{cong2}, using the definition of theta series $\Theta_{\Lambda}(q)=\sum_{\lambda\in\Lambda}q^{\|\lambda\|}$, we have
\allowdisplaybreaks{\begin{eqnarray}
\epsilon_{\Lambda}(\tilde{\sigma})&=&\Theta_{\Lambda^*}\left(2\pi \tilde{\sigma}^2\right)-1 \\
&=&\sum_{\lambda \in \Lambda^*}\exp\left(-2\pi^2\tilde{\sigma}^2\|\lambda\|^2 \right)-1\\
&=&2\sum_{\lambda \in \frac{1}{\eta}\mathbb{Z}_+} \exp\left(-2\pi^2\tilde{\sigma}^2\|\lambda\|^2 \right) \\
&\leq&\frac{2\exp\left(-2 \pi^2\tilde{\sigma}^2\frac{1}{\eta^2}\right)}{1-\exp\left(-2 \pi^2\tilde{\sigma}^2\frac{3}{\eta^2}\right)}\\
&\leq&4\exp\left(-2 \pi^2\tilde{\sigma}^2\frac{1}{\eta^2}\right),
\end{eqnarray}
where $\mathbb{Z}_{+}$ denotes positive integers and the last inequality satisfies for sufficiently small $\eta$.} Let $\frac{1}{\eta^2}=O(N)$ so that $\epsilon_{\Lambda}(\tilde{\sigma})=O\left(e^{-N}\right)$. According to \cite[Lemma 5]{polarlatticeJ},\footnote{A complete proof of this Lemma can be found in Appendix C of the arXiv version of \cite{polarlatticeJ}:	 http://arxiv.org/abs/1411.0187.} the partition chain $\mathbb{Z}/\cdots/2^{r_1}\mathbb{Z}$ with level $r_1=O(\log N)$ can guarantee a capacity loss $\sum_{\ell>r_1}I(Y';X_{\ell}|X_{1:\ell-1})= O\left(e^{-N}\right)$. Finally, the number of levels for partition $\eta\mathbb{Z}/\cdots/\mathbb{Z}/\cdots/2^{r_1}\mathbb{Z}$ satisfies $r=\log(\frac{2^{r_1}}{\eta})=O(\log N)$. Combining this with Theorem \ref{theorem:lingDisGau}, it can be found that the rate of convergence to the rate-distortion bound is $O(e^{-N})$ by using $r$ partition levels.
\end{IEEEproof}

\begin{rem}
Compared with a related work \cite{PolarCodesGaussianLossy}, where two approaches were proposed to form the reconstruction distribution of the Gaussian source. The first one was based on non-uniform scalar quantization and the second was based on the Central Limit Theorem. We note that the rates of convergence for the two approaches are $O\left(\frac{\log N}{N}\right)$ and $O\left(\frac{1}{\log N}\right)$, respectively.
\end{rem}

Note that when the test channel is chosen to be an AWGN channel with noise variance $\Delta$ and the reconstruction alphabet is discrete Gaussian distributed, the distribution of $Y'$ is not exactly a continuous Gaussian distribution. In fact, it is a distribution obtained by adding a continuous Gaussian of variance $\Delta$ to a discrete Gaussian
$D_{\Lambda-\mathrm{c},\sigma_{r}}$, which is expressed as the following convolution
\begin{equation}
f_{Y'}(y')=\frac{1}{f_{\sigma_{r}}(\Lambda-c)}\sum_{\lambda\in\Lambda-c}f_{\sigma_{r}}(\lambda)f_{\sigma}(y'-\lambda),\ \ \ \ y'\in\mathbb{R}^{n},\label{eqn:Y'densi}
\end{equation}
where $\sigma=\sqrt{\Delta}$. For simplicity, we only consider one-dimensional binary partition chain ($n=1$) in this work and hence $Y'$ is also a one-dimensional source. We note that one may follow the same lines and generalize our scheme to multi-dimensional partitions with $n\geq 2$. A design example of $n=2$ can be found in \cite{MultiCmpdAntonio}.

Therefore, we are actually quantizing source $Y'$ instead of $Y$ using the discrete-Gaussian distributed variable $X$. However, when the flatness factor $\epsilon_{\Lambda}(\tilde{\sigma}_{\Delta})$ is small, a good quantizer constructed from polar lattices for the source $Y'$ is also good for source $Y$ because of the following lemma. The relationship
between the quantization of source $Y'$ and $Y$ is shown in Fig. \ref{fig:testchannel}.

\begin{lem}[{\cite[Corollary 1]{LingBel13}}]\label{lem:YY'distance} If $\epsilon=\epsilon_{\Lambda}(\tilde{\sigma}_{\Delta})<\frac{1}{2}$, the variational distance between the density $f_{Y'}$ of source $Y'$ defined in \eqref{eqn:Y'densi} and the Gaussian density $f_{Y}$ satisfies $\mathbb{V}(f_{Y'},f_{Y})\leq2\epsilon$.
\end{lem}

Now we construct polar lattices for quantization. Consider the quantization of source $Y'$ using the reconstruction distribution $D_{\Lambda,\sigma_{r}}$. Since binary partition is used, $X$ can be represented by a binary string $X_{1:r}$, and we have $\lim_{r\rightarrow\infty}P_{X_{1:r}}=P_{X}=D_{\Lambda,\sigma_{r}}$. Because the polar lattices are constructed by ``Construction D\char`\"{}, we are interested in the test channel on each level. Similarly to the setting of shaping for AWGN-good polar lattices \cite{polarlatticeJ}, given the previous $x_{1:\ell-1}$ and the coset $\mathcal{A}_{\ell}$
determined by $x_{1:\ell}$, the channel transition PDF at level $\ell$ is
\begin{eqnarray}
\begin{aligned}\label{eqn:transition1}P_{Y'|X_{\ell},X_{1:\ell-1}} & (y'|x_{\ell},x_{1:\ell-1})\\
 & =\frac{\sum_{a\in\mathcal{A}_{\ell}(x_{1:\ell})}P(a)P_{Y'|A}(y'|a)}{{P\{\mathcal{A}_{\ell}(x_{1:\ell})\}}}\\
 & =\text{exp}\left(-\frac{y'^{2}}{2(\sigma_{s}^{2}+\Delta)}\right)\frac{1}{f_{\sigma_{s}}(\mathcal{A}_{\ell}(x_{1:\ell}))}\frac{1}{2\pi\sqrt{\Delta}\sigma_{s}}\sum_{a\in\mathcal{A}_{\ell}(x_{1:\ell})}\text{exp}\left(-\frac{1}{2\tilde{\sigma}_{\Delta}^{2}}\left(|\alpha y'-a|^{2}\right)\right),
\end{aligned}
\end{eqnarray}
where $\alpha=\frac{\sigma_{r}^{2}}{\sigma_{r}^{2}+\Delta}$ is the MMSE coefficient and $\tilde{\sigma}_{\Delta}=\frac{\sigma_{r}\sqrt{\Delta}}{\sigma_{s}}$. Consequently, using $D_{\Lambda,\sigma_{r}}$ as the constellation, the $\ell$-th channel is generally asymmetric with the input distribution $P_{X_{\ell}|X_{1:\ell-1}}$ $(\ell\leq r)$, which can be calculated according to the definition of $D_{\Lambda,\sigma_{r}}$.

The lattice quantization can be viewed as lossy compression for all binary-input test channels from level $1$ to $r$. Here we start with the first level. Let $y'^{1:N}$ denote the realization of $N$ i.i.d copies of source $Y'$. Although $Y'$ is a continuous source with PDF given by \eqref{eqn:Y'densi}, from
now on we will express the distortion measure as well as the variational distance in the form of summation instead of integration, to keep the notations consistent (the Bhattacharyya parameter in \cite{arikan2009channel} was defined as a summation).

Since the test channel on each level is not necessarily symmetric and the reconstruction constellation is not uniformly distributed, we have to consider the lossy compression for nonuniform source and asymmetric distortion measure \cite{aspolarcodes}. The solution turns out to be similar to the construction of capacity-achieving polar codes for asymmetric channels.

For the first level, let $U_{1}^{1:N}=X_{1}^{1:N}G_{N}$, where $G_{N}$ is the $N\times N$ generator matrix of polar codes. We define the information set $\mathcal{I}_{1}$, frozen set $\mathcal{F}_{1}$ and shaping set $\mathcal{S}_{1}$ based on the Bhattacharyya parameter as follows:
\begin{eqnarray}
 &  & \hspace{-2em}\begin{cases}
\begin{aligned} & \mathcal{F}_{1}=\left\{i\in[N]:Z\left(U_{1}^{i}|U_{1}^{1:i-1},Y'^{1:N}\right)\geq1-2^{-N^{\beta}}\right\}\\
 & \mathcal{I}_{1}=\left\{i\in[N]:Z\left(U_{1}^{i}|U_{1}^{1:i-1}\right)>2^{-N^{\beta}}\text{ and }Z\left(U_{1}^{i}|U_{1}^{1:i-1},Y'^{1:N}\right)<1-2^{-N^{\beta}}\right\}\\
 & \mathcal{S}_{1}=\left\{i\in[N]:Z\left(U_{1}^{i}|U_{1}^{1:i-1}\right)\leq2^{-N^{\beta}}\right\}.
\end{aligned}
\end{cases}\label{eqn:asymdefinition1}
\end{eqnarray}
Note the subtle difference from the definition for channel coding in \cite{polarlatticeJ}. The difference is that $\mathcal{I}_{1}$ is made slightly larger to guarantee the desired distortion. We can apply channel symmetrization \cite[Lemma 7]{polarlatticeJ} to the first test channel $X_1 \to Y'$ and obtain a symmetrized channel $\tilde{X}_1 \to (\tilde{X}_1\oplus X_1,Y')$, where $\tilde{X}_1$ is a uniform binary random variable independent of $X_1$. Let $\tilde{U}_1^{1:N}=\tilde{X}_1^{1:N}G_N$. Then, the asymmetric Bhattacharyya parameter $Z(U_{1}^{i}|U_{1}^{1:i-1},Y'^{1:N})$ and $Z(U_{1}^{i}|U_{1}^{1:i-1})$ can be efficiently calculated from symmetric Bhattacharyya parameter $\tilde{Z}(\tilde{U}_{1}^{i}|\tilde{U}_{1}^{1:i-1},X_{1}^{1:N}\oplus\tilde{X}_{1}^{1:N},Y'^{1:N})$ and $\tilde{Z}(\tilde{U}_{1}^{i}|\tilde{U}_{1}^{1:i-1},X_{1}^{1:N}\oplus\tilde{X}_{1}^{1:N})$, respectively \cite[Theorem 3]{polarlatticeJ}. The proportion of set $\mathcal{I}_{1}$ approaches $I(X_{1};Y')$ when $N\rightarrow\infty$ from the above encoding method.

After obtaining sets $\mathcal{F}_{1}$, $\mathcal{I}_{1}$ and $\mathcal{S}_{1}$, the encoder determines $u_{1}^{1:N}$ according to the following rule:
\begin{eqnarray}
\hspace{-3em} &  & u_{1}^{i}=\begin{cases}
\begin{aligned} & 0\,\,\text{w. p.}\,\,\,P_{U_{1}^{i}|U_{1}^{1:i-1},Y'^{1:N}}\left(0|u_{1}^{1:i-1},y'^{1:N}\right)\\
 & 1\,\,\text{w. p.}\,\,\,P_{U_{1}^{i}|U_{1}^{1:i-1},Y'^{1:N}}\left(1|u_{1}^{1:i-1},y'^{1:N}\right)
\end{aligned}
\,\,\,\text{if}\,\,\,i\in\mathcal{I}_{1},\end{cases}\label{eqn:lossyencoder1}
\end{eqnarray}
and
\begin{eqnarray}
\hspace{-2em} &  & u_{1}^{i}=\begin{cases}
\begin{aligned} & \bar{u}_{1}^{i}\,\,\,\,\,\text{if}\,\,\,\,\,i\in\mathcal{F}_{1}\\
 & \argmax_{u}P_{U_{1}^{i}|U_{1}^{1:i-1}}\left(u|u_{1}^{1:i-1}\right)\,\,\,\,\,\text{if}\,\,\,\,\,i\in\mathcal{S}_{1}.
\end{aligned}
\end{cases}\label{eqn:lossyencoder2}
\end{eqnarray}
Here $\bar{u}_{1}^{i}$ is a uniformly random bit generated before lossy compression. The output of the encoder at level $1$ is $u_{1}^{\mathcal{I}_{1}}=\{u_{1}^{i},i\in\mathcal{I}_{1}\}$. To reconstruct $x_{1}^{1:N}$, the decoder uses the shared $u_{1}^{\mathcal{F}_{1}}$ and the received $u_{1}^{\mathcal{I}_{1}}$ to recover $u_{1}^{\mathcal{S}_{1}}$ according to $\underset{u}{\argmax}\;P_{U_{1}^{i}|U_{1}^{1:i-1}}\left(u|u_{1}^{1:i-1}\right)$
and then obtain $x_{1}^{1:N}=u_{1}^{1:N}G_{N}$.\footnote{Since $G_N=G_N^{-1}$\cite{arikan2009channel}, the relation $u_{1}^{1:N}=x_{1}^{1:N}G_{N}$ also holds.} The probability $P_{U_{1}^{i}|U_{1}^{1:i-1}}$ and $P_{U_{1}^{i}|U_{1}^{1:i-1},Y'^{1:N}}$ can both be calculated efficiently by the successive cancellation
algorithm with complexity $O(N\log N)$ \cite{polarlatticeJ}.

\begin{ther} \label{Thm:Smalldistance1}
Let $Q_{U_{1}^{1:N},Y'^{1:N}}\left(u_{1}^{1:N},y'^{1:N}\right)$ denote the resulted joint distribution for $U_{1}^{1:N}$ and ${Y'}^{1:N}$
according to the encoding rule described in \eqref{eqn:lossyencoder1} and \eqref{eqn:lossyencoder2}. Consider another encoder using the encoding rule \eqref{eqn:lossyencoder1} for all $i\in[N]$ and let
$P_{U_{1}^{1:N},Y'^{1:N}}\left(u_{1}^{1:N},y'^{1:N}\right)$ denote the resulted joint distribution for $U_{1}^{1:N}$ and ${Y'}^{1:N}$ after using rule \eqref{eqn:lossyencoder1}. For any $\beta'<\beta<1/2$ satisfying \eqref{eqn:asymdefinition1} and $R_{1}=\frac{|\mathcal{I}_{1}|}{N}>I(X_{1};Y')$,
\begin{eqnarray}
\mathbb{V}\left(P_{U_{1}^{1:N},Y'^{1:N}},Q_{U_{1}^{1:N},Y'^{1:N}}\right)=O\left(2^{-N^{\beta'}}\right).
\end{eqnarray}
\end{ther}
The same statement has been given in \cite{aspolarcodes} without proof. Here we prove the theorem in Appendix \ref{appendix1} for completeness.

Now we introduce the construction for higher levels. Taking the second level as an example, to make up the reconstruction constellation distribution, the input distribution at level 2 should be $P_{X_{2}|X_{1}}$. Based
on the quantization results $(U_{1}^{1:N},Y'^{1:N})$ given by the encoder at level $1$, some $U_{2}^{i}$$\left(U_{2}^{1:N}=X_{2}^{1:N}G_{N}\right)$ is almost deterministic given $\left(U_{2}^{1:i-1},U_{1}^{1:N}\right)$. Since there is a one-to-one mapping between $X_{1}^{1:N}$ and $U_{1}^{1:N}$, conditioning on $\left(U_{2}^{1:i-1},U_{1}^{1:N}\right)$ is the same as conditioning on $\left(U_{2}^{1:i-1},X_{1}^{1:N}\right)$. We define the information set $\mathcal{I}_{2}$, frozen set $\mathcal{F}_{2}$
and shaping set $\mathcal{S}_{2}$ as follows:
\begin{eqnarray}
 &  & \hspace{-2em}\begin{cases}
\begin{aligned} & \mathcal{F}_{2}=\left\{i\in[N]:Z\left(U_{2}^{i}|U_{2}^{1:i-1},X_{1}^{1:N},Y'^{1:N}\right)\geq1-2^{-N^{\beta}}\right\}\\
 & \mathcal{I}_{2}=\left\{i\in[N]:Z\left(U_{2}^{i}|U_{2}^{1:i-1},X_{1}^{1:N}\right)>2^{-N^{\beta}}\text{ and} Z\left(U_{2}^{i}|U_{2}^{1:i-1},X_{1}^{1:N},Y'^{1:N}\right)<1-2^{-N^{\beta}}\right\}\\
 & \mathcal{S}_{2}=\left\{i\in[N]:Z\left(U_{2}^{i}|U_{2}^{1:i-1},X_{1}^{1:N}\right)\leq2^{-N^{\beta}}\right\}.
\end{aligned}
\end{cases}\label{eqn:asymdefinition2}
\end{eqnarray}
The proportion of $\mathcal{I}_{2}$ approaches $I(X_{2};Y|X_{1})$ when $N$ is sufficiently large \cite[Lemma 9]{polarlatticeJ}. For a given source sequence pair $\left(u_{1}^{1:N},y'^{1:N}\right)$ or $\left(x_{1}^{1:N},y'^{1:N}\right)$,
the encoder at level 2 determines $u_{2}^{1:N}$ according to the following rule:
\begin{eqnarray}
\hspace{-2em} &  & u_{2}^{i}=\begin{cases}
\begin{aligned} & 0\,\,\,\text{w.p.}\,\,\,P_{U_{2}^{i}|U_{2}^{1:i-1},X_{1}^{1:N},Y'^{1:N}}\left(0|u_{2}^{1:i-1},x_{1}^{1:N},y'^{1:N}\right)\\
 & 1\,\,\,\text{w.p.}\,\,\,P_{U_{2}^{i}|U_{2}^{1:i-1},X_{1}^{1:N},Y'^{1:N}}\left(1|u_{2}^{1:i-1},x_{1}^{1:N},y'^{1:N}\right)
\end{aligned}
\end{cases}\,\,\,\text{if}\,\,\,i\in\mathcal{I}_{2},\label{eqn:lossyencoder4}
\end{eqnarray}
and
\begin{eqnarray}
\hspace{-2em} &  & u_{2}^{i}=\begin{cases}
\begin{aligned} & \bar{u}_{2}^{i}\,\,\,\,\,\text{if}\,\,\,\,\,i\in\mathcal{F}_{2}\\
 & \argmax_{u}P_{U_{2}^{i}|U_{2}^{1:i-1},X_{1}^{1:N}}\left(u|u_{2}^{1:i-1},x_{1}^{1:N}\right)\,\,\,\,\,\text{if}\,\,\,\,\,i\in\mathcal{S}_{2}.
\end{aligned}
\end{cases}\label{eqn:lossyencoder5}
\end{eqnarray}
We thus extend Theorem \ref{Thm:Smalldistance1} to the second level.
\begin{ther}\label{Thm:Smalldistance2}
Let $Q_{U_{2}^{1:N},U_{1}^{1:N},Y'^{1:N}}\left(u_{2}^{1:N},u_{1}^{1:N},y'^{1:N}\right)$ denote the resulted joint distribution of $U_{2}^{1:N}$ and $\left(U_{1}^{1:N},Y'^{1:N}\right)$ according to the encoding rule described in \eqref{eqn:lossyencoder4}
and \eqref{eqn:lossyencoder5}. Consider another encoder using the encoding rule \eqref{eqn:lossyencoder4} for all $i\in[N]$ and let $P_{U_{2}^{1:N},U_{1}^{1:N},Y'^{1:N}}\left(u_{2}^{1:N},u_{1}^{1:N},y'^{1:N}\right)$ denote the resulted joint distribution of $U_{2}^{1:N}$ and $\left(U_{1}^{1:N},Y'^{1:N}\right)$ after using rule \eqref{eqn:lossyencoder4}. For any $\beta'<\beta<1/2$ satisfying \eqref{eqn:asymdefinition2} and $R_{2}=\frac{|\mathcal{I}_{2}|}{N}>I(X_{2};Y'|X_{1})$,
\begin{eqnarray}
\mathbb{V}\left(P_{U_{2}^{1:N},U_{1}^{1:N},Y'^{1:N}},Q_{U_{2}^{1:N},U_{1}^{1:N},Y'^{1:N}}\right)=O\left(2^{-N^{\beta'}}\right).
\end{eqnarray}
\end{ther}

Note that Theorem \ref{Thm:Smalldistance2} is based on the assumption that $\mathbb{V}(P_{U_{1}^{1:N},Y'^{1:N}},Q_{U_{1}^{1:N},Y'^{1:N}})=O\left(2^{-N^{\beta'}}\right)$, which means that we also need $R_{1}>I(X_{1};Y')$. Therefore, we have $\sum_{i=1}^{2}R_{i}>I(X_{1}X_{2};Y')$.

By induction, for level $\ell$ ($\ell\leq r$), we define the three sets $\mathcal{F}_{\ell}$, $\mathcal{I}_{\ell}$ and $\mathcal{S}_{\ell}$ in the same form as \eqref{eqn:asymdefinition2} with $X_{1:\ell-1}^{1:N}$ replacing $X_{1}^{1:N}$ and $U_{\ell}$ replacing $U_{2}$. Similarly, the encoder determines $u_{\ell}^{1:N}\left(u_{\ell}^{1:N}=x_{\ell}^{1:N}G_{N}\right)$ according to the rule given by \eqref{eqn:lossyencoder4} and \eqref{eqn:lossyencoder5}, with $X_{1:\ell-1}^{1:N}$ and $x_{1:\ell-1}^{1:N}$ replacing $X_{1}^{1:N}$ and $x_{1}^{1:N}$, respectively. Let $Q_{U_{1:\ell}^{1:N},Y'^{1:N}}\left(u_{1:\ell}^{1:N},y'^{1:N}\right)$ denote the associate joint distribution resulted from this encoder and $P_{U_{1:\ell}^{1:N},Y'^{1:N}}(u_{1:\ell}^{1:N},y'^{1:N})$ denote the one that resulted from an encoder only using \eqref{eqn:lossyencoder4} for all $i\in[N]$. We have $\mathbb{V}\left(P_{U_{1:\ell}^{1:N},Y'^{1:N}},Q_{U_{1:\ell}^{1:N},Y'^{1:N}}\right)=O\left(\ell\cdot2^{-N^{\beta'}}\right)$ for any rate $R_{\ell}=\frac{|\mathcal{I}_{\ell}|}{N}>I(X_{\ell};Y'|X_{1:\ell-1})$. Specifically, at level $r$, for any rate $R_{r}>I(X_{r};Y'|X_{1:r-1})$ and $\sum_{i=1}^{r}R_{i}>I(X_{1:r};Y')$, we have
\begin{eqnarray}
\mathbb{V}\left(P_{U_{1:r}^{1:N},Y'^{1:N}},Q_{U_{1:r}^{1:N},Y'^{1:N}}\right)=O\left(r\cdot2^{-N^{\beta'}}\right).\label{eqn:variantialbound}
\end{eqnarray}
By \cite{polarlatticeJ}, $I(X_{1:r};Y')$ can be arbitrarily close to $I(X;Y')$ as $N\to\infty$ if $r=O(\log\log N)$. Here, to achieve sub-exponentially decaying excess distortion, we further require $r=O(\log N)$ (see Proposition \ref{prop:morelevel}), which gives $\mathbb{V}\left(P_{U_{1:r}^{1:N},Y'^{1:N}},Q_{U_{1:r}^{1:N},Y'^{1:N}}\right)=O\left(2^{-N^{\beta''}}\right)$ for $0<\beta''<\beta'<1/2$.

As a result, the source vector $y^{1:N}$ is eventually compressed to $u_\ell^{\mathcal{I}_\ell}$ for $1\leq \ell \leq r$. We also note that $u_\ell^{\mathcal{F}_\ell}$ ($1\leq \ell \leq r$) can be randomly generated and pre-shared between the source encoder and the source decoder. For the reconstruction, the bits $u_\ell^{\mathcal{S}_\ell}$ are determined by $u_\ell^{\mathcal{F}_\ell}$ and $U_\ell^{\mathcal{I}_\ell}$ according to the distribution $P_{X_\ell|X_{1:\ell-1}}$. After obtaining $u^{1:N}_\ell$ for $1 \leq \ell \leq r$, the realization $x^{1:N}$ of $X^{1:N}$ can be recovered from $u_\ell^{1:N}$ according to the following equation
\begin{eqnarray}\label{eqn:uN2xN}
\chi = \sum_{\ell=1}^{r}2^{\ell-1}\left[\sum_{i\in\mathcal{I}_{\ell}}u_{\ell}^{i}\psi(\mathbf{g}_{i})+\sum_{i\in\mathcal{S}_{\ell}}u_{\ell}^{i}\psi(\mathbf{g}_{i})+\sum_{i\in\mathcal{F}_{\ell}}u_{\ell}^{i}\psi(\mathbf{g}_{i})\right],
\label{constructionD-finite-power}
\end{eqnarray}
where $\mathbf{g}_{i}$ denotes the $i$-th row of the polarization matrix $G_N$ and $\psi$ is the natural embedding. Clearly, $x^{1:N}$ is drawn from $D_{2^r\mathbb{Z}^N+\chi,\sigma_a}$. For each dimension, when $r$ is sufficiently large, the probability of choosing a constellation point outside the interval $[-2^{r-1},2^{r-1})$ is negligible. Therefore, there exists only one point within $[-2^{r-1},2^{r-1})$ with probability close to $1$ and $x^{1:N}$ can be approximated by $\chi \mod 2^r$.

Now we present the main theorem of this section. The proof is given in Appendix \ref{appendix3}.

\begin{ther}\label{Thm:quantizationMain}
Given a Gaussian source $Y$ with variance $\sigma_{s}^{2}$ and an average distortion $\Delta\leq\sigma_{s}^{2}$, there exists a multilevel polar code with sum rate $R>\frac{1}{2}\log\left(\frac{\sigma_{s}^{2}}{\Delta}\right)$ and number of levels $r=O(\log N)$ such that the expected distortion \footnote{The expectation for the average distortion is computed over distribution of source sequences with the codebook held fixed.} is arbitrarily close to $\Delta$ as $N\to\infty$.
\end{ther}

\begin{rem}
The proof shows that this multilevel polar code is actually a shifted polar lattice $L+\mathrm{c}$ for certain shift $\mathrm{c}$, constructed from lattice partition $\Lambda/\Lambda_{1}/\cdots/\Lambda_{r-1}/\Lambda_{r}$ with discrete Gaussian distribution $D_{\Lambda,\sigma_{r}}$, where $\sigma_{r}=\sqrt{\sigma_{s}^{2}-\Delta}$ and the partition chain is scaled such that $\epsilon_{\Lambda}\left(\frac{\sigma_{r}\sqrt{D}}{\sigma_{s}}\right)\to0$ sub-exponentially fast in $N$. Meanwhile, the number of level can be reduced to $r=O(\log \log N)$ if we do not require sub-exponentially fast convergence of the distortion.
\end{rem}

%\begin{rem}\label{rem:Qzrmk}
%From the proof of Theorem \ref{Thm:quantizationMain}, it seems that $R$ could be slightly smaller than $\frac{1}{2}\log\frac{\sigma_{s}^{2}}{\Delta}$ (since $R>I(X;Y')\geq\frac{1}{2}\log\frac{\sigma_{s}^{2}}{\Delta}-\frac{5\epsilon_{\Lambda}(\tilde{\sigma}_{\Delta})}{n}\log(e)$)
%to reach an average distortion $\Delta$, which would contradict Shannon's rate-distortion theory. However, this is not the case. When $R<\frac{1}{2}\log\frac{\sigma_{s}^{2}}{\Delta}$, an arbitrarily small $\epsilon_{\Lambda}(\tilde{\sigma}_{\Delta})$ cannot be guaranteed, which means that the resulted distortion cannot
%be arbitrarily close to $\Delta$. To achieve the desired distortion, we need $R>\frac{1}{2}\log\frac{\sigma_{s}^{2}}{\Delta}-\frac{5\epsilon_{\Lambda}(\tilde{\sigma}_{\Delta})}{n}\log(e)$
%for all possibly small $\epsilon_{\Lambda}(\tilde{\sigma}_{\Delta})$, which leads to $R>\frac{1}{2}\log\frac{\sigma_{s}^{2}}{\Delta}$.
%\end{rem}

\section{Simulation Results}\label{sec:simu}
In this section, we compress a Gaussian source with standard deviation $\sigma_{s}=3$ and target distortion from $0.1$ to $2.5$. The number of levels is chosen to be $r=6$ to guarantee a negligible variational distance $\mathbb{V}(f_{Y'},f_{Y})$ for all target distortions. For the construction of polar codes for each level, we employ the method proposed by Tal and Vardy in \cite{tal2011construct} to evaluate the Bhattacharyya parameters for the three sets $\mathcal{F}_\ell$, $\mathcal{I}_\ell$ and $\mathcal{S}_\ell$. Thanks to the idea of channel symmetrization, the Bhattacharyya parameters (e.g., $Z\left(U_1^i|U_1^{1:i-1},Y'^{1:N}\right)$ and $Z\left(U_1^i|U_1^{1:i-1}\right)$ in \eqref{eqn:asymdefinition1}, $Z\left(U_1^i|U_1^{1:i-1},X_1^{1:N},Y'^{1:N}\right)$ and $Z\left(U_1^i|U_1^{1:i-1}, X_1^{1:N}\right)$ in \eqref{eqn:asymdefinition2} are calculated according to $P_{Y'|X_{\ell},X_{1:\ell-1}}$ and $P_{X_{\ell}|X_{1:\ell-1}}$, using the merging method for binary-input symmetric channel with continuous output described in \cite[Sect. VI]{tal2011construct}. For the lossy compressor at each level, the decision probabilities $P_{U_{\ell}^{i}|U_{\ell}^{1:i-1},X_{1:\ell}^{1:N}, Y'^{1:N}}$ and $P_{U_{\ell}^{i}|U_{\ell}^{1:i-1},X_{1:\ell}^{1:N}}$ for the set $\mathcal{I}_\ell$ and $\mathcal{S}_\ell$ are calculated by the standard SC algorithm, which has complexity $O(N\log N)$.

\begin{figure}[ht]
    \centering
    \includegraphics[width=12cm]{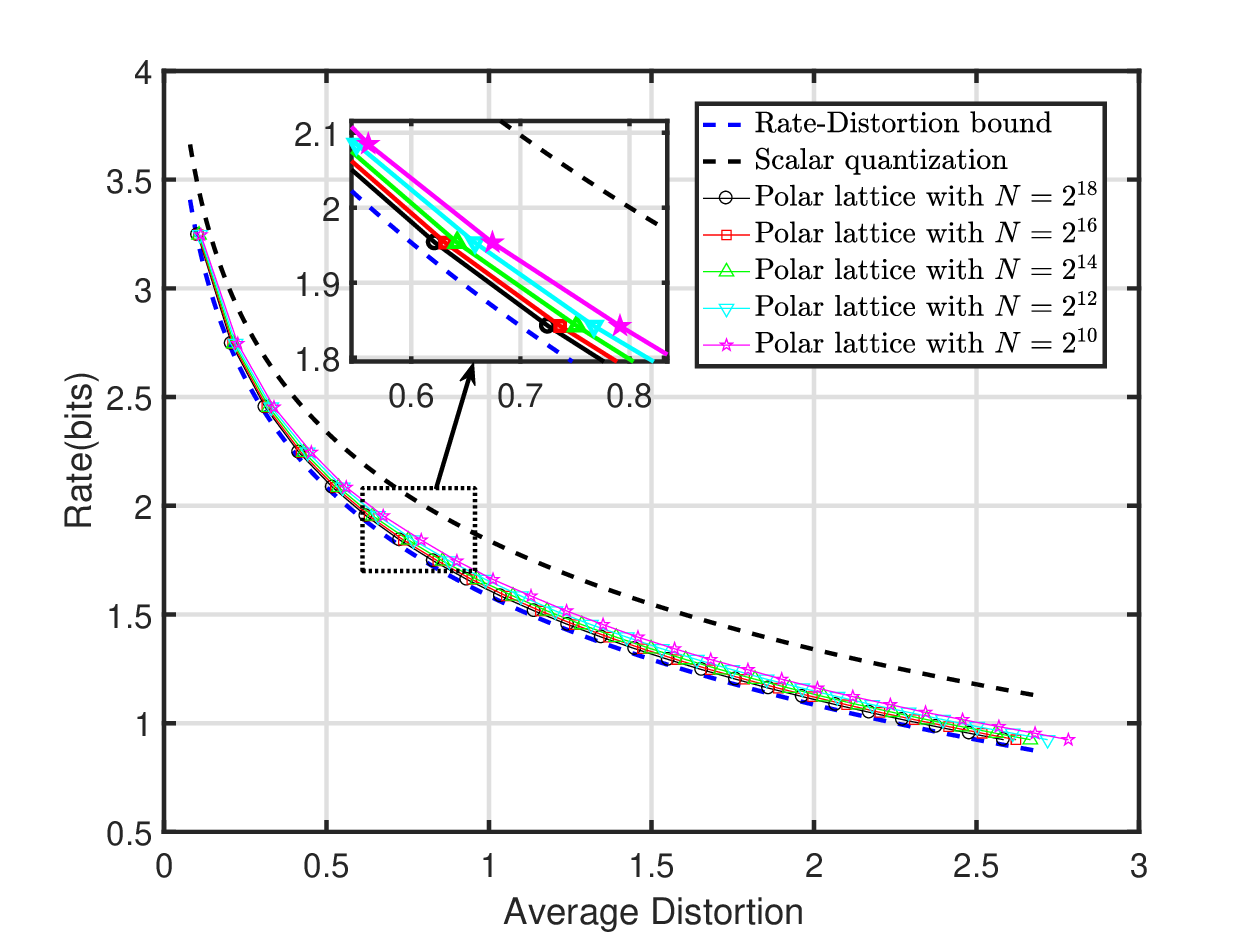}
    \caption{Quantization performance of polar lattices for the Gaussian source with $\sigma_s=3$.}
    \label{fig:RDbound}
\end{figure}

The quantization performance of polar lattices is shown in Fig. \ref{fig:RDbound}, where the expected average distortion is obtained after 1000 simulation rounds. It clearly shows that the rate-distortion bound is approached as the dimension of polar lattices increases from $N=2^{10}$ to $N=2^{18}$. Particularly, when $N=2^{18}$, the gap to the rate-distortion bound is less than $0.2$ dB. \footnote{The gap to the rate-distortion bound is calculated by $10\log_{10}\left(\frac{\Delta_{real}}{\Delta}\right)$ for a same compression rate $R=\log\left(\frac{\sigma_s^2}{\Delta}\right)$, where $\Delta_{real}$ denotes the realized distortion at rate $R$.}

\begin{figure*}[ht]
    \centering
        \subfigure[Comparison of $f_{Y'}$ and $f_{Y}$ when $r=6$ and $\Delta=0.5$.]{%
            \label{fig:YY'a}
            \includegraphics[width=0.45\textwidth]{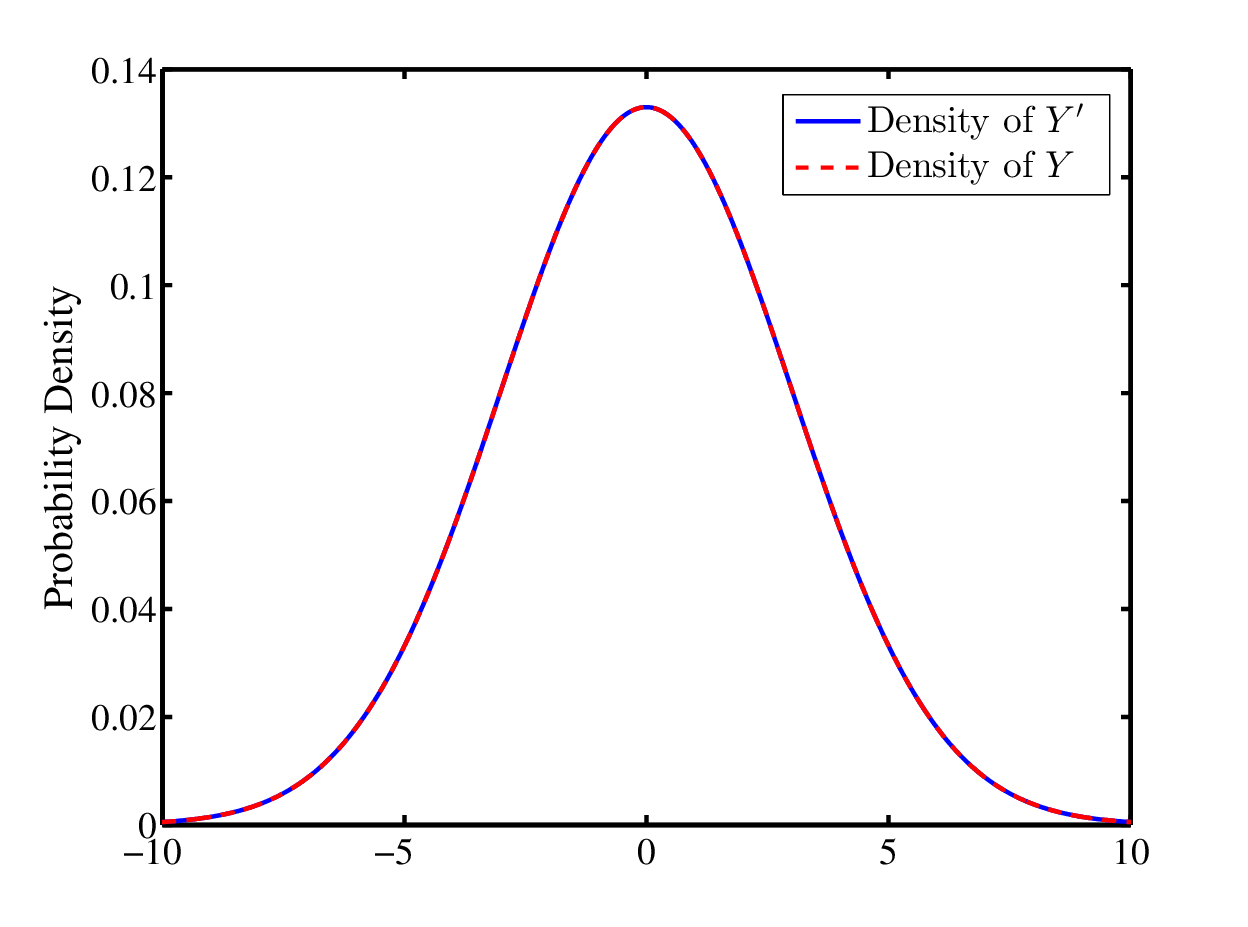}
        }%
        \subfigure[Quantization noise v.s. Gaussian noise when $\Delta=0.1$.]{%
           \label{fig:YY'b}
           \includegraphics[width=0.45\textwidth]{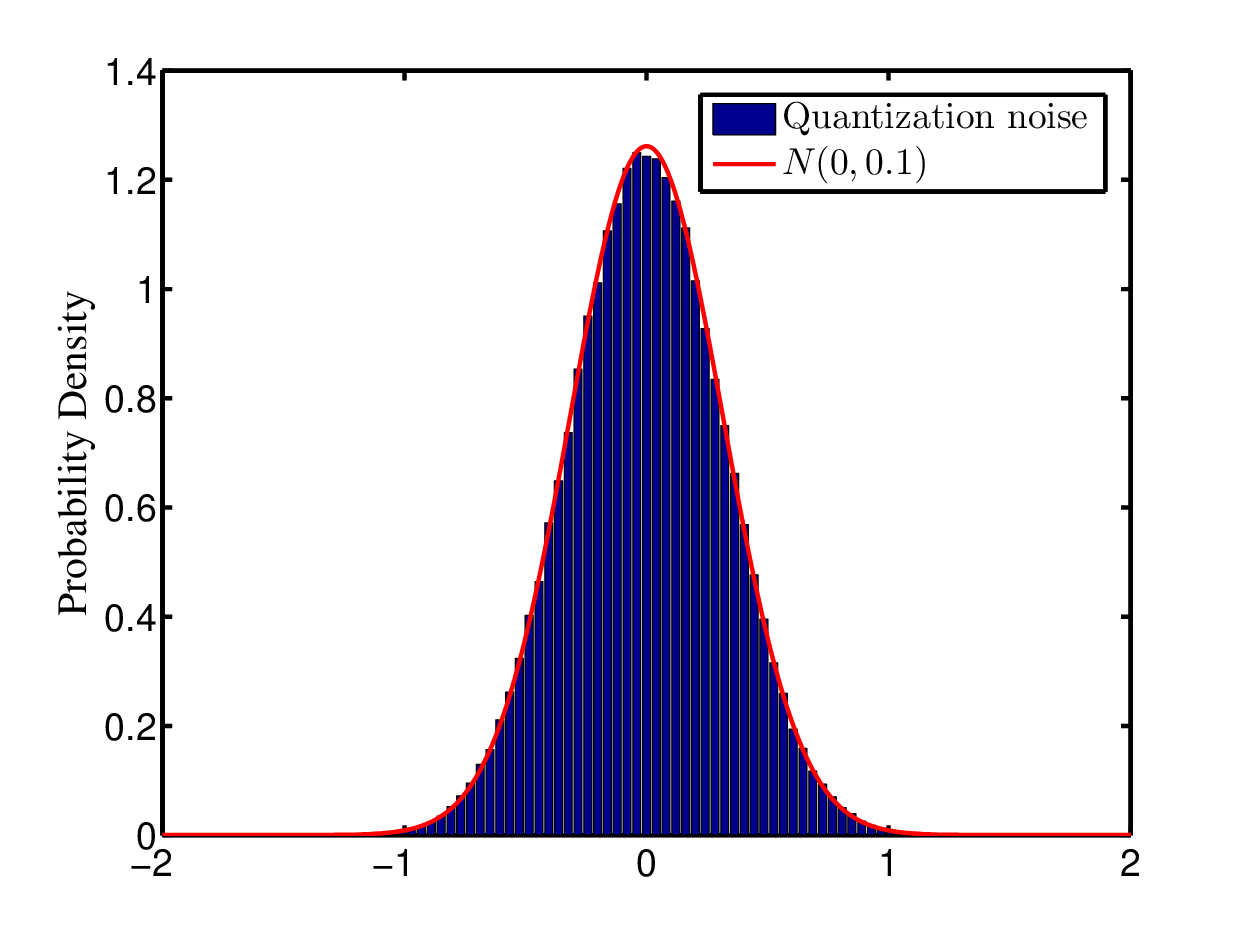}
        }\\ %

    \vspace{-6pt}

    \caption{%
           Comparison of $f_{Y'}$ and $f_{Y}$  and quantization noise for a polar lattice quantizer.
     }%
    \label{fig:YY'}
\end{figure*}

For a target distortion $\Delta=0.5$, the two densities of $Y'$ and $Y$ are compared in Fig. \ref{fig:YY'a}, where negligible difference between $f_{Y}$ and $f_{Y'}$ is found
since $\mathbb{V}(f_{Y'},f_{Y})\approx1.1\times10^{-7}$. Moreover, the quantization noise behaves similarly to a Gaussian noise as shown in Fig. \ref{fig:YY'b}, which will be useful to understand the idea of Gaussian Wyner-Ziv coding and Gelfand-Pinsker coding in the next section.

A performance comparison between the TCQ and polar lattice for quantization is shown in Table \ref{tab:TCQvsPL}. The $\SNR=\frac{\sigma_{s}^{2}}{\Delta}$ is shown in dB. For the TCQ, the dimension is $1000$ and the number
of states is $256$. The performance of the TCQ is taken from \cite[Ch. 3.5]{BK:JPEG2000}. For the quantization polar lattice, the dimension is $1024$. It can be observed that the performance \footnote{The source code of our numerical simulations can be found in the following link. https://github.com/liulingcs/PolarLatticeQuantization.git} of the polar lattice is superior to that of the TCQ with roughly same block length (especially for higher rate). The performance of the Lloyd-Max scalar quantizer is also shown.

\begin{table}[ht]
\begin{center}
\caption{Performance comparison with TCQ for Gaussian source ($\SNR$ in dB).}
\label{tab:TCQvsPL}
\begin{tabular}{|c||c|c|c|c|}\hline
\multirow{1}{*}{} & \multirow{2}{*}{TCQ} & \multirow{2}{*}{Polar Lattice Quantizer} & \multirow{2}{*}{Lloyd-Max Quantizer} &\multirow{2}{*}{Rate-Distortion Bound} \\
\multirow{1}{*}{Rate (bits)} &  &  & & \\
\hline
$1$  &  5.56  &  5.59 &4.40 &6.02   \\\hline
$2$  &  11.04  &  11.55 &9.30& 12.04    \\\hline
$3$  &  16.64  &  17.57 &14.62 &18.06  \\\hline
\end{tabular}
\end{center}
\end{table}

\section{Gaussian Wyner-Ziv Coding}\label{sec:WZ}

\subsection{System model}
\begin{figure}[h]
    \centering
    \includegraphics[width=10cm]{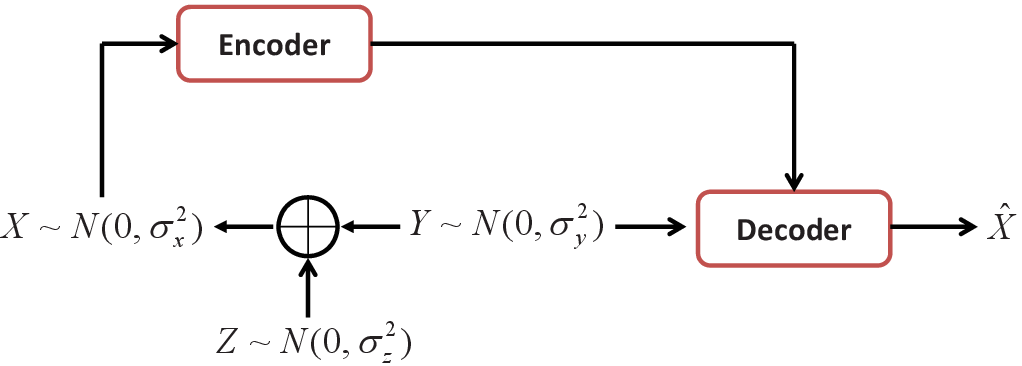}
    \caption{Wyner-Ziv coding for the Gaussian case. The variances of $Z$, $Y$ and $X$ are given by $\sigma_z^2$, $\sigma_y^2$ and $\sigma_x^2=\sigma_y^2+\sigma_z^2$, respectively.}
    \label{fig:WZgraph}
\end{figure}

In this section, we construct polar lattices for the Wyner-Ziv problem. Let $X,Y$ be two joint Gaussian sources and $X=Y+Z$, where $Z$ is a Gaussian noise independent of $Y$ with variance $\sigma_{z}^{2}$.\footnote{For a more general Wyner-Ziv model in the Gaussian case, the relationship between the two joint source can also be $Y=X+Z$, where $Z\sim N(0,\sigma_{z}^{2})$ is a Gaussian noise independent of $X$. In this case, we can perform the MMSE rescaling on $Y$ to make $X=\dot{\alpha}Y+\dot{Z}$, where $\dot{\alpha}=\frac{\sigma_{x}^{2}}{\sigma_{y}^{2}}$ and $\dot{Z}$ is with variance $\frac{\sigma_{z}^{2}\sigma_{x}^{2}}{\sigma_{z}^{2}+\sigma_{x}^{2}}$. Then, the Wyner-Ziv bound is given by $R_{WZ}(\Delta)=\max\Big\{\frac{1}{2}\log\Big(\frac{\sigma_{z}^{2}\sigma_{x}^{2}}{(\sigma_{z}^{2}+\sigma_{x}^{2})\Delta}\Big),0\Big\}$.
Therefore, the system model can still be described by Fig. \ref{fig:WZgraph}, with $Y$ and $Z$ being replaced by $\dot{\alpha}Y$ and $\dot{Z}$, respectively.} A typical system model of Wyner-Ziv coding for the Gaussian case
is shown in Fig. \ref{fig:WZgraph}. Given the side information $Y$, which is only available at the decoder's side, the Wyner-Ziv rate-distortion bound on source $X$ for a target average distortion $\Delta$ between $X$ and its reconstruction $\hat{X}$ is given by
\begin{eqnarray}
R_{WZ}(\Delta)=\max\Big\{\frac{1}{2}\log\Big(\frac{\sigma_{z}^{2}}{\Delta}\Big),0\Big\}.
\end{eqnarray}

\subsection{A solution using continuous auxiliary variable}
To achieve this bound, we assume a continuous auxiliary Gaussian random variable $X'$ which has an average distortion $\Delta'$ from source $X$, i.e., $X'=X+N(0, \Delta')$. Then, we can also obtain that $X'=Y+N(0, \Delta'+\sigma_z^2)$. Letting $\sigma_{x'}^2$ be the variance of $X'$, the difference between the mutual information $I(X';X)$ and $I(X';Y)$ is given by
\begin{eqnarray}
I(X';X)-I(X';Y)&=&\frac{1}{2}\log \frac{\sigma_{x'}^2}{\Delta'}-\frac{1}{2}\log \frac{\sigma_{x'}^2}{\Delta'+\sigma_z^2}  \notag\\
&=& \frac{1}{2}\log \frac{\Delta'+\sigma_z^2}{\Delta'}.
\end{eqnarray}
Let $I(X';X)-I(X';Y)=R_{WZ}(\Delta)$ and assume $\Delta\leq\sigma_{z}^{2}$. Then we have
\begin{eqnarray}
\Delta'=\eta \Delta,
\end{eqnarray}
where $\eta=\frac{\sigma_{z}^{2}}{\sigma_{z}^{2}-\Delta}$. Note that $\eta$ is the reciprocal of the MMSE rescaling parameter in the scenario of quantizing a Gaussian source with variance $\sigma_{z}^{2}$ for a target average distortion $\Delta$.

\begin{figure}[h]
    \centering
    \includegraphics[width=10cm]{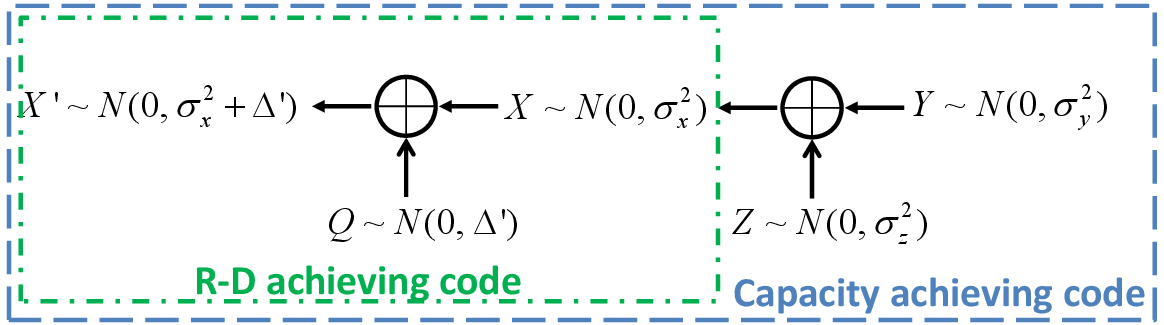}
    \caption{A solution to the Gaussian Wyner-Ziv problem using a continuous Gaussian random variable $X'$.}
    \label{fig:WZsolu}
\end{figure}

The above-mentioned solution for the Gaussian Wyner-Ziv problem, which can be also found in \cite{zamir1}, is depicted by Fig. \ref{fig:WZsolu}. Firstly we design a lossy compression code for source $X$ with Gaussian reconstruction $X'$. The average distortion between $X'$ and $X$ is $\Delta'=\eta\Delta$. We then construct an AWGN capacity achieving code from $Y$ and $X'$. The final reconstruction of $X$ is given by $\hat{X}=Y+\frac{1}{\eta}(X'-Y)$. Clearly $\frac{1}{\eta}(X'-Y)$ is a scaled version of the Gaussian noise, which is independent of $Y$. The variance of $\frac{1}{\eta}(X'-Y)$ is
\begin{eqnarray}
\frac{1}{\eta^{2}}\left(\Delta'+\sigma_{z}^{2}\right) & = & \frac{\Delta}{\eta}+\frac{\sigma_{z}^{2}}{\eta^{2}}\nonumber \\
 & = & \frac{\sigma_{z}^{2}-\Delta}{\sigma_{z}^{2}}\Delta+\frac{\sigma_{z}^{2}-\Delta}{\sigma_{z}^{2}}\left(\sigma_{z}^{2}-\Delta\right)\nonumber \\
 & = & \sigma_{z}^{2}-\Delta.
\end{eqnarray}
We can check that $X=\hat{X}+N(0,\Delta)$, which corresponds to the desired distortion, and the required data rate is $I(X';X)-I(X';Y)=\frac{1}{2}\log\left(\frac{\sigma_{z}^{2}}{\Delta}\right)$. The intuition behind this solution can be described by the concept of binning \cite{zamir1} as follows.

\begin{itemize}
\item Randomly generate $X'$ and map them into bins with distinct indices.
\item Encoding: Quantize $X$ to $X'$ and send only the bin index where $X'$ belongs to. This step roughly requires $2^{n(I(X;X')+\epsilon_{1})}$ codewords.
\item Decoding: Look into the bin and decode $Y$ to $X'$ inside that bin. By channel coding theorem, there are roughly $2^{n(I(Y;X')-\epsilon_{2})}$ codewords inside each bin.
\item Reconstruction: Let $\hat{X}=Y+\frac{1}{\eta}(X'-Y)$. The rate is roughly $R>\frac{1}{n}\log\left(2^{n(I(X;X')+\epsilon_{1})}/2^{n(I(Y;X')-\epsilon_{2})}\right)=I(X';X)-I(X';Y)+\epsilon'$.
\end{itemize}

%In the following section, we will construct a rate-distortion bound achieving polar lattice and a capacity achieving polar lattice for the above encoding and decoding processes, respectively.

\subsection{A practical solution using lattice Gaussian distribution}
\begin{figure}[h]
    \centering
    \includegraphics[width=10cm]{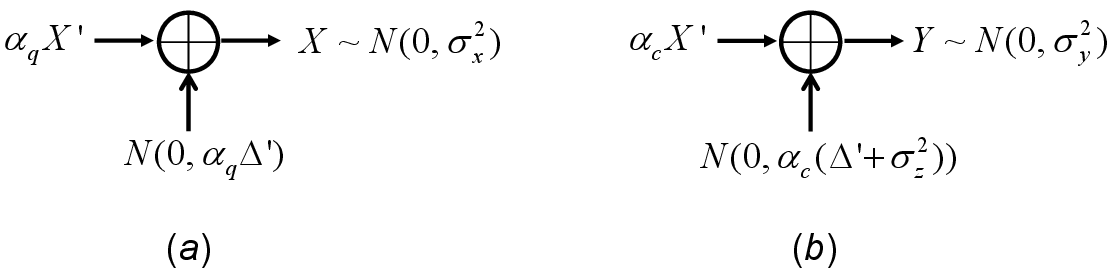}
    \caption{The MMSE rescaled channel blocks $(a)$ and $(b)$ for the Gaussian channels $X \rightarrow X'$ and $Y \rightarrow X'$, respectively.}
    \label{fig:WZMMSE}
\end{figure}
\begin{figure}[h]
    \centering
    \includegraphics[width=10cm]{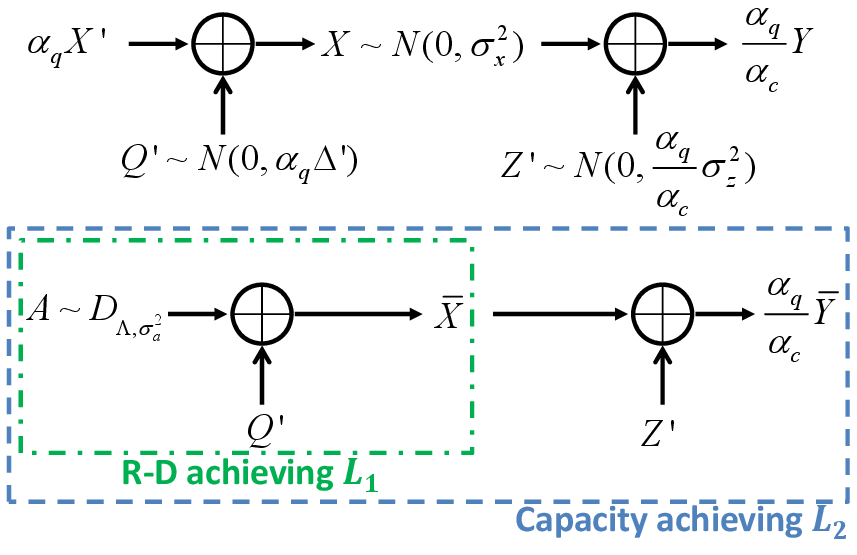}
    \caption{A reverse solution of the Gaussian Wyner-Ziv problem, which is more compatible with lattice Gaussian distribution.}
    \label{fig:WZrvs}
\end{figure}

The problem of the above-mentioned solution is that $X'$ is a continuous Gaussian random variable, which is impractical for the design of lattice codes. In order to utilize the proposed polar lattice coding technique, $X'$ is expected to obey a lattice Gaussian distribution. To this end, we perform MMSE rescaling on $X'$ for the AWGN channels $X\rightarrow X'$ and $Y\rightarrow X'$, respectively. The rescaled channels are shown in Fig. \ref{fig:WZMMSE}, where
\begin{eqnarray}
\alpha_{q}=\frac{\sigma_{x}^{2}}{\sigma_{x}^{2}+\Delta'}=\frac{\sigma_{x}^{2}\left(\sigma_{z}^{2}-\Delta\right)}{\sigma_{x}^{2}\left(\sigma_{z}^{2}-\Delta\right)+\sigma_{z}^{2}\Delta},
\end{eqnarray}
and
\begin{eqnarray}
\alpha_{c}=\frac{\sigma_{y}^{2}}{\sigma_{x}^{2}+\Delta'}=\frac{\left(\sigma_{x}^{2}-\sigma_{z}^{2}\right)\left(\sigma_{z}^{2}-\Delta\right)}{\sigma_{x}^{2}\left(\sigma_{z}^{2}-\Delta\right)+\sigma_{z}^{2}\Delta}.
\end{eqnarray}
Clearly, $\alpha_{c}<\alpha_{q}$. To combine the two blocks in Fig. \ref{fig:WZMMSE} together, block $(b)$ is scaled by $\frac{\alpha_{q}}{\alpha_{c}}$. Consequently, a reversed version of the solution illustrated in Fig. \ref{fig:WZsolu} is obtained and shown in Fig. \ref{fig:WZrvs}. For the reconstruction of $X$, we have the following proposition.

\begin{prop}
To achieve the $R_{WZ}(\Delta)$ bound by the reversed structure shown in Fig. \ref{fig:WZrvs}, the reconstruction of $X$ is given by
\begin{eqnarray}
\hat{X}=\alpha_{q}X'+\gamma\left(\frac{\alpha_{q}}{\alpha_{c}}Y-\alpha_{q}X'\right),\;\;\;\gamma=\frac{\sigma_{y}^{2}\Delta}{\sigma_{x}^{2}\sigma_{z}^{2}}.
\end{eqnarray}
\end{prop}

\begin{IEEEproof}
It suffices to prove that $X=\hat{X}+N(0,\Delta)$. According to Fig. \ref{fig:WZrvs}, we have $X=\alpha_{q}X'+N(0,\alpha_{q}\Delta')$, meaning that showing $\hat{X}=\alpha_{q}X'+N(0,\alpha_{q}\Delta'-\Delta)$
would complete this proof.

Clearly, $\frac{\alpha_{q}}{\alpha_{c}}Y-\alpha_{q}X'$ is a Gaussian random variable with $0$ mean and variance $\alpha_{q}\Delta'+\frac{\alpha_{q}}{\alpha_{c}}\sigma_{z}^{2}$, and it is independent of $X'$. By substituting the parameters $\Delta'$, $\alpha_{q}$ and $\alpha_{c}$, we have
\begin{eqnarray}
\alpha_{q}\Delta'-\Delta & = & \frac{\sigma_{x}^{2}\left(\sigma_{z}^{2}-\Delta\right)}{\sigma_{x}^{2}\left(\sigma_{z}^{2}-\Delta\right)+\sigma_{z}^{2}\Delta}\frac{\sigma_{z}^{2}}{\sigma_{z}^{2}-\Delta}\Delta-\Delta\\
 & = & \frac{\sigma_{y}^{2}\Delta^{2}}{\sigma_{x}^{2}\sigma_{z}^{2}-\sigma_{y}^{2}\Delta},
\end{eqnarray}
and
\begin{eqnarray}
\gamma^{2}\left(\alpha_{q}\Delta'+\frac{\alpha_{q}}{\alpha_{c}}\sigma_{z}^{2}\right) & = & \left(\frac{\sigma_{y}^{2}\Delta}{\sigma_{x}^{2}\sigma_{z}^{2}}\right)^{2}\left(\frac{\sigma_{x}^{2}(\sigma_{z}^{2}-\Delta)}{\sigma_{x}^{2}(\sigma_{z}^{2}-\Delta)+\sigma_{z}^{2}\Delta}\frac{\sigma_{z}^{2}}{\sigma_{z}^{2}-\Delta}\Delta+\frac{\sigma_{x}^{2}}{\sigma_{y}^{2}}\sigma_{z}^{2}\right),\\
 & = & \left(\frac{\sigma_{y}^{2}\Delta}{\sigma_{x}^{2}\sigma_{z}^{2}}\right)^{2}\left[\sigma_{x}^{2}\sigma_{z}^{2}\left(\frac{\Delta}{\sigma_{x}^{2}\sigma_{z}^{2}-\sigma_{y}^{2}\Delta}+\frac{1}{\sigma_{y}^{2}}\right)\right]\\
 & = & \frac{\sigma_{y}^{2}\Delta^{2}}{\sigma_{x}^{2}\sigma_{z}^{2}-\sigma_{y}^{2}\Delta},
\end{eqnarray}
as desired.
\end{IEEEproof}

Now the continuous Gaussian random variable $\alpha_{q}X'$ can be replaced by a lattice Gaussian $A\sim D_{\Lambda,\sigma_{a}^{2}}$, where $\sigma_{a}^{2}=\alpha_{q}^{2}\sigma_{x'}^{2}$. Let $\bar{X}=A+N(0,\alpha_{q}\Delta')$ and $\frac{\alpha_{q}}{\alpha_{c}}\bar{Y}=\bar{X}+N(0,\frac{\alpha_{q}}{\alpha_{c}}\sigma_{z}^{2})$. Let $\bar{B}=\frac{\alpha_{q}}{\alpha_{c}}\bar{Y}$ and $\sigma_{b}^{2}=\frac{\alpha_{q}^{2}}{\alpha_{c}^{2}}\sigma_{y}^{2}$
for convenience. By Lemma \ref{lem:YY'distance}, the distributions of $\bar{X}$ and $\bar{Y}$ can be made arbitrarily close to those of $X$ and $Y$, respectively. Then the polar lattices are designed by treating $\bar{X}$ as the source and $\bar{Y}$ as its side information. A rate-distortion bound achieving polar lattice $L_{1}$ is constructed for source $\bar{X}$ with target distortion $\alpha_{q}\Delta'$, and an AWGN capacity-achieving polar lattice $L_{2}$ is constructed to help the decoder extract some information from $\bar{Y}$, as shown in Fig. \ref{fig:WZrvs}. Finally, the decoder reconstructs $\check{X}=A+\gamma(\bar{B}-A)$. Conceptually, $\bar{B}-A$ is a Gaussian noise which is independent of $A$.\footnote{In fact, when $A$ is reconstructed by the decoder, $\bar{B}-A$ is not exactly a Gaussian noise $N(0,\alpha_{q}\Delta'+\frac{\alpha_{q}}{\alpha_{c}}\sigma_{z}^{2})$, since the quantization noise of $L_{1}$ is not exactly Gaussian distributed. However, according to Theorem \ref{Thm:quantizationMain}, the two
distributions can be made arbitrarily close when $N$ is sufficiently large. See Fig. \ref{fig:YY'b} for an example.} Recall that $\gamma=\frac{\sigma_{y}^{2}\Delta}{\sigma_{x}^{2}\sigma_{z}^{2}}$ scales $\bar{B}-A$ to $N(0,\alpha_{q}\Delta'-\Delta)$. By Lemma \ref{lem:YY'distance} again, the distributions of $\check{X}$ and $\hat{X}$ can be very close, resulting in an average distortion close to $\Delta$.

When lattice Gaussian distribution is utilized, by \cite[Lemma 6]{LiuL15w}, $L_{1}$ and $L_{2}$ are accordingly constructed for the MMSE-rescaled Gaussian noise variance $\tilde{\sigma}_{q}^{2}$ and $\tilde{\sigma}_{c}^{2}$,
where
\begin{eqnarray}
\tilde{\sigma}_{q}^{2}=\frac{\sigma_{a}^{2}}{\sigma_{x}^{2}}\alpha_{q}\Delta'=\frac{\sigma_{a}^{2}\sigma_{z}^{2}\Delta}{\sigma_{x}^{2}\sigma_{z}^{2}-\sigma_{y}^{2}\Delta},
\end{eqnarray}
and
\begin{eqnarray}
\tilde{\sigma}_{c}^{2}=\frac{\sigma_{a}^{2}}{\sigma_{b}^{2}}\left(\alpha_{q}\Delta'+\frac{\alpha_{q}}{\alpha_{c}}\sigma_{z}^{2}\right)=\frac{\sigma_{a}^{2}\sigma_{z}^{4}}{\sigma_{x}^{2}\sigma_{z}^{2}-\sigma_{y}^{2}\Delta}.
\end{eqnarray}
Since $\Delta\leq\sigma_{z}^{2}$, we also have $\tilde{\sigma}_{q}^{2}\leq\tilde{\sigma}_{c}^{2}$.

\begin{figure}[h]
    \centering
    \includegraphics[width=10cm]{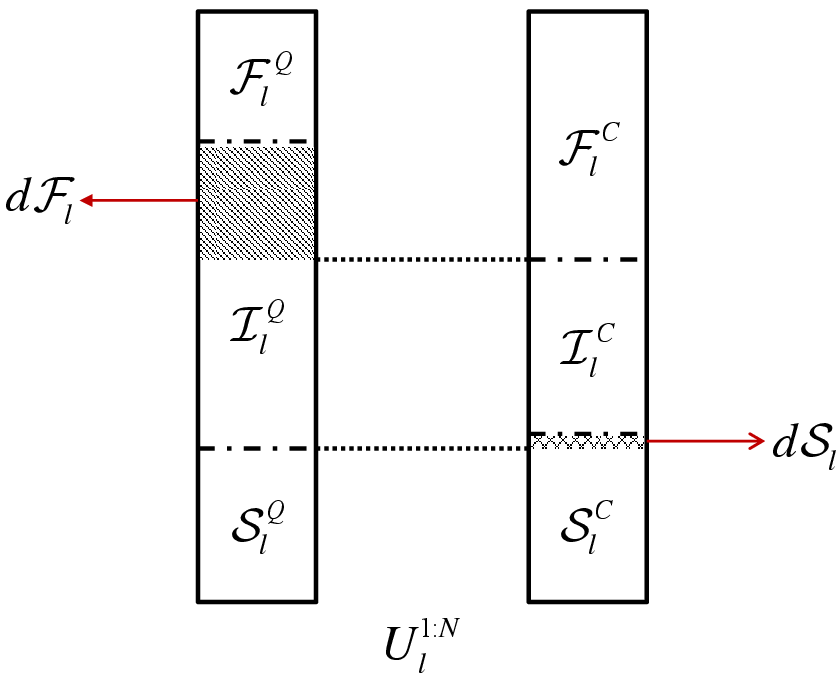}
    \caption{The partitions of $U_{\ell}^{1:N}$ for quantization lattice $L_1$ (left) and channel coding lattice $L_2$ (right). By definition \eqref{eqn:WZlatticeQQ}, $\mathcal{F}_\ell^Q \subseteq \mathcal{F}_\ell^C$, $\mathcal{I}_\ell^C \subseteq \mathcal{I}_\ell^Q$, and $\mathcal{S}_\ell^Q \subseteq \mathcal{S}_\ell^C$. Let $d\mathcal{F}_\ell$, $d\mathcal{I}_\ell$ and $d\mathcal{S}_\ell$ denote the sets $\mathcal{F}_\ell^C \setminus \mathcal{F}_\ell^Q$, $\mathcal{I}_\ell^Q \setminus \mathcal{I}_\ell^C$ and $\mathcal{S}_\ell^C \setminus \mathcal{S}_\ell^Q$, respectively. Without the side information, $U_\ell^{\mathcal{I}_\ell^Q}$ should be sent to achieve the target distortion. With the side information, however, $U_\ell^{\mathcal{I}_\ell^C}$ can be decoded and hence only $U_\ell^{d\mathcal{I}_\ell}$ need to be sent.}
    \label{fig:WZpolar}
\end{figure}

Now it is ready to give the polar lattice coding scheme. We choose a good constellation $D_{\Lambda, \sigma_a^2}$ such that the flatness factor $\epsilon_{\Lambda}(\tilde{\sigma}_q)$ is negligible. Let $\Lambda/\Lambda_1/\cdots/\Lambda_{r-1}/\Lambda'/\cdots$ be a one-dimensional binary partition chain labeled by bits $A_1/A_2/\cdots/A_{r-1}/A_r/\cdots$. Then, $P_{A_{1:r}}$ and $A_{1:r}$ approach $D_{\Lambda, \sigma_a^2}$ and $A$, respectively, as $r \to \infty$. Consider $N$ i.i.d. copies of $A$. Let $U_{\ell}^{1:N}=A_{\ell}^{1:N} G_N$ for each $1 \leq \ell \leq r$. The partitions of $U_{\ell}^{1:N}$ for both $L_1$ and $L_2$ are shown in Fig. \ref{fig:WZpolar}, where the left block is for the quantization lattice $L_1$ and the right one for the channel coding lattice $L_2$. According to Section \ref{sec:qzgood} and \cite{polarlatticeJ}, for $0<\beta<\frac{1}{2}$, the frozen set $\mathcal{F}_\ell^Q$ $\left(\mathcal{F}_\ell^C\right)$, information set $\mathcal{I}_\ell^Q$ $\left(\mathcal{I}_\ell^C\right)$ and the shaping set $\mathcal{S}_\ell^Q$ $\left(\mathcal{S}_\ell^C\right)$ for lattice $L_1$ $\left(L_2\right)$ are given by
\begin{eqnarray}\label{eqn:WZlatticeQQ}
\begin{cases}
\begin{aligned}
&\mathcal{F}_\ell^Q=\left\{i\in[N]:Z\left(U_\ell^i|U_\ell^{1:i-1},A_{1:\ell-1}^{1:N},\bar{X}^{1:N}\right)\geq1-2^{-N^{\beta}}\right\}\\
&\mathcal{I}_\ell^Q=\left\{i\in[N]:Z\left(U_\ell^i|U_\ell^{1:i-1},A_{1:\ell-1}^{1:N}\right)>2^{-N^{\beta}}\text{ and }Z\left(U_\ell^i|U_\ell^{1:i-1},A_{1:\ell-1}^{1:N},\bar{X}^{1:N}\right)<1-2^{-N^{\beta}}\right\}\\
&\mathcal{S}_\ell^Q=\left\{i\in[N]:Z\left(U_\ell^i|U_\ell^{1:i-1},A_{1:\ell-1}^{1:N}\right)\leq2^{-N^{\beta}}\right\},
\end{aligned}
\end{cases}
\end{eqnarray}
and
\begin{eqnarray}\label{eqn:WZlatticeCC}
\begin{cases}
\begin{aligned}
&\mathcal{F}_\ell^C=\left\{i\in[N]:Z\left(U_\ell^i|U_\ell^{1:i-1},A_{1:\ell-1}^{1:N},\bar{B}^{1:N}\right)\geq1-2^{-N^{\beta}}\right\}\\
&\mathcal{I}_\ell^C=\left\{i\in[N]:Z\left(U_\ell^i|U_\ell^{1:i-1},A_{1:\ell-1}^{1:N}\right)\geq 1-2^{-N^{\beta}}\text{ and }Z\left(U_\ell^i|U_\ell^{1:i-1},A_{1:\ell-1}^{1:N},\bar{B}^{1:N}\right)\leq 2^{-N^{\beta}}\right\}\\
&\mathcal{S}_\ell^C=\left\{i\in[N]:Z\left(U_\ell^i|U_\ell^{1:i-1},A_{1:\ell-1}^{1:N}\right)<1-2^{-N^{\beta}}\text{ or }2^{-N^{\beta}}<Z\left(U_\ell^i|U_\ell^{1:i-1},A_{1:\ell-1}^{1:N},\bar{B}^{1:N}\right)<1-2^{-N^{\beta}}\right\}.
\end{aligned}
\end{cases}
\end{eqnarray}

By channel degradation, we have $\mathcal{F}_\ell^Q \subseteq \mathcal{F}_\ell^C$. Let $d\mathcal{F}_\ell$ denote the set $\mathcal{F}_\ell^C\setminus \mathcal{F}_\ell^Q$. Meanwhile, we have $\mathcal{S}_\ell^Q \subseteq \mathcal{S}_\ell^C$ by definition. Denoting by $d\mathcal{S}_\ell$ the set $\mathcal{S}_\ell^C\setminus \mathcal{S}_\ell^Q$, $d\mathcal{S}_\ell$ can be written as
\begin{eqnarray}\label{eqn:setdS}
d\mathcal{S}_\ell=\Big\{i\in[N]:2^{-N^{\beta}}<Z\left(U_\ell^i|U_\ell^{1:i-1},A_{1:\ell-1}^{1:N}\right)<1-2^{-N^{\beta}}\text{ or } \notag\\
2^{-N^{\beta}}<Z\left(U_\ell^i|U_\ell^{1:i-1},A_{1:\ell-1}^{1:N},\bar{B}^{1:N}\right)<1-2^{-N^{\beta}}\Big\},
\end{eqnarray}
and the proportion $\frac{|d\mathcal{S}_\ell|}{N} \to 0$ as $N\to \infty$. Also observe that $d\mathcal{I}_\ell=\mathcal{I}_\ell^Q\setminus \mathcal{I}_\ell^C=d\mathcal{F}_\ell \cup d\mathcal{S}_\ell$.

For the above two partitions, we have the following lemma.
\begin{lem}
Let $L_{1}$ and $L_{2}$ be two polar lattices constructed according to the above two partition rules, respectively. $L_{2}$ is nested within $L_{1}$, i.e., $L_{2}\subseteq L_{1}$.
\end{lem}
\begin{IEEEproof}
Both $L_1$ and $L_2$ follow the multilevel lattice structure \eqref{eqn:multilvl}. Let $\{C_1^q, ..., C_\ell^q,..., C_r^{q}\}$ and $\{C_1^c, ..., C_\ell^c,..., C_r^{c}\}$ denote the multilevel codes for $L_1$ and $L_2$, respectively. When shaping is not involved, the generator matrixes of $C_\ell^q$ and $C_\ell^c$ correspond to the sets of row indices $\mathcal{I}_\ell^Q \cup \mathcal{S}_\ell^Q$ and $\mathcal{I}_\ell^C \cup \mathcal{S}_\ell^Q$, respectively. By the relationship $\tilde{\sigma}_q^2 \leq \tilde{\sigma}_c^2$ and \cite[Lemma 3]{polarlatticeJ}, the subchannel $C(\Lambda_{\ell-1}/\Lambda_\ell, \tilde{\sigma}_c^2)$ is degraded with respect to $C(\Lambda_{\ell-1}/\Lambda_\ell, \tilde{\sigma}_q^2)$. Then by the equivalence lemma \cite[Lemma 10]{polarlatticeJ}, we have $\mathcal{I}_\ell^C \subseteq \mathcal{I}_\ell^Q$, meaning that $C_\ell^c \subseteq C_\ell^q$ for $1\leq \ell \leq r$. As a result, $L_2 \subseteq L_1$.
\end{IEEEproof}

Given an $N$-dimensional realization vector $x^{1:N}$ of $X^{1:N}$, the encoder evaluates $u_{\ell}^{1:N}$ from level $1$ to level $r$ successively according to the randomized rounding quantization rules given in Section \ref{sec:qzgood} (see \eqref{eqn:lossyencoder1}, \eqref{eqn:lossyencoder2}, \eqref{eqn:lossyencoder4} and \eqref{eqn:lossyencoder5}.) Recall that treating $x^{1:N}$ as a realization of $\bar{X}^{1:N}$ is safe because $X$ and $\bar{X}$ are similarly distributed. $u_\ell^{d\mathcal{I}_\ell}$ is then sent to the decoder for each level. For the decoder, the realization vector $y^{1:N}$ of $Y^{1:N}$ is scaled to $b^{1:N}=\frac{\alpha_q}{\alpha_c} y^{1:N}$. Notice that $u_\ell^{\mathcal{F}_\ell^Q}$ is shared between the encoder and decoder before transmission. After receiving $u_\ell^{d\mathcal{I}_\ell}$ and $u_\ell^{\mathcal{I}_\ell^C}$, $u_\ell^{\mathcal{S}_\ell^Q}$ can be decoded with vanishing error probability because their associate Bhattacharyya parameters are arbitrarily small when $N \to \infty$. The probabilities $P_{U_\ell^i|U_\ell^{1:i-1},A_{1:\ell-1}^{1:N}}$, $P_{U_\ell^i|U_\ell^{1:i-1},A_{1:\ell-1}^{1:N},\bar{X}^{1:N}}$ and $P_{U_\ell^i|U_\ell^{1:i-1},A_{1:\ell-1}^{1:N},\bar{B}^{1:N}}$ can be evaluated with $O(N\log N)$ complexity. Usually $u_\ell^{\mathcal{S}_\ell^C}$ is covered by a pre-shared random mapping. However, it is possible to replace the random mapping with MAP decision for $u_\ell^{\mathcal{S}_\ell^Q}$ (see, e.g., \cite{7447169,7429782}). Then, the whole vector $u_\ell^{1:N}$ can be recovered with high probability. Similarly to the reconstruction process in \eqref{eqn:uN2xN}, after obtaining $u_\ell^{1:N}$ for $1 \leq \ell \leq r$, the realization $a^{1:N}$ of $A^{1:N}$ can be recovered from $u_\ell^{1:N}$ according to the following equation
\begin{eqnarray}\label{eqn:uN2xNwz}
\chi = \sum_{\ell=1}^{r}2^{\ell-1}\left[\sum_{i\in\mathcal{I}_{\ell}}u_{\ell}^{i}\psi(\mathbf{g}_{i})+\sum_{i\in\mathcal{S}_{\ell}}u_{\ell}^{i}\psi(\mathbf{g}_{i})+\sum_{i\in\mathcal{F}_{\ell}}u_{\ell}^{i}\psi(\mathbf{g}_{i})\right],
\label{constructionD-finite-power}
\end{eqnarray}
where $\mathbf{g}_{i}$ denotes the $i$-th row of the polarization matrix $G_N$ and $\psi$ is the natural embedding. Please notice that $a^{1:N}$ is drawn from $D_{2^r\mathbb{Z}^N+\chi,\sigma_a}$. When $r$ is sufficiently large, the probability of choosing a constellation point outside the interval $[-2^{r-1},2^{r-1})$ is negligible. Therefore, there exists only one point within $[-2^{r-1},2^{r-1})$ with probability close to $1$ and $a^{1:N}$ can be approximated by $\chi \mod 2^r$. Finally, the reconstruction of $x^{1:N}$ is given by $\check{x}^{1:N}=a^{1:N}+\gamma(b^{1:N}-a^{1:N})$.

To sum up, we have the following Wyner-Ziv coding scheme.
\begin{itemize}
\item Encoding: For the $N$-dimensional i.i.d. source vector $X^{1:N}$, the encoder evaluates $U^{\mathcal{I}_\ell^Q}$ by randomized rounding, and then sends $U_\ell^{d\mathcal{I}_\ell}$ to the decoder.
\item Decoding: Using the pre-shared $U^{\mathcal{F}_\ell^Q}$ and the received $U_\ell^{d\mathcal{I}_\ell}$, the decoder recovers $U^{\mathcal{I}_\ell^C}$ and $U^{\mathcal{S}_\ell^Q}$ from the side information $B^{1:N}$. For each level the decoder obtains $U_\ell^{1:N}$, then $A^{1:N}$ can be recovered according to \eqref{eqn:uN2xNwz}.
\item Reconstruction: $\check{X}^{1:N}=A^{1:N}+\gamma(B^{1:N}-A^{1:N})$.
\end{itemize}

With regard to the design rate, by Theorem \ref{Thm:quantizationMain}, the rate $R_{L_1}$ of $L_1$ can be arbitrarily close to $\frac{1}{2}\log \frac{\sigma_x^2}{\alpha_q\Delta'}$. However, the encoder does not need to send that much information to the decoder because of the side information. Meanwhile, the rate $R_{L_2}$ of $L_2$ can be arbitrarily close to $\frac{1}{2}\log \left(\frac{\sigma_b^2}{\alpha_q\Delta'+\frac{\alpha_q}{\alpha_c}\sigma_z^2}\right)$. After some tedious calculation, we have
\begin{eqnarray}
R_{L_1} \to {\frac{1}{2}\log \bigg(\frac{\sigma_x^2\sigma_z^2-\sigma_y^2\Delta}{\sigma_z^2\Delta}\bigg)}^+,
\end{eqnarray}
and
\begin{eqnarray}
R_{L_2} \to {\frac{1}{2}\log \bigg(\frac{\sigma_x^2\sigma_z^2-\sigma_y^2\Delta}{\sigma_z^4}\bigg)}^-,
\end{eqnarray}
meaning that the transmission rate $R_{L_1}-R_{L_2} \to \frac{1}{2}{\log\left(\frac{\sigma_z^2}{\Delta}\right)}^+$.

The following theorem is proved in Appendix \ref{appendix4}.
\begin{ther}\label{Thm:WZ}
Let $X$ be a Gaussian source and $Y$ be another Gaussian source correlated to $X$ as $X=Y+Z$, where $Z\sim N(0,\sigma_z^2)$ is an independent Gaussian noise. Consider a target distortion $0 \leq \Delta \leq \sigma_z^2$ for source $X$ when $Y$ is only available for the decoder. Let $\Lambda/\Lambda_1/\cdots/\Lambda_{r}$ be a one-dimensional binary partition chain such that $\epsilon_{\Lambda}(\tilde{\sigma}_q)=O\left(e^{-N}\right)$ and $r=O(\log N)$. For any $0<\beta''<\beta'<\frac{1}{2}$, there exists two nested polar lattices $L_1$ and $L_2$ with a differential rate $R=R_{L_1}-R_{L_2}$ arbitrarily close to $\frac{1}{2}\log\left(\frac{\sigma_z^2}{\Delta}\right)$ such that the expect distortion $\Delta_Q$ satisfies
\begin{eqnarray}
\Delta_Q \leq \Delta+O\left(N^22^{-N^{\beta''}}\right),
\end{eqnarray}
and the block error probability satisfies
\begin{eqnarray}
P_e^{WZ} \leq O\left(2^{-N^{\beta''}}\right).
\end{eqnarray}
\end{ther}

\begin{figure}[ht]
    \centering
    \includegraphics[width=12cm]{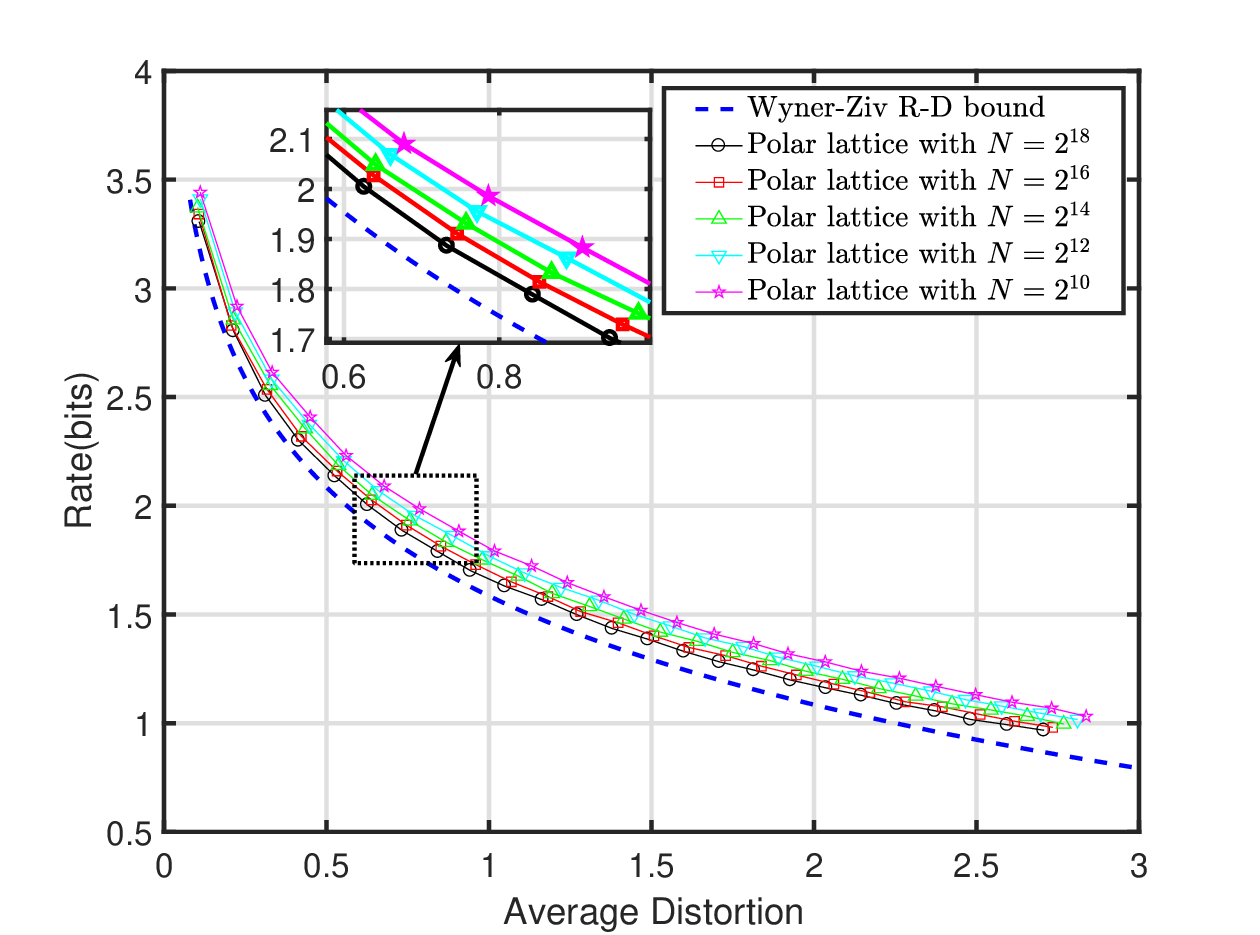}
    \caption{The performance of polar lattices for the Gaussian Wyner-Ziv problem with $\sigma_z=3$ and $\sigma_y=1$.}
    \label{fig:RDbound_WZ}
\end{figure}

To test the performance of our scheme, we run numerical simulation for the Gaussian Wyner-Ziv problem with $\sigma_y=1$ and $\sigma_z=3$ as shown in Fig. \ref{fig:WZgraph}. We note that in this case the Wyner-Ziv rate distortion bound is the same as the rate distortion bound for the Gaussian source with $\sigma_s=3$ in Sect. \ref{sec:simu}. For comparison, we choose the same partition level $r=6$ and construct polar lattices for a target distortion from $0.1$ to $2.5$. Similarly, the Bhattacharyya parameters in \eqref{eqn:WZlatticeQQ} and \eqref{eqn:WZlatticeCC} can be evaluated by the method proposed by Tal and Vardy in \cite{tal2011construct}. The probabilities $P_{U_\ell^i|U_\ell^{1:i-1},A_{1:\ell-1}^{1:N}}$, $P_{U_\ell^i|U_\ell^{1:i-1},A_{1:\ell-1}^{1:N},\bar{X}^{1:N}}$ and $P_{U_\ell^i|U_\ell^{1:i-1},A_{1:\ell-1}^{1:N},\bar{B}^{1:N}}$ are calculate by standard SC algorithm with complexity $O(N\log N)$. The realized performance of polar lattices for $N=2^{10}, 2^{12}, 2^{14}, 2^{16}$ and $2^{18}$ is shown in Fig. \ref{fig:RDbound_WZ}. Compared with the performance presented in Fig. \ref{fig:RDbound}, a certain level of performance degradation can be observed, which is mainly due to the decoding error of the channel coding polar lattice $L_2$ at the side of decoder. In order to guarantee an acceptable block error probability of $L_2$ ($10^{-4}$ in our simulation), we have to suffer some rate loss, which increases the gap to the optimal bound. For $N=2^{18}$, the maximum gap increases to approximately $0.6$ dB for relatively large distortion. However, the tendency still shows that the Wyner-Ziv rate distortion bound can be approached by our scheme as the dimension of polar lattices grows. We also note the performance can be further improved by employing more sophisticated decoding algorithms of polar codes such as \cite{ListPolar}.

\begin{rem}
Compared with a related work \cite{PolarCodesWZ}, where the authors proposed to use nested $q$-ary polar codes for the Gaussian Wyner-Ziv problem, our polar-lattice based scheme has no specific restriction on the $\SNR$ \footnote{In \cite{PolarCodesWZ}, the $\SNR$ is defined as $\frac{P}{N}$, which in fact is $\frac{\sigma_x^2}{\sigma_z^2}$ in our setting.} of the system. For the first scheme (Scheme A in \cite{PolarCodesWZ}), the side-information noise variance is required to be much higher than the source variance to achieve the optimal bound. Whereas for the second scheme (Scheme B in \cite{PolarCodesWZ}), the side-information noise variance is assumed to be much lower than the source variance.
\end{rem}

\section{Gaussian Gelfand-Pinsker coding}
\begin{figure}[h]
    \centering
    \includegraphics[width=12cm]{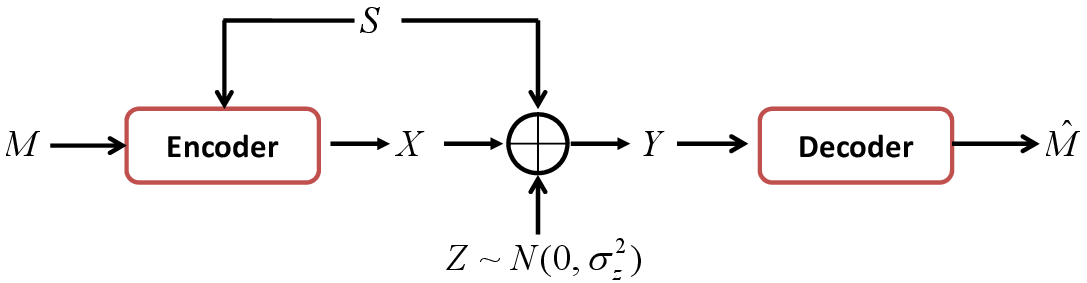}
    \caption{Gelfand-Pinsker coding for the Gaussian case.}
    \label{fig:GPgraph}
\end{figure}
\begin{figure}[ht]
    \centering
    \includegraphics[width=10cm]{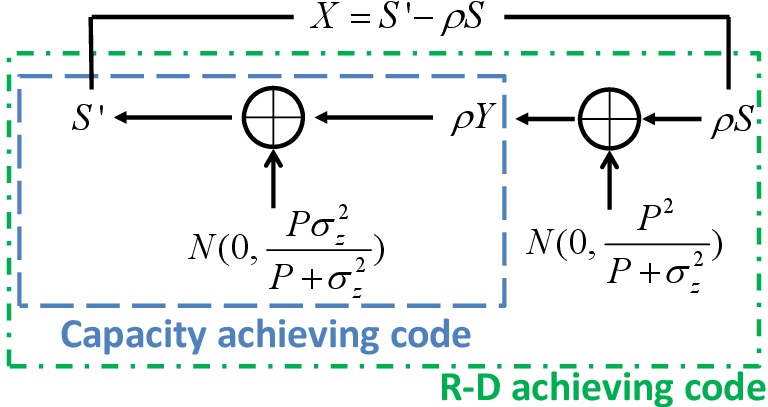}
    \caption{A solution to the Gaussian Gelfand-Pinsker problem using continuous Gaussian random variable $S'$.}
    \label{fig:GPsolu}
\end{figure}

For the Gelfand-Pinsker problem, with some abuse of notations, consider the channel described by $Y=X+S+Z$, where $X$ and $Y$ are the channel input and output, respectively, $Z$ is an unknown additive Gaussian noise with variance $\sigma_z^2$ and $S$ is an interference Gaussian signal with variance $\sigma_i^2$ known only to the encoder. A diagram of Gelfand-Pinsker coding is shown in Fig. \ref{fig:GPgraph}. Message $M$ is encoded into $X$ which satisfies the power constraint $E\left[\|X\|^2\right] \leq P$. The channel capacity of this Gaussian Gelfand-Pinsker model \cite{CostaGP, GelPinsker} is given by
\begin{eqnarray}
C_{GP}=\frac{1}{2}\log\left(1+\frac{P}{\sigma_z^2}\right). \notag\
\end{eqnarray}
To achieve this capacity, the roles of quantization lattice and channel coding lattice are reversed. To see this, we still start with a continuous auxiliary variable and then replace it with a discrete Gaussian. Letting $\rho=\frac{P}{P+\sigma_z^2}$, we firstly design a lossy compression code for $\rho S$ with Gaussian reconstruction alphabet $S'$.
The distortion between $S'$ and $\rho S$ is targeted to be $P$, i.e., $S'=\rho S+N(0, P)$. Then the encoder transmits $X=S'-\rho S$ ($X$ is independent of $S$), which satisfies the power constraint. Moreover, the relationship between $Y$ and $S'$ is given by
\begin{eqnarray}
\begin{aligned}
S'&=X+\rho S\\
&=X+ \rho(Y-X-Z) \\
&=\rho Y+(1-\rho) X- \rho Z.
\end{aligned}
\end{eqnarray}
We can check the expectation
\begin{eqnarray}
E\left[Y\cdot[(1-\rho)X-\rho Z]\right]=(1-\rho)E\left[X^2\right]-\rho E\left[Z^2\right]=0,
\end{eqnarray}
meaning that $(1-\rho) X- \rho Z$ is independent of $Y$, which gives $S'=\rho Y+ N\left(0, \frac{P\sigma_z^2}{P+\sigma_z^2}\right)$.
Then, we construct an AWGN capacity-achieving code to recover $S'$ from $\rho Y$. Without the power constraint, the maximum data rate that can be sent is actually $I(S';\rho Y)$. However, when power constraint is taken into consideration, some bits should be selected according to the realization of $S$ since $S'$ and $S$ are correlated. The maximum data rate becomes $I(S';\rho Y)-I(S';\rho S)=\frac{1}{2}\log\left(1+\frac{P}{\sigma_z^2}\right)=C_{GP}$. A diagram of this solution is shown in Fig. \ref{fig:GPsolu}, where
\begin{eqnarray}
\sigma_{s'}^2=\rho^2\sigma_i^2+P,
\end{eqnarray}
and
\begin{eqnarray}
\sigma_{y}^2=\frac{1}{\rho^2}\frac{P^2}{P+\sigma_z^2}+\sigma_i^2=\sigma_i^2+P+\sigma_z^2
\end{eqnarray}
are the variances of $S'$ and $Y$, respectively. It is worth noting that the idea of using nested binning scheme for this Gaussian Gelfand-Pinsker problem was given in \cite{zamir1}.

\begin{figure}[ht]
    \centering
    \includegraphics[width=10cm]{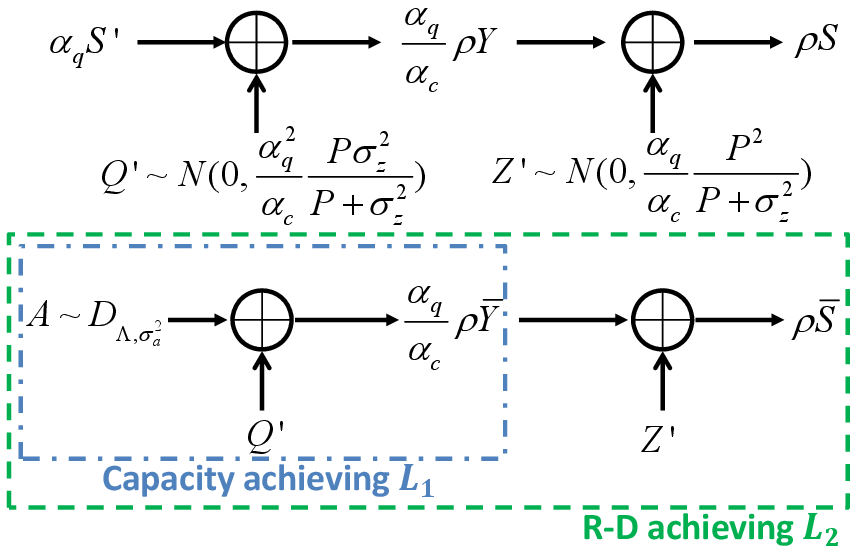}
    \caption{A reverse solution of the Gaussian Gelfand-Pinsker problem.}
    \label{fig:GPMMSE}
\end{figure}

Similarly, we use a discrete Gaussian version of $S'$ to approach this capacity. The idea is to perform MMSE rescaling on $S'$ to get a reversed version of the model shown in Fig. \ref{fig:GPsolu}. The analysis is similar to that presented in Section \ref{sec:WZ} and is omitted here for brevity. Finally, the reversed solution is given in Fig. \ref{fig:GPMMSE}.

With some abuse of notation, let $A$ denote the discrete version of $\alpha_qS'$. The MMSE rescaling factor $\alpha_c$ for channel coding and $\alpha_q$ for quantization are given by
\begin{eqnarray}
\alpha_c=\frac{P\sigma_y^2}{P\sigma_i^2+(P+\sigma_z^2)^2},
\end{eqnarray}
and
\begin{eqnarray}
\alpha_q=\frac{P\sigma_i^2}{P\sigma_i^2+(P+\sigma_z^2)^2},
\end{eqnarray}
respectively. The variance $\sigma_a^2$ for $D_{\Lambda, \sigma_a^2}$ is chosen to be $\alpha_q^2 \sigma_{s'}^2$. Polar lattices $L_1$ and $L_2$ are accordingly constructed for Gaussian noise variance $\tilde{\sigma}_c^2$ and $\tilde{\sigma}_q^2$, where
\begin{eqnarray}
\tilde{\sigma}_c^2 =\frac{\alpha_q^2\alpha_c}{\rho}\cdot\frac{\sigma_z^2\sigma_{s'}^2}{\sigma_y^2},
\end{eqnarray}
and
\begin{eqnarray}
\tilde{\sigma}_q^2 = \frac{\alpha_q^3}{\rho^2} \cdot \frac{P\sigma_{s'}^2}{\sigma_i^2}.
\end{eqnarray}
Check that $\frac{\tilde{\sigma}_c^2}{\tilde{\sigma}_q^2} = \frac{\sigma_z^2}{P+\sigma_z^2}\leq1$. Recall that $X=S'-\rho S$=$(1-\alpha_q) S'+\alpha_q S'-\rho S$. When $\alpha_q S'$ is replaced by $A$, the encoded signal, denoted by $\bar{X}$, is given by
\begin{eqnarray}\label{eqn:GPenc}
\bar{X} = \frac{1-\alpha_q}{\alpha_q} A+A-\rho S.
\end{eqnarray}
Note that the distributions of $S$ and $\bar{S}$ can be arbitrarily close when $\epsilon_{\Lambda}(\tilde{\sigma}_c) \to 0$. Clearly, $A-\rho \bar{S}$ is a Gaussian random variable independent of $A$ with distribution $N(0, \alpha_q P)$. By Lemma \ref{lem:YY'distance}, $\bar{X}$ can be very close to a Gaussian random variable with distribution $N(0, P)$.\footnote{check that $\left(\frac{1-\alpha_q}{\alpha_q}\right)^2\sigma_a^2=(1-\alpha_q)P$.} Thus, the power constraint can be satisfied.

\begin{figure}[ht]
    \centering
    \includegraphics[width=10cm]{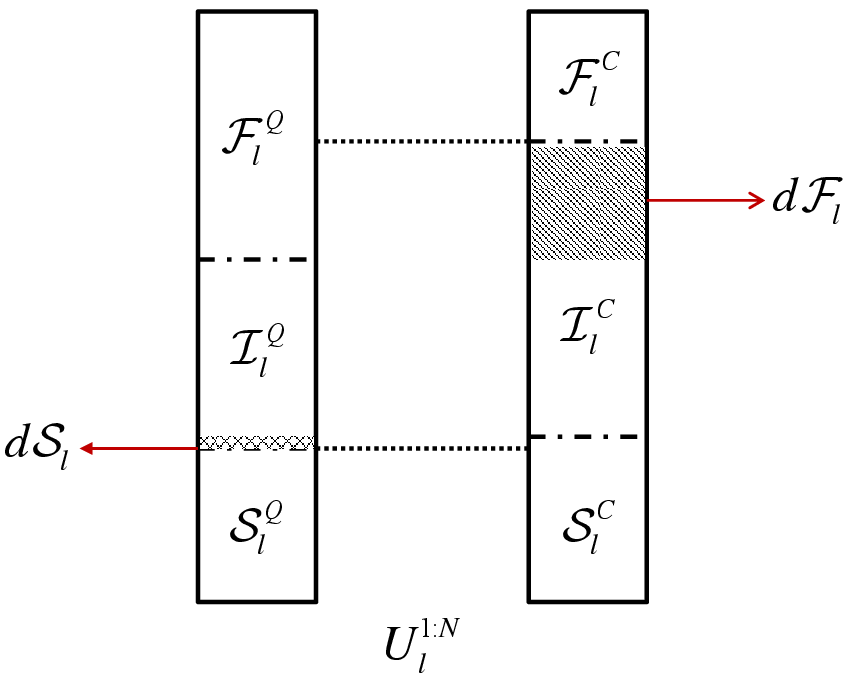}
    \caption{The partitions of $U_{\ell}^{1:N}$ for quantization lattice $L_2$ (left) and channel coding lattice $L_1$ (right). $\mathcal{F}_\ell^C \subseteq \mathcal{F}_\ell^Q$ and $\mathcal{S}_\ell^Q \subseteq \mathcal{S}_\ell^C$. Without the power constraint, $U_\ell^{\mathcal{I}_\ell^C}$ can be sent as message bits. With the power constraint, however, $U_\ell^{\mathcal{I}_\ell^Q}$ should be selected according to the interference $S^{1:N}$ and hence only $U_\ell^{d\mathcal{F}_\ell}$ can be fed with the message bits.}
    \label{fig:GPpolar}
\end{figure}

Again with some abuse of notation, let $B=\frac{\alpha_q}{\alpha_c}\rho Y$, $\bar{B}=\frac{\alpha_q}{\alpha_c}\rho \bar{Y}$, $T=\rho S$, and  $\bar{T}=\rho \bar{S}$ for convenience. We choose a good constellation $D_{\Lambda, \sigma_a^2}$ such that the flatness factor $\epsilon_{\Lambda}(\tilde{\sigma}_ c)$ is negligible. Let $\Lambda/\Lambda_1/\cdots/\Lambda_{r-1}/\Lambda'/\cdots$ be a one-dimensional binary partition chain labeled by bits $A_1/A_2/\cdots/A_{r-1}/A_r/\cdots$. Then, $P_{A_{1:r}}$ and $A_{1:r}$ approach $D_{\Lambda, \sigma_a^2}$ and $A$, respectively, as $r \to \infty$. Consider $N$ i.i.d. copies of $A$. Let $U_{\ell}^{1:N}=A_{\ell}^{1:N} G_N$ for each $1 \leq \ell \leq r$. The partition of $U_{\ell}^{1:N}$ is shown in Fig. \ref{fig:GPpolar}, where the left block is for the quantization lattice $L_2$ and the right one for the channel coding lattice $L_1$. For $0<\beta<\frac{1}{2}$, the frozen set $\mathcal{F}_\ell^Q$ $\left(\mathcal{F}_\ell^C\right)$, information set $\mathcal{I}_\ell^Q$ $(\mathcal{I}_\ell^C)$ and the shaping set $\mathcal{S}_\ell^Q$ $\left(\mathcal{S}_\ell^C\right)$ for lattice $L_2$ ($L_1$) are given by

\begin{eqnarray}\label{eqn:GPlatticeQQ}
\begin{cases}
\begin{aligned}
&\mathcal{F}_\ell^Q=\left\{i\in[N]:Z\left(U_\ell^i|U_\ell^{1:i-1},A_{1:\ell-1}^{1:N},\bar{T}^{1:N}\right)\geq1-2^{-N^{\beta}}\right\}\\
&\mathcal{I}_\ell^Q=\left\{i\in[N]:Z\left(U_\ell^i|U_\ell^{1:i-1},A_{1:\ell-1}^{1:N}\right)>2^{-N^{\beta}}\text{and }  Z\left(U_\ell^i|U_\ell^{1:i-1},A_{1:\ell-1}^{1:N},\bar{T}^{1:N}\right)<1-2^{-N^{\beta}}\right\}\\
&\mathcal{S}_\ell^Q=\left\{i\in[N]:Z\left(U_\ell^i|U_\ell^{1:i-1},A_{1:\ell-1}^{1:N}\right)\leq2^{-N^{\beta}}\right\},
\end{aligned}
\end{cases}
\end{eqnarray}
and
\begin{eqnarray}\label{eqn:GPlatticeCC}
\begin{cases}
\begin{aligned}
&\mathcal{F}_\ell^C=\left\{i\in[N]:Z\left(U_\ell^i|U_\ell^{1:i-1},A_{1:\ell-1}^{1:N},\bar{B}^{1:N}\right)\geq1-2^{-N^{\beta}}\right\}\\
&\mathcal{I}_\ell^C=\left\{i\in[N]:Z\left(U_\ell^i|U_\ell^{1:i-1},A_{1:\ell-1}^{1:N}\right)\geq 1-2^{-N^{\beta}}\text{ and }Z\left(U_\ell^i|U_\ell^{1:i-1},A_{1:\ell-1}^{1:N},\bar{B}^{1:N}\right)\leq 2^{-N^{\beta}}\right\}\\
&\mathcal{S}_\ell^C=\left\{i\in[N]:Z\left(U_\ell^i|U_\ell^{1:i-1},A_{1:\ell-1}^{1:N}\right)<1-2^{-N^{\beta}}\text{ or }2^{-N^{\beta}}<Z\left(U_\ell^i|U_\ell^{1:i-1},A_{1:\ell-1}^{1:N},\bar{B}^{1:N}\right)<1-2^{-N^{\beta}}\right\}.
\end{aligned}
\end{cases}
\end{eqnarray}
By channel degradation, we have $\mathcal{F}_\ell^C \subseteq \mathcal{F}_\ell^Q$. Let $d\mathcal{F}_\ell$ denote the set $\mathcal{F}_\ell^Q\setminus \mathcal{F}_\ell^C$. Meanwhile, we also have $\mathcal{S}_\ell^Q \subseteq \mathcal{S}_\ell^C$. The difference $d\mathcal{S}_\ell=\mathcal{S}_\ell^C\setminus \mathcal{S}_\ell^Q$ can also be written as \eqref{eqn:setdS}, and the proportion $\frac{|d\mathcal{S}_\ell|}{N} \to 0$ as $N\to \infty$.

Given an $N$-dimensional realization vector $s^{1:N}$ of $S^{1:N}$, the encoder scales $s^{1:N}$ to $t^{1:N}=\rho s^{1:N}$ and evaluates $u_{\ell}^{1:N}$ from level $1$ to level $r$ successively according to the randomized rounding quantization rules. Note that $u_\ell^{\mathcal{F}_\ell^C}$ is uniformly random and known to the decoder, and $u_\ell^{d\mathcal{F}_\ell}$ is fed with message bits which are also uniform. Recall that treating $t^{1:N}$ as a realization of $\bar{T}^{1:N}$ is reasonable because $T$ and $\bar{T}$ are similarly distributed. Then $u_{\ell}^{1:N}$ can be obtained for $1 \leq \ell \leq r$. When $r$ is sufficiently large, the lattice points outside $[-2^{r-1},2^{r-1})$ occur with almost $0$ probability, the realization $a^{1:N}$ of $A^{1:N}$ can be determined by $u_{1:r}^{1:N}$ according to \eqref{eqn:uN2xN}. Then, $x^{1:N}=\frac{1}{\alpha_q} a^{1:N}-\rho s^{1:N}$ is the encoded signal as discussed in \eqref{eqn:GPenc}.

For the encoder, the realization vector $y^{1:N}$ of $Y^{1:N}$ is scaled to $b^{1:N}=\frac{\alpha_q}{\alpha_c} \rho y^{1:N}$. The task is to recover $u_\ell^{\mathcal{I}_\ell^C}$ on each level and hence message $u_\ell^{d\mathcal{F}_\ell}$ can be reliably recovered. Note that $u_\ell^{\mathcal{F}_\ell^Q}$ is shared between the encoder and decoder before transmission, and $u_\ell^{\mathcal{S}_\ell^Q}$ can be decoded with vanishing error probability using the bit-wise MAP rule. The decoder still needs to know the unpolarized bits $u_\ell^{d\mathcal{S}_\ell}$ since the Bhattacharyya parameters of those indices are not necessarily vanishing. Therefore, a code with negligible rate is needed to send $u_\ell^{\mathcal{S}_\ell^Q}$ to the decoder in advance on each level. In this sense, $L_2$ is not exactly nested within $L_1$ because of those unpolarized indices. When $u_\ell^{d\mathcal{S}_\ell}$ is also available, $u_\ell^{\mathcal{I}_\ell^C}$ can be decoded with very small error probability \cite[Theorem 3]{LiuL15w} with $O(N\log N)$ complexity.

The Gaussian Gelfand-Pinsker coding scheme is summarized as follows.
\begin{itemize}
\item Encoding: According to the $N$-dimensional i.i.d. interference vector $S^{1:N}$, the encoder evaluates $U^{\mathcal{I}_\ell^Q}$ by randomized rounding, and then feeds $U_\ell^{d\mathcal{F}_\ell}$ with message bits. $U^{\mathcal{F}_\ell^C}$ is pre-shared and $U^{\mathcal{S}_\ell^Q}$ is determined by other bits according to $D_{\Lambda, \sigma_a^2}$. For each level, the encoder obtains $U_\ell^{1:N}$ for $1 \leq \ell \leq r$, then $A^{1:N}$ is recovered from $U_{1:r}^{1:N}$. The encoded signal is given by
    \begin{eqnarray}
    \bar{X}^{1:N}=\frac{1}{\alpha_q}A^{1:N}-\rho S^{1:N}.
    \end{eqnarray}
\item Decoding: Using the pre-shared $U^{\mathcal{F}_\ell^Q}$ and the bits $U_\ell^{d\mathcal{S}_\ell}$ by the two phases transmission, the decoder recovers $U^{\mathcal{I}_\ell^C}$ including the message bits and $U^{\mathcal{S}_\ell^Q}$ from the received signal.
\end{itemize}

For the rate of lattice codes, we have
\begin{eqnarray}
R_{L_1} \to {\frac{1}{2}\log \bigg(\frac{P\sigma_i^2+(P+\sigma_z^2)^2}{\sigma_z^2(P+\sigma_z^2)}\bigg)}^-,
\end{eqnarray}
and
\begin{eqnarray}
R_{L_2} \to {\frac{1}{2}\log \bigg(\frac{P\sigma_i^2+(P+\sigma_z^2)^2}{(P+\sigma_z^2)^2}\bigg)}^+,
\end{eqnarray}
indicating that the $R_{L_1}-R_{L_2} \to C_{GP}$.

The proof of the following theorem is given in Appendix \ref{appendix5}.

\begin{ther}\label{Thm:GP}
Let $S$ be a Gaussian noise known to the encoder, and $Z$ be another independent and unknown Gaussian noise with variance $\sigma_z^2$. Consider a power constraint $P$ for the encoded signal. Let $\Lambda/\Lambda_1/\cdots/\Lambda_{r}$ be a one-dimensional binary partition chain such that $\epsilon_{\Lambda}(\tilde{\sigma}_c)=O\left(e^{-N}\right)$ and $r=O(\log N)$. For any $0<\beta''<\beta'<0.5$, there exist two nested polar lattices $L_1$ and $L_2$ with a differential rate $R=R_{L_1}-R_{L_2}$ arbitrarily close to $\frac{1}{2}\log(1+\frac{P}{\sigma_z^2})$ such that the expect transmit power $P_T$ satisfies
\begin{eqnarray}
P_T \leq P+O\left(N^22^{-N^{\beta''}}\right),
\end{eqnarray}
and the block error probability satisfies
\begin{eqnarray}
P_e^{GP} \leq O\left(2^{-N^{\beta''}}\right).
\end{eqnarray}
\end{ther}

\begin{figure}[ht]
    \centering
    \includegraphics[width=10cm]{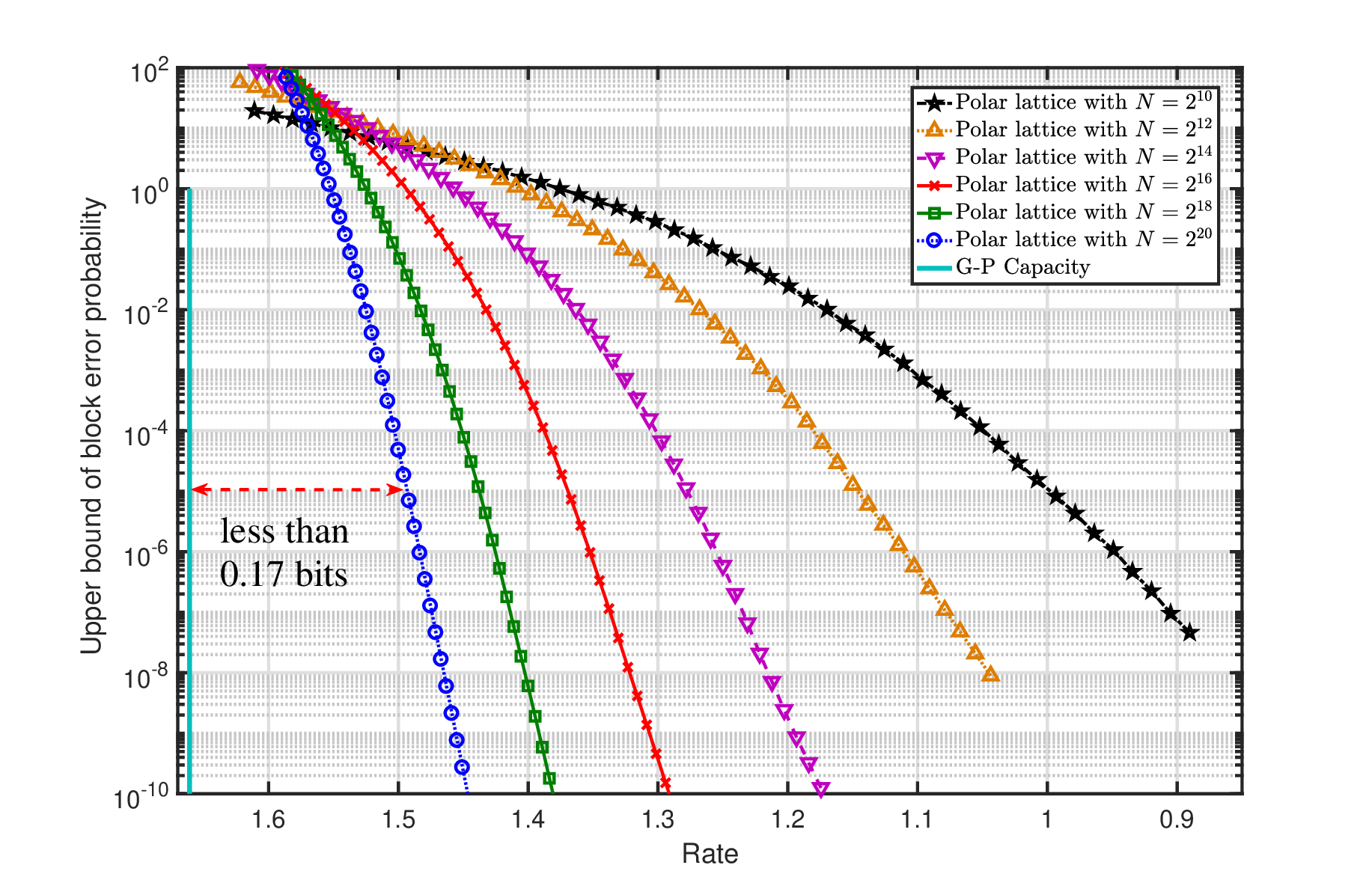}
    \caption{The upper-bounds of the block error probability of polar lattices under the SC decoding when $N=2^{10}, 2^{12}, ..., 2^{20}$.}
    \label{fig:GPBound}
\end{figure}

\begin{figure}[hb]
    \centering
    \includegraphics[width=8cm]{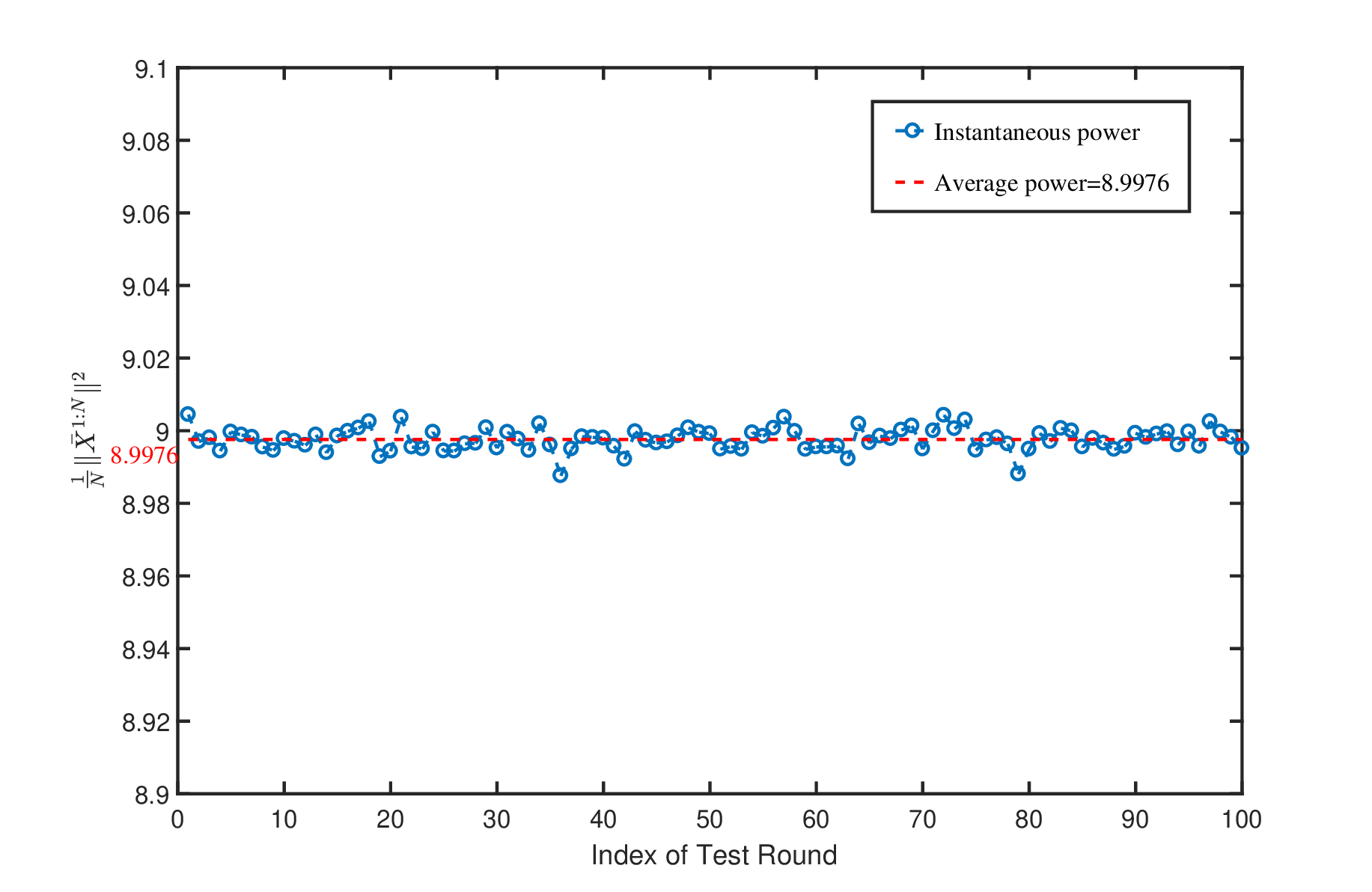}
    \caption{Samples of instantaneous transmission power when the shaping bits $U_\ell^{\mathcal{S}^C_\ell}$ are chosen according to $D_{\Lambda, \sigma_a^2}$.}
    \label{fig:GPPower}
\end{figure}

To give an example, we run numerical simulation for the case when $P=9$, $\sigma_i=3$, and $\sigma_z=1$. The capacity of the Gaussian Gelfand-Pinsker channel with these parameters is $C_{GP}=1.6610$ bits. According to the above analysis, we obtain $\rho=0.9$, $\alpha_c=0.9448$, $\alpha_q=0.4475$, and $\sigma_a^2=3.2622$. We choose the number of levels $r=5$ and $\eta=1$. The noise variances of $Q'$ and $Z'$ in Fig. \ref{fig:GPMMSE} are $0.1908$ and $3.8368$, respectively. Then, the best achievable rates for the two lattices are given by $R_{L_1}=2.0889$ and $R_{L_2}=0.4280$, respectively. The gap $R_{L_1}-R_{L_2}=1.6609$ is very close to $C_{GP}$, which verifies our settings of the binary partition chain. For the channel coding lattice $L_1$, the best achievable rates $\frac{|\mathcal{I}_\ell^C|}{N}$ are 0.2609, 0.9264, 0.7958, 0.1047 and 0.0001 for $\ell=1, 2, ..., 5$, respectively. For the quantization lattice $L_2$, the best achievable rates $\frac{|\mathcal{I}_\ell^Q|}{N}$ are 0, 0.0165, 0.3186, 0.0928 and 0 for $\ell=1, 2, ..., 5$, respectively. For both $L_1$ and $L_2$, the optimal proportion $\frac{|\mathcal{S}_\ell^C|}{N}$ of the shaping set at each level is $0$, $0.0005$, $0.2042$, $0.8953$ and $0.9999$, respectively. For finite block length $N$, we have to carefully adjust the size of $|\mathcal{S}_\ell^C|$ to make sure the power constraint is satisfied, which causes some rate loss. For the construction of polar codes, the Bhattacharyya parameters in \eqref{eqn:GPlatticeQQ} and \eqref{eqn:GPlatticeCC} are calculated by the method of Tal and Vardy in \cite{tal2011construct}. We then plot the upper-bound $\sum_{\ell=1}^{5}\sum_{i \in \mathcal{I}_\ell^C} Z(U_\ell^i|U_\ell^{1:i-1},A_{1:\ell-1}^{1:N},\bar{B}^{1:N})$ of the block error probability under the SC decoding in Fig. \ref{fig:GPBound}. Particularly, the gap between the realized rate and the capacity is smaller than 0.17 bits for a block error probability $10^{-5}$ when $N=2^{20}$. To show the power constraint is satisfied, we also plot the instantaneous transmission power $\frac{1}{N} \|\bar{X}^{1:N}\|^2$ for 100 test rounds in Fig. \ref{fig:GPPower} when $N=2^{18}$. The average transmission power is $8.9976$, slightly smaller than $P$.

\section{Lossy Gray-Wyner Coding for Gaussian RVs \label{subsec:CI for joint Gaussian} }
\begin{figure}
\begin{centering}
\includegraphics[width=6cm]{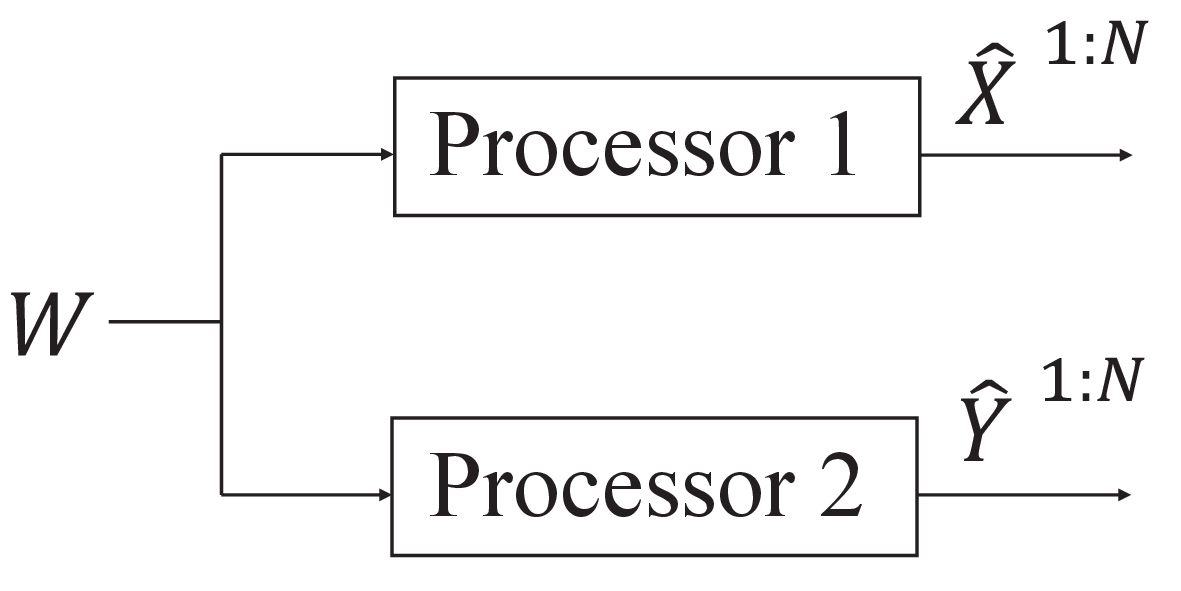}
\par\end{centering}
\caption{Gray-Wyner source coding. \label{fig:The-RV-generators}}
\end{figure}

In this section, we propose an explicit construction of polar lattices to extract Wyner's CI of two or more Gaussian sources presented in \cite{GeXulossyCI}.

First, we briefly review the original definition of Wyner's CI or the lossless CI, i.e., $C\left(X,Y\right)$ given by (\ref{eq:WynerCI}). The Gray-Wyner network \cite{gray1974source}, as depicted in Fig. \ref{fig:The-RV-generators}, demonstrates Wyner's interpretation for $C\left(X,Y\right)$. In this model, a common message $W\in\mathcal{W}$ is sent to two independent processors that generate output sequences individually according to the distributions $P_{X|W}(x|w)$ and $P_{Y|W}(y|w)$. The output sequences $\hat{X}^{1:N}$ and $\hat{Y}^{1:N}$ form a joint probability
\begin{eqnarray}
P_{\hat{X}^{1:N}\hat{Y}^{1:N}}(\hat{x}^{1:N},\hat{y}^{1:N})=\sum_{w\in\mathcal{W}}\frac{1}{|\mathcal{W}|}P_{X^{1:N}|W}(\hat{x}^{1:N}|w)P_{Y^{1:N}|W}(\hat{y}^{1:N}|w). \notag
\end{eqnarray}
Wyner showed that $C(X,Y)$ equals the minimum rate of the shared message, under the condition that the joint distribution $P_{\hat{X}^{1:N}\hat{Y}^{1:N}}(\hat{x}^{1:N},\hat{y}^{1:N})$
is arbitrarily close to $P_{X^{1:N}Y^{1:N}}(x^{1:N},y^{1:N})$.

The Gray-Wyner problem is then extended to lossy CI $C\left(\Delta_{1},\Delta_{2}\right)$, where the output sequences $(\hat{X}^{1:N},\hat{Y}^{1:N})$ have certain
distortions $\left(\Delta_{1},\Delta_{2}\right)$ with respect to $(X^{1:N},Y^{1:N})$, respectively \cite{viswanatha2014lossy}.
Consider the distortion measure $d^{N}(y^{1:N},x^{1:N})\triangleq\sum_{i=1}^{N}d(y^{i},x^{i})$, where $d(y,x)$ denotes the Euclidean distance.
The rate-distortion function for a pair of joint sources $(X,Y)\in \left(\mathcal{X},\mathcal{Y}\right)$ is defined as
\begin{eqnarray}
R_{XY}\left(\Delta_{1},\Delta_{2}\right)=\min_{P_{X'Y'|XY}(x'y'|xy):\mathsf{\mathit{E}}d(X',X)\leq\Delta_{1},\mathsf{\mathit{E}}d(Y',Y)\leq\Delta_{2}}I(X',Y';X,Y),
\notag
\end{eqnarray}
where the minimum is taken over all $P_{X'Y'|XY}(x'y'|xy)$ such that
$\mathsf{\mathit{E}}d(X',X)\leq\Delta_{1}$ and $\mathsf{\mathit{E}}d(Y',Y)\leq\Delta_{2}$.
The lossy CI is then given as follows \cite{viswanatha2014lossy}.

\begin{deft}
Given a pair of joint sources $\left(X,Y\right)\sim P_{XY}(x,y)$, for any $\Delta_{1},\Delta_{2}\geq0$, the lossy CI reads
\begin{eqnarray}
C\left(\Delta_{1},\Delta_{2}\right)=\inf I\left(X,Y;W\right), \notag
\end{eqnarray}
where the infimum is taken over all joint distributions for $X$, $Y$, $X'$, $Y'$, $W$ such that
\begin{equation}
\begin{aligned}X'-W-Y',\\
\left(X,Y\right)-\left(X',Y'\right)-W,
\end{aligned}
\label{eq:Character_lossyCI}
\end{equation}
and $\left(X',Y'\right)$ achieves $R_{XY}\left(\Delta_{1},\Delta_{2}\right)$.
\end{deft}

Note that this characterization of CI is more general than that of $C\left(X,Y\right)$ in the lossless case where $X$ and $Y$ are independent given $W$.

\subsection{Extracting CI of two Gaussian sources}
\begin{figure}
\begin{centering}
\includegraphics[scale=1]{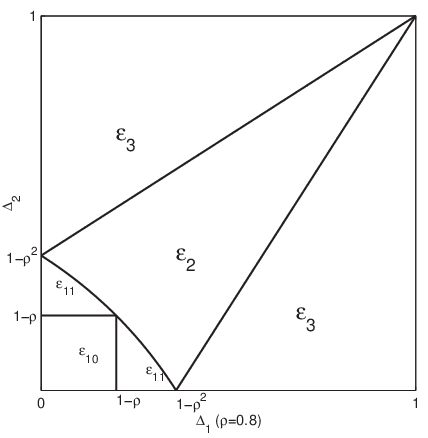}
\par\end{centering}
\caption{The distortion regions for bivariate Gaussian RVs when $\rho=0.8$.
\label{fig:The-distortion-regions-GaussianRVs}}
\end{figure}

Consider bivariate Gaussian RVs $X$, $Y$ with zero mean and covariance matrix
\begin{eqnarray}
K_{2}=\ensuremath{\begin{bmatrix}1 & \rho\\
\rho & 1
\end{bmatrix},} \notag
\end{eqnarray}
with $0<\rho<1$. The lossy CI for $\left(X,Y\right)$ has been given in \cite{GeXulossyCI}:
\begin{equation}
C\left(\Delta_{1},\Delta_{2}\right)=\begin{cases}
C\left(X,Y\right) & \left(\Delta_{1},\Delta_{2}\right)\in\varepsilon_{10},\\
R_{XY}\left(\Delta_{1},\Delta_{2}\right) & \left(\Delta_{1},\Delta_{2}\right)\in\varepsilon_{2}\cup\varepsilon_{3}\\
0 & \left(\Delta_{1},\Delta_{2}\right)\geq\left(1,1\right),
\end{cases},\label{eq:LossyCI_TwoGaussian}
\end{equation}
\begin{eqnarray}
C\left(X,Y\right)\leq C\left(\Delta_{1},\Delta_{2}\right)\leq R_{XY}\left(\Delta_{1},\Delta_{2}\right),\textrm{ }\textrm{ }\textrm{ }\left(\Delta_{1},\Delta_{2}\right)\in\varepsilon_{11}, \notag
\end{eqnarray}
where
\begin{align*}
\varepsilon_{10} & =\left\{ \left(\Delta_{1},\Delta_{2}\right):0\leq\Delta_{i}\leq1-\rho,\textrm{ }i=1,2\right\} ,\\
\varepsilon_{11} & =\varepsilon_{10}^{c}\cap\left\{ \left(\Delta_{1},\Delta_{2}\right):\Delta_{1}+\Delta_{2}-\Delta_{1}\Delta_{2}\leq1-\rho^{2}\right\} ,\\
\varepsilon_{2} & =\varepsilon_{10}^{c}\cap\varepsilon_{11}^{c}\cap\left\{ \left(\Delta_{1},\Delta_{2}\right):\min\left\{ \frac{1-\Delta_{1}}{1-\Delta_{2}},\frac{1-\Delta_{2}}{1-\Delta_{1}}\right\} \leq\rho^{2}\right\} ,\\
\varepsilon_{3} & =\varepsilon_{10}^{c}\cap\varepsilon_{11}^{c}\cap\varepsilon_{2}^{c}\cap\left\{ \left(\Delta_{1},\Delta_{2}\right):\Delta_{i}\leq1,\textrm{ }i=1,2\right\} .
\end{align*}
These distortion regions are illustrated in Fig. \ref{fig:The-distortion-regions-GaussianRVs}. An explicit expression of the rate-distortion function for $\left(X,Y\right)$ is given by \cite{nayak2010successive}:
\begin{eqnarray}
\begin{aligned} & R_{XY}\left(\Delta_{1},\Delta_{2}\right)=\begin{cases}
\frac{1}{2}\log\frac{1-\rho^{2}}{\Delta_{1}\Delta_{2}}, & \left(\Delta_{1},\Delta_{2}\right)\in\varepsilon_{1},\\
\frac{1}{2}\log\frac{1-\rho^{2}}{\Delta_{1}\Delta_{2}-\left(\rho-\sqrt{\left(1-\Delta_{1}\right)\left(1-\Delta_{2}\right)}\right)^{2}}, & \left(\Delta_{1},\Delta_{2}\right)\in\varepsilon_{2},\\
\frac{1}{2}\log\frac{1}{\min\left\{ \Delta_{1},\Delta_{2}\right\} }, & \left(\Delta_{1},\Delta_{2}\right)\in\varepsilon_{3}, \notag
\end{cases}\end{aligned}
\end{eqnarray}
where $\varepsilon_{1}=\varepsilon_{10}\cup\varepsilon_{11}$.

Next we show how to extract the lossy CI of distortion regions $\varepsilon_{10}$, $\varepsilon_{2}$ and $\varepsilon_{3}$, where the characterizations of $W$ (\ref{eq:Character_lossyCI}) can be applied.
The key step of our coding scheme is to use a discretized version of $W$, denoted by $\bar{W}$, to convey the lossy CI, according to the system model depicted in Fig. \ref{fig:The-RV-generators}.

\subsubsection{Lossy CI for region $\varepsilon_{10}$}

For $\left(\Delta_{1},\Delta_{2}\right)\in\varepsilon_{10}$, the lossy CI of $(X,Y)$ is conveyed by a Gaussian RV $W$ with mean $0$ and variance $\rho$ such that
\begin{equation}
\begin{cases}
X=W+\sqrt{1-\rho}N_{1}\\
Y=W+\sqrt{1-\rho}N_{2}
\end{cases}\label{eq:JointGaussianTestChannel}
\end{equation}
where $N_{1}$ and $N_{2}$ are standard Gaussian RVs and $N_{1},N_{2},W$ are independent of each other \cite{GeXulossyCI}. Accordingly, the lossy CI is given by
\begin{equation}
I\left(X,Y;W\right)=\frac{1}{2}\log\frac{1+\rho}{1-\rho}. \notag
\end{equation}

\begin{lem}\label{lem:Lemma Gaussian}
Let $\bar{W}$ be a RV which follows a discrete Gaussian distribution $D_{\Lambda,\sqrt{\rho}}$. Consider two continuous RVs $\bar{X}$ and $\bar{Y}$
\begin{equation}
\begin{cases}
\bar{X}=\bar{W}+\sqrt{1-\rho}N_{1}\\ \notag
\bar{Y}=\bar{W}+\sqrt{1-\rho}N_{2}
\end{cases}
\end{equation}
where $N_{1}$ and $N_{2}$ are the same as in (\ref{eq:JointGaussianTestChannel}). Let $f_{\bar{X},\bar{Y}}(x,y)$ and $f_{X,Y}(x,y)$ denote the joint PDF of $(\bar{X},\bar{Y})$ and $(X,Y)$, respectively. If $\epsilon=\epsilon_{\Lambda}\Big(\sqrt{\frac{\rho(1-\rho)}{1+\rho}}\Big)<\frac{1}{2}$, the variation distance between $f_{\bar{X},\bar{Y}}(x,y)$ and $f_{X,Y}(x,y)$ is upper-bounded by
\begin{equation}
\frac{1}{2}\int_{\mathbb{R}^{2}}\left|f_{\bar{X},\bar{Y}}(x,y)-f_{X,Y}(x,y)\right|dxdy\leq2\epsilon, \notag
\end{equation}
and the mutual information $I(\bar{X},\bar{Y};\bar{W})$ satisfies
\begin{equation}
I(\bar{{X}},\bar{{Y}};\bar{{W}})\geq\frac{1}{2}\log\frac{1+\rho}{1-\rho}-5\epsilon\log(e). \notag
\end{equation}
Therefore, $\bar{W}$ is an eligible candidate of the common CI of $(X,Y)$ when $\epsilon\to0$.
\end{lem}

\begin{IEEEproof}
Since $\bar{X}-\bar{W}-\bar{Y}$ is a Markov chain, we have
\begin{equation}
\begin{aligned} & f_{\bar{X},\bar{Y}}(x,y)\\
 & =\sum_{a\in\Lambda}f_{\bar{X},\bar{Y},\bar{W}}(x,y,a)\\
 & =\sum_{a\in\Lambda}f_{\bar{W}}(a)f_{\bar{X}|\bar{W}}(x|a)f_{\bar{Y}|\bar{W}}(y|a)\\
 & =\frac{1}{f_{\sqrt{\rho}}(\Lambda)}\sum_{a\in\Lambda}\frac{1}{\sqrt{2\pi\rho}}\exp\bigg(-\frac{a^{2}}{2\rho}\bigg)\frac{1}{\sqrt{2\pi(1-\rho)}}\exp\bigg(-\frac{(x-a)^{2}}{2(1-\rho)}\bigg)\frac{1}{\sqrt{2\pi(1-\rho)}}\exp\bigg(-\frac{(y-a)^{2}}{2(1-\rho)}\bigg)\\
 & =\frac{1}{2\pi\sqrt{1-\rho^{2}}}\exp\bigg(-\frac{x^{2}+y^{2}-2\rho xy}{2(1-\rho^{2})}\bigg)\frac{\sum_{a\in\Lambda}\frac{1}{\sqrt{2\pi\frac{\rho(1-\rho)}{1+\rho}}}\exp\bigg(-\frac{(a-\frac{\rho(x+y)}{1+\rho})^{2}}{2\cdot\frac{\rho(1-\rho)}{1+\rho}}\bigg)}{f_{\sqrt{\rho}}(\Lambda)},
\end{aligned}
\label{eq:GaussianJointProbXY}
\end{equation}
 where $\frac{1}{2\pi\sqrt{1-\rho^{2}}}\exp\bigg(-\frac{x^{2}+y^{2}-2\rho xy}{2(1-\rho^{2})}\bigg)=f_{X,Y}(x,y)$
is the PDF of two joint Gaussian RVs $X$ and $Y$. By the definition of the flatness factor (\ref{eq:flatnessFactor}), we have
\begin{equation}
\begin{aligned}\left|V(\Lambda)\sum_{a\in\Lambda}\frac{1}{\sqrt{2\pi\frac{\rho(1-\rho)}{1+\rho}}}\exp\bigg(-\frac{(a-\frac{\rho(x+y)}{1+\rho})^{2}}{2\frac{\rho(1-\rho)}{1+\rho}}\bigg)-1\right|\leq\epsilon_{\Lambda}\Bigg(\sqrt{\frac{\rho(1-\rho)}{1+\rho}}\Bigg)=\epsilon.\end{aligned}
\label{eq:GaussianFlatFactor}
\end{equation}
Since $\epsilon_{\Lambda}(\sigma)$ is a monotonically decreasing function of $\sigma$ (see \cite[Remark 2]{cong2}), we have $\epsilon_{\Lambda}(\sqrt{\rho})\leq\epsilon$ and hence
\begin{equation}
\left|V(\Lambda)f_{\sqrt{\rho}}(\Lambda)-1\right|\leq\epsilon.\label{eq:GaussianFlatFactor2}
\end{equation}

Combining (\ref{eq:GaussianJointProbXY}), (\ref{eq:GaussianFlatFactor}) and (\ref{eq:GaussianFlatFactor2}) gives us
\begin{equation}
f_{X,Y}(x,y)(1-2\epsilon)\leq f_{X,Y}(x,y)\cdot\frac{1-\epsilon}{1+\epsilon}\leq f_{\bar{X},\bar{Y}}(x,y), \notag
\end{equation}
and
\begin{equation}
f_{\bar{X},\bar{Y}}(x,y)\leq f_{X,Y}(x,y)\cdot\frac{1+\epsilon}{1-\epsilon}\leq f_{X,Y}(x,y)(1+4\epsilon), \notag
\end{equation}
when $\epsilon<\frac{1}{2}$. Finally,
\begin{align*}
\int_{\mathbb{R}^{2}}\left|f_{\bar{X},\bar{Y}}(x,y)-f_{X,Y}(x,y)\right|dxdy\leq4\epsilon\int_{\mathbb{R}^{2}}f_{X,Y}(x,y)dxdy=4\epsilon.
\end{align*}

Similarly, the Kullback-Leibler divergence between $f_{\bar{X},\bar{Y}}(x,y)$ and $f_{X,Y}(x,y)$ can be upper-bounded as
\begin{equation}
\begin{aligned}\mathbb{D}(f_{\bar{X},\bar{Y}}\|f_{X,Y}) & =\int_{\mathbb{R}^{2}}f_{\bar{X},\bar{Y}}(x,y)\log\frac{f_{\bar{X},\bar{Y}}(x,y)}{f_{X,Y}(x,y)}dxdy\\
 & \leq\int_{\mathbb{R}^{2}}f_{\bar{X},\bar{Y}}(x,y)\log(1+4\epsilon)dxdy\\
 & =\log(1+4\epsilon).
\end{aligned}
\label{eq:GaussianDistance}
\end{equation}
For any $\sqrt{\frac{\rho(1-\rho)}{1+\rho}}>0$, $\epsilon$ can be made arbitrarily small by scaling $\Lambda$. Therefore, when $\epsilon\to0$, $\bar{W}$ can be viewed as the common message sent to the processors
in Fig. \ref{fig:The-RV-generators}. To see that $I(\bar{X},\bar{Y};\bar{W})$ can be arbitrarily close to the lossy CI, we rewrite $\mathbb{D}(f_{\bar{X},\bar{Y}}\|f_{X,Y})$ as
\begin{equation}
\begin{aligned} & \mathbb{D}(f_{\bar{X},\bar{Y}}\|f_{X,Y})\\
 & =\int_{\mathbb{R}^{2}}f_{\bar{X},\bar{Y}}(x,y)\log\frac{f_{\bar{X},\bar{Y}}(x,y)}{f_{X,Y}(x,y)}dxdy\\
 & =-\int_{\mathbb{R}^{2}}f_{\bar{X},\bar{Y}}(x,y)\log f_{X,Y}(x,y)dxdy-h(\bar{X},\bar{Y})\\
 & =-\int_{\mathbb{R}^{2}}f_{\bar{X},\bar{Y}}(x,y)\log\Bigg(\frac{1}{2\pi\sqrt{1-\rho^{2}}}\exp\bigg(-\frac{x^{2}+y^{2}-2\rho xy}{2(1-\rho^{2})}\bigg)\Bigg)dxdy-h(\bar{X},\bar{Y})\\
 & =\log\big(2\pi\sqrt{1-\rho^{2}}\big)+\frac{\mathbb{\mathit{E}}_{\bar{X},\bar{Y}}[x^{2}+y^{2}-2\rho xy]}{2(1-\rho^{2})}\log(e)-h(\bar{X},\bar{Y})\\
 & =\log\big(2\pi\sqrt{1-\rho^{2}}\big)+\frac{1+\mathbb{\mathit{E}}_{\bar{W}}[w^{2}]}{1+\rho}\log(e)-h(\bar{X},\bar{Y}). \notag
\end{aligned}
\end{equation}
Note that $\epsilon_{\Lambda}(\sqrt{\rho})\leq\epsilon$. By \cite[Lemma 5]{LingBel13} and \cite[Remark 3]{LingBel13}, $\mathsf{\mathit{E}}_{\bar{W}}\left[w^{2}\right]\geq\rho(1-2\epsilon)$. Then we have
\begin{equation}
\begin{aligned}\mathbb{D}(f_{\bar{X},\bar{Y}}\|f_{X,Y}) & \geq\log\big(2\pi\sqrt{1-\rho^{2}}\big)+(1-\epsilon)\log(e)-h(\bar{X},\bar{Y})\\ \notag
 & =h(X,Y)-h(\bar{X},\bar{Y})-\epsilon\log(e).
\end{aligned}
\end{equation}

Using (\ref{eq:GaussianDistance}), we obtain
\begin{equation}
\begin{aligned}I(X,Y;W)-I(\bar{X},\bar{Y};\bar{W}) & =h(X,Y)-h(\bar{X},\bar{Y})\\ \notag
 & \leq\log(1+4\epsilon)+\epsilon\log(e)\\
 & \leq5\epsilon\log(e).
\end{aligned}
\end{equation}
\end{IEEEproof}

Hence, using $D_{\Lambda,\sqrt{\rho}}$ as the reconstruction distribution, we can design a quantization lattice to extract the lossy CI. The next theorem shows that the construction of polar lattices for this problem can be reduced to that for quantizing
a single Gaussian source.

\begin{ther}\label{thm:GaussianSource}
The construction of a polar lattice for extracting the lossy CI of a pair of joint Gaussian sources $(X,Y)$ in distortion region $\varepsilon_{10}$ is equivalent to the construction of a polar lattice achieving the rate-distortion bound of a single Gaussian source $\frac{X+Y}{2}$.
\end{ther}
\begin{IEEEproof}
Let $\bar{W}$ be labeled by bits $\bar{W}_{1:r}=\{\bar{W}_{1},\cdots,\bar{W}_{r}\}$ according to a binary partition chain $\Lambda/\Lambda_{1}/\cdots/\Lambda_{r-1}/\Lambda_r$. Then, $D_{\Lambda,\sqrt{\rho}}$ induces a distribution $P_{\bar{W}_{1:r}}$ whose limit corresponds to $D_{\Lambda,\sqrt{\rho}}$ as $r\rightarrow\infty$.
By the chain rule of mutual information,
\begin{eqnarray}
I(\bar{X},\bar{Y};\bar{W}_{1:r})=\sum_{\ell=1}^{r}I(\bar{X},\bar{Y};\bar{W}_{\ell}|\bar{W}_{1:\ell-1}), \notag
\end{eqnarray}
we obtain $r$ binary-input test channels $V_{\ell}$ for $1\leq\ell\leq r$. Proposition \ref{prop:morelevel} shows that there exists $r=O(\log N)$ such that the approximation error of the above mutual information tends to zero.

Given the realization $w_{1:\ell-1}$ of $\bar{W}_{1:\ell-1}$, denote by $\mathcal{A}_{\ell}(w_{1:\ell})$ the coset of $\Lambda_{\ell}$ indexed by $w_{1:\ell-1}$ and $w_{\ell}$. According to \cite[(5)]{multilevel}, the channel transition PDF of the $\ell$-th channel $V_{\ell}$ is given by
\begin{eqnarray}
\begin{aligned} & f_{\bar{X},\bar{Y}|\bar{W}_{\ell},\bar{W}_{1:\ell-1}}(x,y|w_{\ell},w_{1:\ell-1})\\
 & =\frac{1}{f_{\sqrt{\rho}}(\mathcal{A}_{\ell}(w_{1:\ell}))}\sum_{a\in\mathcal{A}_{\ell}(w_{1:\ell})}f_{\sqrt{\rho}}(a)f_{\bar{X},\bar{Y}|\bar{W}}(x,y|a)\\ \notag
 & =\frac{1}{f_{\sqrt{\rho}}(\mathcal{A}_{\ell}(w_{1:\ell}))}\sum_{a\in\mathcal{A}_{\ell}(w_{1:\ell})}\frac{1}{\sqrt{2\pi\rho}}\exp\bigg(-\frac{a^{2}}{2\rho}\bigg)\frac{1}{\sqrt{2\pi(1-\rho)}}\exp\bigg(-\frac{(x-a)^{2}}{2(1-\rho)}\bigg)\frac{1}{\sqrt{2\pi(1-\rho)}}\exp\bigg(-\frac{(y-a)^{2}}{2(1-\rho)}\bigg)\\
 & =\frac{1}{2\pi\sqrt{1-\rho^{2}}}\exp\bigg(-\frac{x^{2}+y^{2}-2\rho xy}{2(1-\rho^{2})}\bigg)\frac{1}{f_{\sqrt{\rho}}(\mathcal{A}_{\ell}(w_{1:\ell}))}\sum_{a\in\mathcal{A}_{\ell}(w_{1:\ell})}\frac{1}{\sqrt{2\pi\frac{\rho(1-\rho)}{1+\rho}}}\exp\bigg(-\frac{(a-\frac{\rho(x+y)}{1+\rho})^{2}}{2\cdot\frac{\rho(1-\rho)}{1+\rho}}\bigg).
\end{aligned}
\end{eqnarray}

Let $\tilde{V}_{\ell}$ be a symmetrized channel with input $\tilde{W}_{\ell}$ (assumed to be uniformly distributed) and output $\left(\bar{X},\bar{Y},\bar{W}_{1:\ell-1},\bar{W}_{\ell}\oplus\tilde{W}_{\ell}\right)$, built from the asymmetric channel $V_{\ell}$. Then the joint PDF of $V_{\ell}$ can be represented by the transition PDF of $\tilde{V}_{\ell}$ (see \cite{polarlatticeJ} for more details), as shown in the following:
\begin{equation}
\begin{aligned} & f_{\tilde{V}_{\ell}}(x,y,w_{1:\ell-1},w_{\ell}\oplus\tilde{w}_{\ell}|\tilde{w}_{\ell})\\
 & =f_{\bar{X},\bar{Y},\bar{W}_{1:\ell}}(x,y,w_{1:\ell})\\
 & =\frac{1}{2\pi\sqrt{1-\rho^{2}}}\exp\bigg(-\frac{x^{2}+y^{2}-2\rho xy}{2(1-\rho^{2})}\bigg)\frac{1}{f_{\sqrt{\rho}}(\Lambda)}\sum_{a\in\mathcal{A}_{\ell}(w_{1:\ell})}\frac{1}{\sqrt{2\pi\frac{\rho(1-\rho)}{1+\rho}}}\exp\bigg(-\frac{(a-\frac{\rho(x+y)}{1+\rho})^{2}}{2\cdot\frac{\rho(1-\rho)}{1+\rho}}\bigg).
\end{aligned}
\label{eq:GaussianSymmetrimizedF}
\end{equation}

It is readily verified that the symmetrized channel (\ref{eq:GaussianSymmetrimizedF}) is equivalent to a $\Lambda_{\ell-1}/\Lambda_{\ell}$ channel with noise variance $\frac{\rho(1-\rho)}{1+\rho}$. To construct polar lattices, we are interested in the likelihood ratio, which is only affected by the sum term of (\ref{eq:GaussianSymmetrimizedF}). It turns out that the same likelihood ratio can be obtained by quantizing a single Gaussian source $\frac{X+Y}{2}$ using the reconstruction
distribution $D_{\Lambda,\sqrt{\rho}}$.

Recall that $X$, $Y$ are bivariate Gaussian with zero mean and covariance matrix $\ensuremath{\left[\begin{smallmatrix}1 & \rho\\
\rho & 1
\end{smallmatrix}\right]}$. Therefore, $\frac{X+Y}{2}$ is Gaussian with zero mean and variance
\begin{eqnarray}
\begin{aligned}\sigma^{2}\left(\frac{X+Y}{2}\right) & \mathsf{=\mathit{E}}\left[\left(\frac{X+Y}{2}\right)^{2}\right]-\left(\mathit{E}\left[\frac{X+Y}{2}\right]\right)^{2}\\
 & =\frac{1}{4}\left(\mathit{E}\left[X^{2}+Y^{2}+2XY\right]\right)-0\\ \notag
 & =\frac{1+\rho}{2}.
\end{aligned}
\end{eqnarray}
Consider the construction of a polar lattice to quantize $\frac{X+Y}{2}$ using the reconstruction distribution $D_{\Lambda,\sqrt{\rho}}$. Denote the variance of the source and the reconstruction by $\sigma_{s}^{2}=\frac{1+\rho}{2}$ and $\sigma_{r}^{2}=\rho$, respectively. Thus, the variance of the noise is $\sigma_{z}^{2}=\sigma_{s}^{2}-\sigma_{r}^{2}=\frac{1-\rho}{2}$. Then we perform MMSE scaling. By definition, the MMSE coefficient $\alpha$ and noise variance $\tilde{\sigma}_{z}^{2}$ are given by \cite{BK:Zamir}
\begin{eqnarray}
\alpha=\frac{\sigma_{r}^{2}}{\sigma_{s}^{2}}=\frac{2\rho}{1+\rho}, \notag
\end{eqnarray}
\begin{eqnarray}
\tilde{\sigma}_{z}^{2}=\alpha\cdot\sigma_{z}^{2}=\frac{\rho(1-\rho)}{1+\rho}, \notag
\end{eqnarray}
which are the same as those in the sum term of (\ref{eq:GaussianSymmetrimizedF}).
\end{IEEEproof}

The result of Lemma \ref{lem:Lemma Gaussian} and Theorem \ref{thm:GaussianSource} will be generalized to multivariate Gaussian sources in Section \ref{subsec:MultiGaussian}. The CI of multivariate Gaussian
RVs can also be conveyed by a single discretized RV. Moreover, the construction of polar lattices can be designed in the same way as that of a single Gaussian source, given by the arithmetic mean of multiple Gaussian sources.

So far, we have shown how to extract the lossy CI for region $\varepsilon_{10}$. Next we show how to achieve the distortions $\left(\Delta_{1},\Delta_{2}\right)\in\varepsilon_{10}$ from $\bar{W}$ and the Gaussian sources $\left(X,Y\right)$.

First, the conditional rate-distortion function $R_{X|W}\left(\Delta_{1}\right)$ is defined by \cite{GrayConditionalRD}
\begin{eqnarray}
R_{X|W}\left(\Delta_{1}\right)=\min_{P_{\left(X'|X,W\right)}\left(x'|x,w\right):\mathsf{\mathit{E}}d(X',X)\leq\Delta_{1}}I\left(X;X'|W\right). \notag
\end{eqnarray}
In region $\varepsilon_{10}$, the conditional distribution of $X$ given $W$ is Gaussian with variance $1-\rho$ from the test channel $\left(\ref{eq:JointGaussianTestChannel}\right)$, therefore
\begin{eqnarray}
\begin{cases}
R_{X|W}\left(\Delta_{1}\right)=\frac{1}{2}\log\frac{1-\rho}{\Delta_{1}}\\ \notag
R_{Y|W}\left(\Delta_{2}\right)=\frac{1}{2}\log\frac{1-\rho}{\Delta_{2}}.
\end{cases}
\end{eqnarray}
Hence the condition \cite{GeXulossyCI}
\begin{eqnarray}
R_{X|W}\left(\Delta_{1}\right)+R_{Y|W}\left(\Delta_{2}\right)+I\left(X,Y;W\right)=R_{XY}\left(\Delta_{1},\Delta_{2}\right) \notag
\end{eqnarray}
is satisfied for region $\varepsilon_{10}$.

Note that the distributions of $X-\bar{W}$ and $\bar{X}-\bar{W}$ can be made arbitrarily close to each other, since $\mathbb{V}\left(f_{X}\left(x\right),f_{\bar{X}}\left(x\right)\right)\leq2\epsilon$ \cite{LingBel13} and $\epsilon=O(e^{-{N}})$ if $r=O(\log N)$. $\bar{X}-\bar{W}$ can be regarded as another Gaussian source $\bar{X}-\bar{W}=\sqrt{1-\rho}N_{1}$ according to $\left(\ref{eq:JointGaussianTestChannel}\right)$. Therefore, we can apply lossy compression to source $\sqrt{1-\rho}N_{1}$ with distortion $\Delta_{1}$. Then, the reconstruction RV can be represented as $\sqrt{1-\rho-\Delta_{1}}\bar{N}$, satisfying a discrete Gaussian distribution. Next we shall use $\sqrt{1-\rho-\Delta_{1}}\bar{N}$ to reconstruct $\bar{X'}$ through either the shared channel or the private channel. More explicitly, the reconstruction $\bar{X'}$ can be derived by
\begin{eqnarray}
\bar{X'}=\bar{W}+\sqrt{1-\rho-\Delta_{1}}\bar{N}. \notag
\end{eqnarray}
The distortion between $X$ and $\bar{X'}$ approaches $\Delta_{1}$, when the compression rate $R>\frac{1}{2}\log\frac{1-\rho}{\Delta_{1}}$, $N\rightarrow\infty$ and $\epsilon\rightarrow0$. According to the
same arguments, distortion $\Delta_{2}$ for the source $Y$ can be achieved as well.

\subsubsection{Lossy CI for region $\varepsilon_{2}\cup\varepsilon_{3}$}

For region $\varepsilon_{2}$, the lossy CI of $\left(X,Y\right)$ equals the optimal rate for a certain distortion pair of the joint Gaussian sources. It is shown in \cite{GeXulossyCI} that the RV $W$ in (\ref{eq:Character_lossyCI}) satisfying
\begin{eqnarray}
I\left(X,Y;W\right)=I\left(X,Y;X'\right)=I\left(X,Y;X',Y'\right), \notag
\end{eqnarray}
where $\left(X',Y'\right)$ achieve $R_{XY}\left(\Delta_{1},\Delta_{2}\right)$. Therefore, the extraction of lossy CI can be regarded as lossy compression that achieves $R_{XY}\left(\Delta_{1},\Delta_{2}\right)$ for a pair of Gaussian sources with zero-mean and covariance matrix $K_{2}$. Authors in \cite{nayak2010successive} proposed an optimal
backward test channel for region $\varepsilon_{2}$, which is given by
\begin{equation}
\begin{bmatrix}X\\
Y
\end{bmatrix}=\begin{bmatrix}X'\\
Y'
\end{bmatrix}+\begin{bmatrix}Z_{1}\\
Z_{2}
\end{bmatrix},\label{eq:TwoGaussianRegion2}
\end{equation}
where both $\left[X',Y'\right]^{\mathrm{T}}$ and $\left[Z_{1},Z_{2}\right]^{\mathrm{T}}$ are Gaussian vectors independent of each other and their covariance matrices are respectively given by
\begin{eqnarray}
\begin{aligned}K_{X',Y'} & =\begin{bmatrix}\delta_{1} & \sqrt{\delta_{1}\delta_{2}}\\ \notag
\sqrt{\delta_{1}\delta_{2}} & \delta_{2}
\end{bmatrix},\\
K_{Z_{1},Z_{2}} & =\begin{bmatrix}\Delta_{1} & \rho-\sqrt{\delta_{1}\delta_{2}}\\
\rho-\sqrt{\delta_{1}\delta_{2}} & \Delta_{2}
\end{bmatrix},
\end{aligned}
\end{eqnarray}
for $\left(\Delta_{1},\Delta_{2}\right)\in\varepsilon_{2}$. We use the notation
\begin{eqnarray}
\delta_{i}\triangleq1-\Delta_{i},\textrm{ }i=1,2. \notag
\end{eqnarray}
Since $K_{X',Y'}$ is singular in this region, the relation between $X'$ and $Y'$ is
\begin{eqnarray}
Y'=\sqrt{\frac{\delta_{2}}{\delta_{1}}}X'. \notag
\end{eqnarray}
Let $\bar{X'}\sim D_{\Lambda,\sqrt{\delta_{1}}}$ and $\bar{Y'}\sim D_{\Lambda,\sqrt{\delta_{2}}}$. As a result, the covariance matrix of $\left(\bar{X'},\bar{Y'}\right)$ is the same as $K_{X',Y'}$, therefore, $\left(\bar{X'},\bar{Y'}\right)$ also has the relation $\bar{Y'}=\sqrt{\frac{\delta_{2}}{\delta_{1}}}\bar{X'}$.

\begin{lem}\label{lem:TwoGaussianRegion2}
Consider two continuous RVs $\bar{X}$ and $\bar{Y}$
\begin{eqnarray}
\begin{bmatrix}\bar{X}\\ \notag
\bar{Y}
\end{bmatrix}=\begin{bmatrix}\bar{X'}\\
\bar{Y'}
\end{bmatrix}+\begin{bmatrix}Z_{1}\\
Z_{2}
\end{bmatrix},
\end{eqnarray}
where $Z_{1}$ and $Z_{2}$ are the same as that given in (\ref{eq:TwoGaussianRegion2}). Let $f_{\bar{X},\bar{Y}}(x,y)$ and $f_{X,Y}(x,y)$ denote the joint PDF of $(\bar{X},\bar{Y})$ and $(X,Y)$, respectively. If $\epsilon=\epsilon_{\Lambda}\left(\sqrt{\frac{\delta_{1}\left(\Delta_{1}\Delta_{2}-\left(\rho-\sqrt{\delta_{1}\delta_{2}}\right)^{2}\right)}{1-\rho^{2}}}\right)<\frac{1}{2}$,
the variation distance between $f_{\bar{X},\bar{Y}}(x,y)$ and $f_{X,Y}(x,y)$ is upper-bounded by
\begin{eqnarray}
\frac{1}{2}\int_{\mathbb{R}^{2}}\left|f_{\bar{X},\bar{Y}}(x,y)-f_{X,Y}(x,y)\right|dxdy\leq2\epsilon, \notag
\end{eqnarray}
and the mutual information $I(\bar{X},\bar{Y};\bar{X'})$ satisfies
\begin{eqnarray}
I(\bar{{X}},\bar{{Y}};\bar{X'})\geq I(X,Y;X')-5\epsilon\log(e). \notag
\end{eqnarray}
Therefore, $\bar{X'}$ is an eligible candidate of the CI of $(X,Y)$ when $\epsilon\to0$.
\end{lem}
\begin{IEEEproof}
See Appendix \ref{sec:Proof-of-Lemma-TwoGaussianRegion2}.
\end{IEEEproof}

\begin{ther}\label{thm:Gaussian_Region2_Them}
Given two correlated Gaussian sources $\left(X,Y\right)$ with zero mean and covariance matrix $K_{2}$ and an average distortion pair $\left(\Delta_{1},\Delta_{2}\right)\in\varepsilon_{2}$, there exists a polar lattice with rate $R>\frac{1}{2}\log\frac{1-\rho^{2}}{\Delta_{1}\Delta_{2}-\left(\rho-\sqrt{\left(1-\Delta_{1}\right)\left(1-\Delta_{2}\right)}\right)^{2}}$ such that the distortions are arbitrarily close to $\left(\Delta_{1},\Delta_{2}\right)$ if $r=O\left(\log N\right)$ and $N\rightarrow\infty$.
\end{ther}
\begin{IEEEproof}
See Appendix \ref{sec:Proof-of-Theorem-Achivability}.
\end{IEEEproof}
\begin{rem}

\label{rem:Region2}Similar to Theorem \ref{thm:GaussianSource},
the construction of polar lattices for extracting the lossy CI of Gaussian
sources $\left(X,Y\right)$ in region $\varepsilon_{2}$ is equivalent to
the construction
for a Gaussian source
\begin{eqnarray}
U=\frac{\left(\delta_{1}-\rho\sqrt{\delta_{1}\delta_{2}}\right)X+\left(\sqrt{\delta_{1}\delta_{2}}-\rho\delta_{1}\right)Y}{\delta_{1}+\delta_{2}-2\rho\sqrt{\delta_{1}\delta_{2}}}.
\notag
\end{eqnarray}
\end{rem}
%This means that the construction for a pair of Gaussian sources can be degenerated to that for a linear combination of the two sources.
It is readily verified that $U$ follows the Gaussian distribution with zero mean and variance
\begin{eqnarray}
\sigma^{2}\left[U\right]=\frac{\delta_{1}\left(1-\rho^{2}\right)}{\delta_{1}+\delta_{2}-2\rho\sqrt{\delta_{1}\delta_{2}}}.\notag
\end{eqnarray}

Consider the construction of a polar lattice to quantize $U$ using the reconstruction distribution $D_{\Lambda,\sqrt{\delta_{1}}}$. The MMSE coefficient and noise variance are respectively given by
\begin{eqnarray}
\begin{aligned}\alpha & =\frac{\delta_{1}+\delta_{2}-2\rho\sqrt{\delta_{1}\delta_{2}}}{1-\rho^{2}}\\
\sigma_{MMSE}^{2} & =\frac{\delta_{1}\left(\Delta_{1}\Delta_{2}-\left(\rho-\sqrt{\delta_{1}\delta_{2}}\right)^{2}\right)}{1-\rho^{2}}\\
 & =\delta_{1}-\frac{\delta_{1}\left(\delta_{1}+\delta_{2}-2\rho\sqrt{\delta_{1}\delta_{2}}\right)}{1-\rho^{2}}\\
 & =\delta_{1}-\sigma^{2}\left[\alpha U\right]. \notag
\end{aligned}
\end{eqnarray}
We omit the details, since the proof is similar to that of Theorem \ref{thm:GaussianSource}.

\begin{figure}
\begin{centering}
\includegraphics[scale=0.9]{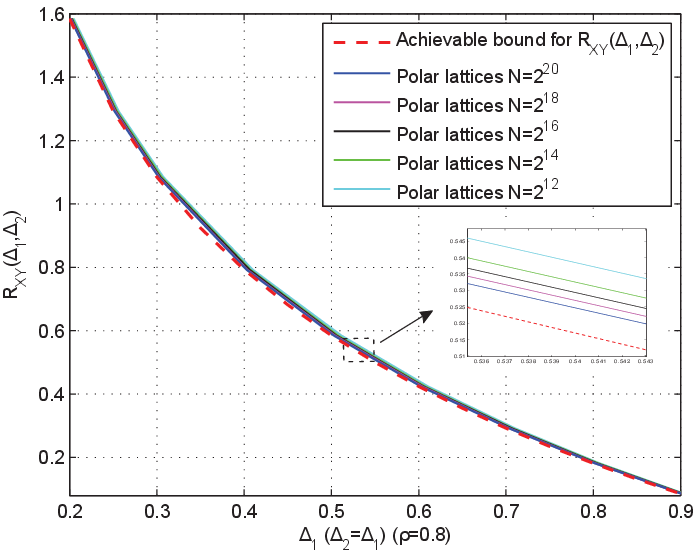}
\par\end{centering}
\caption{Simulation performance for Gray-Wyner coding of two Gaussian sources
in region $\varepsilon_{2}$. We set $\Delta_{1}=\Delta_{2}$, $\rho=0.8$.
The blocklength of polar lattices of each level is given by $N=2^{12}$,
$2^{14}$, $2^{16}$, $2^{18}$, $2^{20}$. \label{fig:Simulation-performance-for-GaussianRVs}}
\end{figure}

The simulation results of region $\varepsilon_{2}$ are depicted in Fig. \ref{fig:Simulation-performance-for-GaussianRVs}. The dashed line is the achievable bound $R_{XY}\left(\Delta_{1},\Delta_{2}\right)$ when $\left(\Delta_{1},\Delta_{2}\right)\in\varepsilon_{2}$ and $\Delta_{1}=\Delta_{2}$. The correlation of Gaussian sources $\left(X,Y\right)$ is set to $\rho=0.8$. Therefore, we have a wider distortion range where $\Delta_{1}=\Delta_{2}\in\left(0.2,1\right)$. As for the lines of simulation results with $N=2^{12},2^{14},2^{16},2^{18},2^{20}$, the horizontal axis refers to the average distortion between the practical $\Delta_{1}$ and $\Delta_{2}$. Fig. \ref{fig:Simulation-performance-for-GaussianRVs} indicates the performances of polar lattices approach the achievable
bound as the blocklength increases.

The region $\varepsilon_{3}$ is a degenerated region. If $\frac{\delta_{2}}{\delta_{1}}<\rho^{2}$,
$R_{XY}\left(\Delta_{1},\Delta_{2}\right)=\frac{1}{2}\log\frac{1}{\Delta_{1}}$, which coincides with the rate-distortion function of a single Gaussian source. This means that the optimal coding strategy is to ignore $Y$
and simply compress $X$. Then $Y'$ can be optimally estimated from $X'$ by $Y'=\rho X'$. The case where $\frac{\delta_{1}}{\delta_{2}}<\rho^{2}$ can be solved similarly.

\subsection{Extracting CI of multiple Gaussian sources\label{subsec:MultiGaussian}}
Let $\mathbf{X}_{L}\triangleq\{X_{1},X_{2},\ldots X_{L}\}$ be $L$ dependent RVs that take values in some arbitrary space $\mathcal{X}_{1}\times\mathcal{X}_{2}\times\cdots\times\mathcal{X}_{L}$. The joint distribution of $\mathbf{X}_{L}$ is denoted by $P_{\mathbf{X}_{L}}(\mathbf{x}_{L})$, which is either a probability mass function or a PDF.
\begin{deft}
Wyner's CI of multiple Gaussian sources $\mathbf{X}_{L}$ is defined as \cite{GeXulossyCI}
\begin{eqnarray}
C\left(\mathbf{X}_{L}\right)\triangleq\inf I\left(\mathbf{X}_{L};W\right), \notag
\end{eqnarray}
where the infimum is taken over all the joint distributions of $\left(\mathbf{X}_{L},W\right)$ such that
\begin{itemize}
\item the marginal distribution for $\mathbf{X}_{L}$ is $P_{\mathbf{X}_{L}}(\mathbf{x}_{L})$,
\item $\mathbf{X}_{L}$ are conditionally independent given $W$.
\end{itemize}
\end{deft}

Next, we show the construction of polar lattices to extract Wyner's CI of multiple joint Gaussian sources. For $L$ joint Gaussian RVs $\mathbf{X}_{L}=\{X_{1},X_{2},\ldots X_{L}\}$ with covariance matrix
\begin{equation}
K_{L}=\begin{bmatrix}1 & \rho & \cdots & \rho\\
\rho & 1 & \cdots & \rho\\
\vdots & \vdots & \ddots & \vdots\\
\rho & \rho & \cdots & 1
\end{bmatrix}\label{eq:covariance matrix KL}
\end{equation}
the CI of $\mathbf{X}_{L}$ is conveyed by a Gaussian RV $W$ with mean $0$ and variance $\rho$ such that
\begin{equation}
X_{i}=W+\sqrt{1-\rho}N_{i},\textrm{ }\label{eq:N gaussian TC}
\end{equation}
where $i\in\left[L\right]$. Besides, $N_{i}$ are standard Gaussian RVs independent of each other and of $W$. The CI is given by \cite{GeXulossyCI}
\begin{eqnarray}
I(\mathbf{X}_{L};W)=\frac{1}{2}\log\left(1+\frac{L\rho}{1-\rho}\right). \notag
\end{eqnarray}

We construct polar lattices by using a discretized $W$ to represent the CI of multiple Gaussian sources. The next lemma indicates that the CI of $\bar{W}$ is very close to the CI of $W$, when the flatness factor is negligible.
\begin{lem}\label{lem:Lemma L Gaussian source }
Let $\bar{W}$ be a RV which follows a discrete Gaussian distribution $D_{\Lambda,\sqrt{\rho}}$. Consider $L$ continuous RVs $\mathbf{\bar{X}}_{L}=\{X_{1},X_{2},\ldots X_{L}\}$ and relations
\begin{eqnarray}
\bar{X}_{i}=\bar{W}+\sqrt{1-\rho}N_{i},\textrm{ } \notag
\end{eqnarray}
where $i\in\left[L\right]$ and $N_{i}$ are the same as that given in (\ref{eq:N gaussian TC}). Let $f_{\mathbf{\bar{X}}_{L}}(\mathbf{x}_{L})$ and $f_{\mathbf{X}_{L}}(\mathbf{x}_{L})$ denote the joint PDF of $\mathbf{\bar{X}}_{L}=\{\bar{X}_{1},\bar{X}_{2},\ldots\bar{X}_{L}\}$ and $\mathbf{X}_{L}=\{X_{1},X_{2},\ldots X_{L}\}$, respectively. If $\epsilon=\epsilon_{\Lambda}\left(\sqrt{\frac{\rho(1-\rho)}{1+(L-1)\rho}}\right)<\frac{1}{2}$, the variation distance between $f_{\mathbf{\bar{X}}_{L}}(\mathbf{x}_{L})$ and $f_{\mathbf{X}_{L}}(\mathbf{x}_{L})$ is upper-bounded by
\begin{eqnarray}
\frac{1}{2}\int_{\mathbb{R}^{L}}\left|f_{\mathbf{\bar{X}}_{L}}(\mathbf{x}_{L})-f_{\mathbf{X}_{L}}(\mathbf{x}_{L})\right|dx_{1}dx_{2}\ldots dx_{L}\leq2\epsilon, \notag
\end{eqnarray}
and the mutual information $I(\mathbf{\bar{X}}_{L};\bar{W})$ satisfies
\begin{eqnarray}
I(\mathbf{\bar{X}}_{L};\bar{W})\geq\frac{1}{2}\log\left(1+\frac{L\rho}{1-\rho}\right)-5\epsilon\log\left(e\right). \notag
\end{eqnarray}
Therefore, $\bar{W}$ is an eligible candidate of the CI of \textup{$\mathbf{X}_{L}=\{X_{1},X_{2},\ldots X_{L}\}$ when $\epsilon\rightarrow0$.}
\end{lem}
\begin{IEEEproof}
See Appendix \ref{sec:Proof-of-Lemma-LGaussian}.
\end{IEEEproof}

Therefore, the reconstruction follows the distribution $D_{\Lambda,\sqrt{\rho}}$. However, it is a complicated problem to process a number of sources together. The next theorem shows that the construction of polar lattices for extracting CI of multiple Gaussian sources is again the same as that for a single Gaussian source.

\begin{ther}\label{thm:Theorem L Gaussian source}
The construction of a polar lattice for extracting the CI of $L$ joint Gaussian sources $\mathbf{X}_{L}=\{X_{1},X_{2},\ldots X_{L}\}$ is equivalent to the construction of a
polar lattice achieving the rate-distortion bound for a Gaussian source $\frac{X_{1}+X_{2}+\ldots+X_{L}}{L}$.
\end{ther}
\begin{IEEEproof}
See Appendix \ref{sec:Proof-of-Theorem-LGaussian}.
\end{IEEEproof}

\section{Concluding Remarks}
In this work, we have presented an explicit construction of polar lattices which are good for lossy compression. Compared with the original idea given in \cite{zamir3}, the entropy encoder is integrated with lattice quantization in our proposed lattice codes. They were utilized to solve the Gaussian version of the Wyner-Ziv and the Gelfand-Pinsker problems. For lossy compression of Gaussian sources, the complexity of encoding and decoding is $O(N \log^2 N)$ for a sub-exponentially decaying excess distortion. Since the same complexity holds for polar lattices to achieve the capacity of AWGN channels, the complexities for both the Wyner-Ziv and the Gelfand-Pinsker problems are $O(N \log^2 N)$.

The construction of polar lattices was further extended to extract the common information of Gaussian sources for each best-known distortion region. More importantly, it was found that the construction for a pair of Gaussian sources is equivalent to that for a single Gaussian source. Therefore, a polar lattice designed for a Gaussian source can be directly used to extract the common information of a pair of Gaussian sources or even multiple Gaussian sources.
We have not addressed region $\varepsilon_{11}$ though, since its distortion region is unknown in literature; this is left as an open problem. The case of multiple Gaussian sources with a general covariance matrix is another open problem for future research.

%\section*{Acknowledgments}
%The authors would like to thank Prof. Jean-Claude Belfiore for helpful discussions.
%\appendices
%\section{Proof of Theorem} \label{appendix0}
\appendices

\section{Proof of Theorem \ref{Thm:Smalldistance1}}\label{appendix1}
\begin{proof}
Firstly, we change the encoding rule for the $u_1^i$ in $i \in \mathcal{S}_1$ and \eqref{eqn:lossyencoder2} is modified to
\begin{eqnarray}\label{eqn:lossyencoder3}
&&\hspace{-2em}u_1^i=
\begin{cases}
\begin{aligned}
&\,\,\,\bar{u}_1^i \,\,\,\,\,\text{if} \,\,\,\,\, i \in \mathcal{F}_1\\
&\begin{cases}
\begin{aligned}
&0 \,\,\,\,\,\text{w. p.} \,\,\,\,\, P_{U_1^i|U_1^{1:i-1}}\left(0|u_1^{1:i-1}\right)\\
&1 \,\,\,\,\,\text{w. p} \,\,\,\,\, P_{U_1^i|U_1^{1:i-1}}\left(1|u_1^{1:i-1}\right)
\end{aligned}
\end{cases}
\text{if} \,\,\,\,\, i \in \mathcal{S}_1.
\end{aligned}
\end{cases}
\end{eqnarray}

Let $Q'_{U_1^{1:N},Y'^{1:N}}\left(u_1^{1:N},y'^{1:N}\right)$ denote the associate joint distribution for $U_1^{1:N}$ and $Y_1^{1:N}$ according to the encoding rule described in \eqref{eqn:lossyencoder1} and \eqref{eqn:lossyencoder3}. Then the variational distance between $P_{U^{1:N},Y^{1:N}}$ and $Q'_{U^{1:N},Y^{1:N}}$ can be bounded as follows.
%\allowdisplaybreaks{
\begin{eqnarray}\label{eqn:longone}
\begin{aligned}
&\hspace{-3em}2\mathbb{V}\left(P_{U_1^{1:N},Y'^{1:N}},Q'_{U_1^{1:N},Y'^{1:N}}\right) \\
&\hspace{-3em}= \sum_{u_1^{1:N},y'^{1:N}} \left|Q'(u_1^{1:N},y'^{1:N})-P(u_1^{1:N},y'^{1:N})\right|\\
&\hspace{-3em}\stackrel{(a)}=\sum_{u_1^{1:N},y'^{1:N}}\left|\sum_i\left(Q'(u_1^i|u_1^{1:i-1},y'^{1:N})-P(u_1^i|u_1^{1:i-1},y'^{1:N})\right)\left(\prod_{j=1}^{i-1}P(u_1^j|u_1^{1:j-1},y'^{1:N})\right)
\left(\prod_{j=i+1}^{N}Q'(u_1^j|u_1^{1:j-1},y'^{1:N})\right)P(y'^{1:N})\right|\\
&\hspace{-3em}\stackrel{(b)}\leq\sum_{i\in\mathcal{F}_1\cup\mathcal{S}_1}\sum_{u_1^{1:N},y'^{1:N}}\left|Q'(u_1^i|u_1^{1:i-1},y'^{1:N})-P(u_1^i|u_1^{1:i-1},y'^{1:N})\right|\left(\prod_{j=1}^{i-1}P(u_1^j|u_1^{1:j-1},y'^{1:N})\right)
\left(\prod_{j=i+1}^{N}Q'(u_1^j|u_1^{1:j-1},y'^{1:N})\right)P(y'^{1:N})\\
&\hspace{-3em}=\sum_{i\in\mathcal{F}_1\cup\mathcal{S}_1}\sum_{u_1^{1:i},y'^{1:N}}\left|Q'(u_1^i|u_1^{1:i-1},y'^{1:N})-P(u_1^i|u_1^{1:i-1},y'^{1:N})\right|\left(\prod_{j=1}^{i-1}P(u_1^j|u_1^{1:j-1},y'^{1:N})\right)P(y'^{1:N})\\
&\hspace{-3em}=\sum_{i\in\mathcal{F}_1\cup\mathcal{S}_1} \sum_{u_1^{1:i-1},y'^{1:N}} 2P\left(u_1^{1:i-1},y'^{1:N}\right)\mathbb{V}\left(Q'_{U^i|U_1^{1:i-1}=u_1^{1:i-1},Y'^{1:N}=y'^{1:N}},P_{U_1^i|U_1^{1:i-1}=u_1^{1:i-1},Y'^{1:N}=y'^{1:N}}\right)\\
&\hspace{-3em}\stackrel{(c)}\leq \sum_{i\in\mathcal{F}_1\cup\mathcal{S}_1} \sum_{u_1^{1:i-1},y'^{1:N}} P\left(u_1^{1:i-1},y'^{1:N}\right) \sqrt{2\ln2 D\left(P_{U_1^i|U_1^{1:i-1}=u_1^{1:i-1},Y'^{1:N}=y'^{1:N}}||Q'_{U_1^i|U_1^{1:i-1}=u_1^{1:i-1},Y'^{1:N}=y'^{1:N}}\right)}\\
&\hspace{-3em}\stackrel{(d)}\leq \sum_{i\in\mathcal{F}_1\cup\mathcal{S}_1} \sqrt{2\ln2 \sum_{u_1^{1;i-1},y'^{1:N}}P\left(u_1^{1:i-1},y'^{1:N}\right) D\left(P_{U_1^i|U_1^{1:i-1}=u_1^{1:i-1},Y'^{1:N}=y'^{1:N}}||Q'_{U_1^i|U_1^{1:i-1}=u_1^{1:i-1},Y'^{1:N}=y'^{1:N}}\right)}\\
&\hspace{-3em}\leq \sum_{i\in\mathcal{F}_1\cup\mathcal{S}_1} \sqrt{2\ln2 D\left(P_{U_1^i}||Q'_{U_1^i}|U_1^{1:i-1},Y'^{1:N}\right)}\\
&\hspace{-3em}\stackrel{(e)}\leq \sum_{i\in\mathcal{F}_1} \sqrt{2\ln2\left(1-H(U_1^i|U_1^{1:i-1},Y'^{1:N})\right)}+\sum_{i\in\mathcal{S}_1} \sqrt{2\ln2\left(H(U_1^i|U_1^{1:i-1})-H(U_1^i|U_1^{1:i-1},Y'^{1:N})\right)}\\
&\hspace{-3em}\stackrel{(f)}\leq \sum_{i\in\mathcal{F}_1} \sqrt{2\ln2\left(1-Z(U_1^i|U_1^{1:i-1},Y'^{1:N})^2\right)}+\sum_{i\in\mathcal{S}_1} \sqrt{2\ln2\left(Z(U_1^i|U_1^{1:i-1})-Z(U_1^i|U_1^{1:i-1},Y^{1:N})^2\right)}\\
&\hspace{-3em}\stackrel{(g)}\leq 2N\sqrt{4\ln2\cdot2^{-N^\beta}}=O\left(2^{-N^{\beta'}}\right), \notag\
\end{aligned}
\end{eqnarray}
where $D(\cdot||\cdot)$ is the Kullback-Leibler divergence, and the equalities and the inequalities follow from
\begin{itemize}
\item[] \hspace{-2em} $(a)$ The telescoping expansion \cite{aspolarcodes}.
\item[] \hspace{-2em} $(b)$ $Q'\left(u_1^i|u_1^{1:i-1},y'^{1:N}\right)=P\left(u_1^i|u_1^{1:i-1},y'^{1:N}\right)$ for $i \in \mathcal{I}_1$.
\item[] \hspace{-2em} $(c)$ Pinsker's inequality.
\item[] \hspace{-2em} $(d)$ Jensen's inequality.
\item[] \hspace{-2em} $(e)$ $Q'\left(u_1^i|u_1^{1:i-1}\right)=\frac{1}{2}$ ($\bar{u}_1^i$ is uniformly random) for $i \in \mathcal{F}_1$ and
$Q'_{U_1^i|U_1^{1:i-1},Y'^{1:N}}=P_{U_1^i|U_1^{1:i-1}}$ for $i \in \mathcal{S}_1$.
\item[] \hspace{-2em} $(f)$ $Z(X|Y)^2<H(X|Y)<Z(X|Y)$ \cite{polarsource}.
\item[] \hspace{-2em} $(g)$ Eq. \eqref{eqn:asymdefinition1}.
\end{itemize}

When using the MAP decision in \eqref{eqn:lossyencoder2} for $i \in \mathcal{S}_1$, we obtain the joint distribution $Q_{U_1^{1:N},Y'^{1:N}}$. Following the same fashion,
\begin{eqnarray}
\begin{aligned}
&2\mathbb{V}\left(Q'_{U_1^{1:N},Y'^{1:N}},Q_{U_1^{1:N},Y'^{1:N}}\right)\\\notag\
&\leq \sum_{i\in\mathcal{S}_1} \sqrt{2\ln2 D\left(Q_{U_1^i}||Q'_{U^i}|U_1^{1:i-1},Y'^{1:N}\right)}\\ \notag\
&\stackrel{(h)}\leq\sum_{i\in\mathcal{S}_1} \sqrt{2\ln2\left(H(U_1^i|U_1^{1:i-1})-0\right)}\\ \notag\
&\leq\sum_{i\in\mathcal{S}_1} \sqrt{2\ln2Z\left(U_1^i|U_1^{1:i-1}\right)} \leq N\sqrt{2\text{ln}2\cdot2^{-N^\beta}}=O\left(2^{-N^{\beta'}}\right),
\end{aligned}
\end{eqnarray}
where inequality $(h)$ follows from the MAP decision in \eqref{eqn:lossyencoder2} for $i \in \mathcal{S}_1$, and the relationship $p_0\log(\frac{1}{p_0})+p_1\log(\frac{1}{p_1})-0\log(\frac{0}{p_0})-1\log(\frac{1}{p_1})=p_0\log(\frac{p_1}{p_0}) \geq 0$ when $p_0+p_1=1$ and $p_1\geq p_0 \geq 0$.

Finally, we have
\begin{eqnarray}
\begin{aligned}
&\hspace{-1em}\mathbb{V}\left(P_{U_1^{1:N},Y'^{1:N}},Q_{U_1^{1:N},Y'^{1:N}}\right)  \\\notag\
&\hspace{-1em}\leq \mathbb{V}\left(P_{U_1^{1:N},Y'^{1:N}},Q'_{U_1^{1:N},Y'^{1:N}}\right)+\mathbb{V}\left(Q'_{U_1^{1:N},Y'^{1:N}},Q_{U_1^{1:N},Y'^{1:N}}\right)\\ \notag\
&\hspace{-1em}=O\left(2^{-N^{\beta'}}\right).
\end{aligned}
\end{eqnarray}
Clearly, when $N$ goes to infinity, for any $R>\frac{|\mathcal{I}_1|}{N}=I(X_1;Y')$, $\mathbb{V}\left(P_{U_1^{1:N},Y'^{1:N}},Q_{U_1^{1:N},Y'^{1:N}}\right)$ can be arbitrarily small.
\end{proof}

\section{Proof of Theorem \ref{Thm:Smalldistance2}}\label{appendix2}
\begin{proof}
The variational distance can be upper bounded as follows.
\begin{eqnarray}\label{eqn:longone1}
\begin{aligned}
&2\mathbb{V}\left(P_{U_2^{1:N},U_1^{1:N},Y'^{1:N}},Q_{U_2^{1:N},U_1^{1:N},Y'^{1:N}}\right)\\
&= \sum_{u_2^{1:N},u_1^{1:N},y'^{1:N}} \left|Q(u_2^{1:N},u_1^{1:N},y'^{1:N})-P(u_2^{1:N},u_1^{1:N},y'^{1:N})\right|\\
&=\sum_{u_2^{1:N},u_1^{1:N},y'^{1:N}}\left|P(u_2^{1:N}|u_1^{1:N},y'^{1:N})P(u_1^{1:N},y'^{1:N})-Q(u_2^{1:N}|u_1^{1:N},y'^{1:N})Q(u_1^{1:N},y'^{1:N})\right|\\
&\leq\sum_{u_2^{1:N},u_1^{1:N},y'^{1:N}}\left|P(u_2^{1:N}|u_1^{1:N},y'^{1:N})-Q(u_2^{1:N}|u_1^{1:N},y'^{1:N})\right|P\left(u_1^{1:N},y'^{1:N}\right)\\
&\hspace{+4em}+\sum_{u_2^{1:N},u_1^{1:N},y'^{1:N}}\left|P(u_1^{1:N},y'^{1:N})-Q(u_1^{1:N},y'^{1:N})\right|Q\left(u_2^{1:N}|u_1^{1:N},y'^{1:N}\right).
\end{aligned}
\end{eqnarray}

Treating $\left(U_1^{1:N},Y'^{1:N}\right)$ as a new source with distribution $P\left(u_1^{1:N},y'^{1:N}\right)$, the first summation can be proved to be $O\left(2^{-N^{\beta'}}\right)$ in the same fashion as the proof of Theorem \ref{Thm:Smalldistance1}. For the second summation, we have
\begin{eqnarray}
\begin{aligned}
&\sum_{u_{1:2}^{1:N},y'^{1:N}}\left|P(u_1^{1:N},y'^{1:N})-Q(u_1^{1:N},y'^{1:N})\right|Q\left(u_2^{1:N}|u_1^{1:N},y'^{1:N}\right)\\
&=\sum_{u_1^{1:N},y'^{1:N}}\left|P(u_1^{1:N},y'^{1:N})-Q(u_1^{1:N},y'^{1:N})\right|\\
&=2\mathbb{V}\left(P_{U_1^{1:N},Y'^{1:N}},Q_{U_1^{1:N},Y'^{1:N}}\right)=O\left(2^{-N^{\beta'}}\right).
\end{aligned}
\end{eqnarray}

Finally,
\begin{eqnarray}
\mathbb{V}\left(P_{U_2^{1:N},U_1^{1:N},Y'^{1:N}},Q_{U_2^{1:N},U_1^{1:N},Y'^{1:N}}\right)=O\left(2\cdot 2^{-N^{\beta'}}\right). \notag\
\end{eqnarray}
\end{proof}

\section{Proof of Theorem \ref{Thm:quantizationMain}}\label{appendix3}
\begin{proof}
Firstly, for the source $Y'$, we consider the average performance of the multilevel polar codes with all possible choice of $u_\ell^{\mathcal{F}_\ell}$ on each level. If the encoding rule described in the form of \eqref{eqn:lossyencoder4} is used for all $i \in [N]$ on each level, the resulted average distortion is given by
\begin{eqnarray}
D_{P,Y'}=\frac{1}{N}\sum_{u_{1:r}^{1:N},y'^{1:N}}P_{U_{1:r}^{1:N},Y'^{1:N}}\left(u_{1:r}^{1:N},y'^{1:N}\right)d\left(y'^{1:N},\mathcal{M}(u_{1:r}^{1:N}G_N)\right), \notag\
\end{eqnarray}
where $\mathcal{M}\left(u_{1:r}^{1:N}G_N\right)$ denotes a mapping from $u_{1:r}^{1:N}$ to $x^{1:N}$ according to \eqref{eqn:uN2xN} (remind that $x$ is drawn from $\Lambda$ according to $D_{\Lambda,\sigma_r}$). For instance, let $\Lambda=\mathbb{Z}$ and the partition is given by $\mathbb{Z}/2\mathbb{Z}/...2^{r}\mathbb{Z}$, then we have the coset $\chi^{1:N}=x_1^{1:N}+2x_2^{1:N}+...+2^{r-1}x_r^{1:N}+2^r \mathbb{Z}^N$ with $x_\ell^{1:N}=u_\ell^{1:N}G_N$ for $1\leq \ell \leq r$, and $x^{1:N}= \chi^{1:N} \mod 2^r\mathbb{Z}^N$. When $r \to \infty$, there exists a one-to-one mapping from $u_{1:r}^{1:N}$ to $x^{1:N}$. For a finite $r = O(\log N)$, the mapping works well for $x$ lying in the interval $[-2^{r-1}, 2^{r-1})$. Let $\kappa=2^{r-1}$ to simplify the notations. We divide the lattice points into the two sets $|x| \leq \kappa$ and  $|x| > \kappa$, respectively. We now calculate $D_{P,Y'}$ from the joint distribution $P_{X,Y'}$ directly and consider the penalty caused by the modulo $2^r\mathbb{Z}^N$ in the squared-error distortion measure $d(x,y)$ as follows.
\begin{eqnarray}\label{eqn:DPYbound}
\begin{aligned}
D_{P,Y'}&=\frac{1}{N}\sum_{x^{1:N},y'^{1:N}}P_{X^{1:N},Y'^{1:N}}\left(x^{1:N},y'^{1:N}\right)d\left(y'^{1:N},x^{1:N}\right)\\
&=\frac{1}{N}\cdot N \sum_{x,y'}P_{X,Y'}(x,y')d(x,y') \\
&=\sum_{|x|\leq \kappa, y'} P_{X,Y'}(x,y')d(x,y') + \sum_{|x|> \kappa, y'} P_{X,Y'}(x,y')d(x,y')\\
&\leq \sum_{x \in \Lambda} P_X(x)\int_{-\infty}^{+\infty}\frac{1}{\sqrt{2\pi \Delta}} \exp\left(-\frac{(y'-x)^2}{2\Delta}\right)(y'-x)^2 dy' + \sum_{|x|> \kappa, y'} P_{X,Y'}(x,y')d(x,y')\\
&=\Delta + \sum_{|x|> \kappa, |y'|\leq \kappa} P_{X,Y'}(x,y')d(x,y') + \sum_{|x|> \kappa, |y'|>\kappa} P_{X,Y'}(x,y')d(x,y') \\
&\leq \Delta + \sum_{|x|> \kappa, |y'|\leq \kappa} P_{X,Y'}(x,y')(2\kappa)^2 + \sum_{|x|> \kappa, |y'|>\kappa} P_{X,Y'}(x,y')(2y)^2,
\end{aligned}
\end{eqnarray}
where the first inequality holds because the modulo $2^r\mathbb{Z}$ operation does not affect when $|x|\leq \kappa$ and the squared-error distortion measure is given by $d(x,y')=(y'-x)^2$. For the second inequality, it holds because the modulo $2^r\mathbb{Z}$ operation maps $x$ to the interval $[-\kappa,\kappa)$ when $|x|> \kappa$, we then have $d(x,y')\leq (2\kappa)^2$ and $d(x,y')\leq (2y')^2$ for the two cases $|y'|\leq \kappa$ and $|y'|> \kappa$, respectively.

For the last term in the above inequality, we have the following upper-bound.
\begin{eqnarray}
\begin{aligned}\notag\
\int_\kappa^\infty\frac{1}{\sqrt{2\pi\Delta}}e^{-\frac{-(y-x)^2}{2\Delta}}y^2dy&\overset{t=y-x}{=}\int_{\kappa-x}^\infty \frac{1}{\sqrt{2\pi\Delta}} e^{-\frac{t^2}{2\Delta}}(t+x)^2 dt \\ \notag\
&=\int_{\kappa-x}^\infty \frac{1}{\sqrt{2\pi\Delta}} e^{-\frac{t^2}{2\Delta}}t^2 dt+2x\int_{\kappa-x}^\infty \frac{1}{\sqrt{2\pi\Delta}} e^{-\frac{t^2}{2\Delta}}t dt + x^2\int_{\kappa-x}^\infty \frac{1}{\sqrt{2\pi\Delta}} e^{-\frac{t^2}{2\Delta}} dt \\ \notag\
&\leq \Delta\left(\frac{1}{\sqrt{2\pi}}\frac{\kappa-x}{\sqrt{\Delta}}e^{-\frac{(\kappa-x)^2}{2\Delta}}+e^{-\frac{(\kappa-x)^2}{2\Delta}}\right)+ 2x\frac{\sqrt{\Delta}}{\sqrt{2\pi}}e^{-\frac{(\kappa-x)^2}{2\Delta}} + x^2 e^{-\frac{(\kappa-x)^2}{2\Delta}},
\end{aligned}
\end{eqnarray}
where we used the integration by parts and the Chernoff bound. It can be seen that $x^2 e^{-\frac{(\kappa-x)^2}{2\Delta}}$ is the dominate term and there exists a maximum value for $|x|>\kappa$. Therefore, we have $\int_\kappa^\infty\frac{1}{\sqrt{2\pi\Delta}}e^{-\frac{-(y-x)^2}{2\Delta}}y^2dy \leq 2\kappa^2$ for a large $\kappa$ and similarly $\int_{-\infty}^{-\kappa}\frac{1}{\sqrt{2\pi\Delta}}e^{-\frac{-(y-x)^2}{2\Delta}}y^2dy \leq 2\kappa^2$.

Using the above upper-bound in \eqref{eqn:DPYbound}, we have
\begin{eqnarray}\notag
\begin{aligned}
D_{P,Y'}&\leq \Delta + \sum_{|x|> \kappa, |y'|\leq \kappa} P_{X,Y'}(x,y')(2\kappa)^2 + \sum_{|x|> \kappa, |y'|>\kappa} P_{X,Y'}(x,y')(2y)^2, \\
&\leq \Delta + 4\kappa^2\sum_{|x|> \kappa, y'} P_{X,Y'}(x,y') + 4\sum_{|x|> \kappa} P_{X}(x) \int_{|y'|>\kappa} P_{Y'|X}(y'|x)y^2 dy\\
&\leq \Delta + 20 \kappa^2\sum_{|x|> \kappa} P_{X}(x)
\end{aligned}
\end{eqnarray}

Recall that $X \sim D_{\Lambda, \sigma_r^2}$, where $\Lambda=\eta \mathbb{Z}$ and $\eta$ is a scaling factor making $\epsilon_{\Lambda}(\tilde{\sigma})$ sufficiently small. By the definition of discrete Gaussian distribution, we have
\begin{eqnarray}
\sum_{x \in \Lambda, |x|>\kappa} P_X(x)&=&\frac{2\sum_{x=\eta \mathbb{Z}, x>\kappa} \exp\left(\frac{-x^2}{2\sigma_r^2}\right)}{\sum_{x=\eta \mathbb{Z}} \exp\left(\frac{-x^2}{2\sigma_r^2}\right)}\\
&\leq& \frac{2}{\sum_{x=\eta \mathbb{Z}} \exp\left(\frac{-x^2}{2\sigma_r^2}\right)} \frac{\exp\left(-\frac{\kappa^2}{2\sigma_r^2}\right)}{1-\exp\left(\frac{\eta \kappa}{\sigma_r^2}\right)}\\
&=& O\left(C_1^{-N^2}\right), \label{eq:2region2}
\end{eqnarray}
where $C_1$ is a constant larger than 1, and we used the fact that $r=O(\log N)$ and $\kappa=2^{r-1}$.

Consequently, we arrive at
\begin{eqnarray}\label{eqn:DPYboundFinal}
D_{P,Y'} \leq \Delta + O\left(N^2C_1^{-N^2}\right),
\end{eqnarray}
for a constant $C_1>1$.

The result $\lim_{N\to\infty}D_{P,Y'}=\Delta$ is reasonable since the encoder does not do any compression. Now we consider the average distortion $D_{Q,Y}$. To do so we replace the joint distribution $P_{U_{1:r}^{1:N},Y'^{1:N}}\left(u_{1:r}^{1:N},y'^{1:N}\right)$ with $Q_{U_{1:r}^{1:N},Y^{1:N}}\left(u_{1:r}^{1:N},y'^{1:N}\right)$ and compress $y'^{1:N}$ to $u_\ell^{\mathcal{I}_\ell}$ on each level according to the rule given by \eqref{eqn:lossyencoder4} and \eqref{eqn:lossyencoder5}. Please notice that for a same realization $y^{1:N}$ of $Y^{1:N}$ and $Y'^{1:N}$, we have $Q_{U_{1:r}^{1:N}|Y^{1:N}}\left(u_{1:r}^{1:N}|y^{1:N}\right)=Q_{U_{1:r}^{1:N}|Y'^{1:N}}\left(u_{1:r}^{1:N}|y^{1:N}\right)$ since the same encoder is used. The expected average distortion $D_{Q,Y}$ can be divided into the following two parts according to the $N$-dimensional cube $[-\kappa,\kappa]^N$ as
\begin{eqnarray}\label{eqn:DQY}
\begin{aligned}
D_{Q,Y}&=\frac{1}{N}\sum_{u_{1:r}^{1:N},y^{1:N}}Q_{U_{1:r}^{1:N},Y^{1:N}}\left(u_{1:r}^{1:N},y^{1:N}\right)d\left(y^{1:N},\mathcal{M}(u_{1:r}^{1:N}G_N)\right) \\
&=\frac{1}{N}\sum_{\underset{|y^i|\leq \kappa, \forall i \in [N]}{u_{1:r}^{1:N}, y^{1:N}:}}Q_{U_{1:r}^{1:N},Y^{1:N}}\left(u_{1:r}^{1:N},y^{1:N}\right)d\left(y^{1:N},\mathcal{M}(u_{1:r}^{1:N}G_N)\right)\\
&\hspace{2em}+\frac{1}{N}\sum_{\underset{|y^i|>\kappa, \exists i \in [N]}{u_{1:r}^{1:N}, y^{1:N}:}}Q_{U_{1:r}^{1:N},Y^{1:N}}\left(u_{1:r}^{1:N},y^{1:N}\right)d\left(y^{1:N},\mathcal{M}(u_{1:r}^{1:N}G_N)\right).
\end{aligned}
\end{eqnarray}

For the second term, we can upper-bound it by using a dummy quantizer which maps $y^{1:N}$ to $u_\ell^{1:N}=0^{1:N}$ for all $1\leq \ell\leq r$ if there exists $i \in [N]$ such that $|y^i|>\kappa$. We also note that $x^{1:N}=\mathcal{M}\left(u_{1:r}^{1:N}G_N\right)=0^{1:N}$ in this case. We have
\begin{eqnarray}
\begin{aligned}
\frac{1}{N}\sum_{\underset{|y^i|>\kappa, \exists i \in [N]}{u_{1:r}^{1:N}, y^{1:N}:}}&Q_{U_{1:r}^{1:N},Y^{1:N}}\left(u_{1:r}^{1:N},y^{1:N}\right)d\left(y^{1:N},\mathcal{M}(u_{1:r}^{1:N}G_N)\right) \\
&\leq \frac{1}{N}\sum_{\underset{|y^i|>\kappa, \exists i \in [N]}{ y^{1:N}:}}P_{Y^{1:N}}\left(y^{1:N}\right)d\left(y^{1:N},0^{1:N}\right) \\
&=\frac{1}{N}\int_{\underset{|y^i|>\kappa, \exists i \in [N]}{ y^{1:N}:}} f_{Y^{1:N}}\left(y^{1:N}\right) \sum_{j=1}^N (y^j)^2 dy^{1:N} \\
&\leq \frac{1}{N} \int_{|y^i|>\kappa} (y^i)^2 f_{Y^i}(y^i) dy^i + \frac{1}{N}\sum_{j\neq i} \int_{y^j} (y^j)^2 f_{Y^j}(y^j) dy^j \cdot \int_{|y^i|>\kappa}f_{Y^i}(y^i) dy^i\\
&\leq \frac{1}{N} \int_{|y^i|>\kappa} (y^i)^2 f_{Y^i}(y^i) dy^i + \sigma_s^2 \int_{|y^i|>\kappa}f_{Y^i}(y^i) dy^i,
\end{aligned}
\end{eqnarray}
where we used the fact that $\int_{y^j} (y^j)^2 f_{Y^j}(y^j) dy^j \leq \int_{-\infty}^{\infty} (y^j)^2 f_{Y^j}(y^j) dy^j=\sigma_s^2$. By the Chernoff bound, we have $\int_{t}^\infty f_{Y}(y)dy \leq e^{-\frac{t^2}{2}}$ and $\int_{t}^\infty y^2f_{Y}(y)dy \leq \frac{1}{\sqrt{2\pi}}te^{-\frac{-t^2}{2}}+e^{-\frac{t^2}{2}}$, which give us the following upper-bound.
\begin{eqnarray}\label{eqn:DQYPart1}
\begin{aligned}
\frac{1}{N}\sum_{\underset{|y^i|>\kappa, \exists i \in [N]}{u_{1:r}^{1:N}, y^{1:N}:}}Q_{U_{1:r}^{1:N},Y^{1:N}}&\left(u_{1:r}^{1:N},y^{1:N}\right)d\left(y^{1:N},\mathcal{M}(u_{1:r}^{1:N}G_N)\right) \\
&\leq \frac{2}{N}\left(\frac{\sigma_s \kappa}{\sqrt{2\pi}} e^{-\frac{\kappa^2}{2\sigma_s^2}}+\sigma_s^2e^{-\frac{\kappa^2}{2\sigma_s^2}}\right)+ 2\sigma_s^2 e^{-\frac{\kappa^2}{2\sigma_s^2}}\\
&= O\left(C_2^{-N^2}\right),
\end{aligned}
\end{eqnarray}
for a constant $C_2>1$.

With regarding to the first term in \eqref{eqn:DQY}, we have
\begin{eqnarray}\label{eqn:DQYPart2}
\begin{aligned}
&\frac{1}{N}\sum_{\underset{|y^i|\leq \kappa, \forall i \in [N]}{u_{1:r}^{1:N}, y^{1:N}:}}Q_{U_{1:r}^{1:N},Y^{1:N}}\left(u_{1:r}^{1:N},y^{1:N}\right)d\left(y^{1:N},\mathcal{M}(u_{1:r}^{1:N}G_N)\right) \\
&\leq\frac{1}{N}\sum_{\underset{|y^i|\leq \kappa, \forall i \in [N]}{u_{1:r}^{1:N}, y^{1:N}:}}Q_{U_{1:r}^{1:N},Y'^{1:N}}\left(u_{1:r}^{1:N},y^{1:N}\right)d\left(y^{1:N},\mathcal{M}(u_{1:r}^{1:N}G_N)\right)\\
&\hspace{2em}+\frac{1}{N}\sum_{\underset{|y^i|\leq \kappa, \forall i \in [N]}{u_{1:r}^{1:N}, y^{1:N}:}}\left|Q_{U_{1:r}^{1:N},Y'^{1:N}}(u_{1:r}^{1:N},y^{1:N})-Q_{U_{1:r}^{1:N},Y^{1:N}}(u_{1:r}^{1:N},y^{1:N})\right|d\left(y^{1:N},\mathcal{M}(u_{1:r}^{1:N}G_N)\right)\\
&\leq\frac{1}{N}\sum_{\underset{|y^i|\leq \kappa, \forall i \in [N]}{u_{1:r}^{1:N}, y^{1:N}:}}P_{U_{1:r}^{1:N},Y'^{1:N}}\left(u_{1:r}^{1:N},y^{1:N}\right)d\left(y^{1:N},\mathcal{M}(u_{1:r}^{1:N}G_N)\right)\\
&\hspace{2em}+\frac{1}{N}\sum_{\underset{|y^i|\leq \kappa, \forall i \in [N]}{u_{1:r}^{1:N}, y^{1:N}:}}\left|P_{U_{1:r}^{1:N},Y'^{1:N}}(u_{1:r}^{1:N},y^{1:N})-Q_{U_{1:r}^{1:N},Y'^{1:N}}(u_{1:r}^{1:N},y^{1:N})\right|d\left(y^{1:N},\mathcal{M}(u_{1:r}^{1:N}G_N)\right)\\
&\hspace{4em}+\frac{1}{N}\sum_{\underset{|y^i|\leq \kappa, \forall i \in [N]}{u_{1:r}^{1:N}, y^{1:N}:}}\left|Q_{U_{1:r}^{1:N},Y'^{1:N}}(u_{1:r}^{1:N},y^{1:N})-Q_{U_{1:r}^{1:N},Y^{1:N}}(u_{1:r}^{1:N},y^{1:N})\right|d\left(y^{1:N},\mathcal{M}(u_{1:r}^{1:N}G_N)\right)
\end{aligned}
\end{eqnarray}

It can be seen that the first term in the above inequality is upper-bounded by $D_{P, Y'}$. For the second term, we note that the mapping function $\mathcal{M}(\cdot)$ picks the single point within the interval $[-\kappa, \kappa)$, and we have $d(y',x) \leq 4\kappa^2$. By using the bound \eqref{eqn:variantialbound}, the second term can be upper-bounded by $\frac{2}{N}\mathbb{V}(P_{U_{1:r}^{1:N},Y'^{1:N}}(u_{1:r}^{1:N},y'^{1:N}),Q_{U_{1:r}^{1:N},Y'^{1:N}}(u_{1:r}^{1:N},y'^{1:N}))\cdot N \cdot 4\kappa^2$. Similarly, for the third term, since $Q_{U_{1:r}^{1:N}|Y^{1:N}}(u_{1:r}^{1:N}|y^{1:N})=Q_{U_{1:r}^{1:N}|Y'^{1:N}}(u_{1:r}^{1:N}|y^{1:N})$, it can be upper-bounded by $\frac{2}{N}\mathbb{V}(P_{Y'^{1:N}}(y^{1:N}),P_{Y^{1:N}}(y^{1:N}))\cdot N \cdot 4\kappa^2$. Again, by the telescoping expansion,
\begin{eqnarray}
\begin{aligned}
&\hspace{-2em}\sum_{y^{1:N}}\left|P_{Y^{1:N}}(y^{1:N})-P_{Y'^{1:N}}(y^{1:N})\right|\\
&\hspace{-2em}=\sum_{y^{1:N}}\left|\sum_{i=1}^{N}\left(P_{Y^i}(y^i)-P_{Y'^i}(y^i)\right)P_{Y^{1:i-1}}\left(y^{1:i-1}\right)P_{Y'^{i+1:N}}\left(y^{i+1:N}\right)\right|\\ \notag\
&\hspace{-2em}\leq\sum_{i=1}^{N}\sum_{y^i}\left|P_{Y^i}(y^i)-P_{Y'^i}(y^i)\right|\\
&\hspace{-3em}\underset{\mathrm{Lemma}\,\ref{lem:YY'distance}}{\leq} N\cdot 4 \epsilon_{\Lambda}(\tilde{\sigma}_\Delta).
\end{aligned}
\end{eqnarray}

As a result, \eqref{eqn:DQYPart2} can be re-written as
\begin{eqnarray}\label{eqn:DQYPart22}
\begin{aligned}
&\frac{1}{N}\sum_{\underset{|y^i|\leq \kappa, \forall i \in [N]}{u_{1:r}^{1:N}, y^{1:N}:}}Q_{U_{1:r}^{1:N},Y^{1:N}}\left(u_{1:r}^{1:N},y^{1:N}\right)d\left(y^{1:N},\mathcal{M}(u_{1:r}^{1:N}G_N)\right) \\
&\leq D_{P,Y'} + \frac{2}{N}\mathbb{V}\left(P_{U_{1:r}^{1:N},Y'^{1:N}}\left(u_{1:r}^{1:N},y'^{1:N}\right), Q_{U_{1:r}^{1:N},Y'^{1:N}}\left(u_{1:r}^{1:N},y'^{1:N}\right)\right)\cdot N \cdot 4\kappa^2 + 4 \epsilon_{\Lambda}(\tilde{\sigma}_\Delta) \cdot N \cdot 4\kappa^2.
\end{aligned}
\end{eqnarray}

%\begin{eqnarray}
%&&\frac{1}{N}\sum_{x^{1:N},y'^{1:N}}|P_{X^{1:N},Y'^{1:N}}(x^{1:N},y'^{1:N})-Q_{X^{1:N},Y'^{1:N}}(x^{1:N},y'^{1:N})|d(y'^{1:N},x^{1:N}) \label{eq:2region1}\\
%&&=\frac{1}{N}\sum_{\underset{|x^i|\leq N, |y'^i|\leq N, \forall i \in [N] }{x^{1:N},y'^{1:N}:}}|P_{X^{1:N},Y'^{1:N}}(x^{1:N},y'^{1:N})-Q_{X^{1:N},Y'^{1:N}}(x^{1:N},y'^{1:N})|d(y'^{1:N},x^{1:N})\\
%&&+\frac{1}{N}\sum_{\underset{|x^i|> N \text{or} \; |y'^i|> N, \forall i \in [N] }{x^{1:N},y'^{1:N}:}}|P_{X^{1:N},Y'^{1:N}}(x^{1:N},y'^{1:N})-Q_{X^{1:N},Y'^{1:N}}(x^{1:N},y'^{1:N})|d(y'^{1:N},x^{1:N})
%\end{eqnarray}

%For the first summation, since $|x^i| \leq N$ and $|y'^i| \leq N$, we have $d(y'^i,x^i) \leq 4N^2$ for $i \in [N]$. For the second summation, we have the following upper-bound.
%\begin{eqnarray}
%&&\frac{1}{N}\sum_{\underset{|x^i|> N \text{or}\; |y'^i|> N, \forall i \in [N] }{x^{1:N},y'^{1:N}:}}|P_{X^{1:N},Y'^{1:N}}(x^{1:N},y'^{1:N})-Q_{X^{1:N},Y'^{1:N}}(x^{1:N},y'^{1:N})|d(y'^{1:N},x^{1:N})\\
%&&\leq  \frac{1}{N} \sum_{\underset{|x^i|> N \text{or}\; |y'^i|> N, \forall i \in [N] }{x^{1:N},y'^{1:N}:}} P_{X^{1:N},Y'^{1:N}}(x^{1:N},y'^{1:N}) d(y'^{1:N},x^{1:N}) \\
%&&\leq \sum_{\underset{|x|> N \text{or}\; |y'|> N }{x,y':}} P_{X,Y'}(x,y') d(y',x) \\
%&&\leq 2\sum_{x \in \Lambda, |x|>N} P_X(x) \int_{-\infty}^{\infty} P_{Y'|X}(y'|x)(y'-x)^2 dy'\\
%&&= 2\Delta \sum_{x \in \Lambda, |x|>N} P_X(x),
%\end{eqnarray}
%where in the last inequality we use the symmetry of the $(X, Y')$ plane.

Combining \eqref{eqn:DPYboundFinal}, \eqref{eqn:DQYPart1} and \eqref{eqn:DQYPart22}, we then have
\begin{eqnarray}\label{eqn:distortionBound}
D_{Q,Y}&\leq& D_{P,Y'}+\frac{2}{N}\mathbb{V}\left(P_{U_{1:r}^{1:N},Y'^{1:N}},Q_{U_{1:r}^{1:N},Y'^{1:N}}\right)\cdot N \cdot 4\kappa^2 + O\left(C_2^{-N^2}\right) + 4 \epsilon_{\Lambda}(\tilde{\sigma}_\Delta) \cdot N \cdot 4\kappa^2\\ \notag\
&=& \Delta+O\left(N^2C_1^{-N^2}\right)+O\left(N^22^{-N^{\beta''}}\right)+O\left(C_2^{-N^2}\right)+ 4 \epsilon_{\Lambda}(\tilde{\sigma}_\Delta) \cdot N \cdot 4\kappa^2,
\end{eqnarray}
where $C_1>1$, $C_2>1$, and $0<\beta''<\frac{1}{2}$.

Since $r=O(\log N)$, we can scale down $\Lambda$ such that $\epsilon_{\Lambda}(\tilde{\sigma}_\Delta)=O\left(e^{-N}\right)$ (see Proposition \ref{prop:morelevel}), and the last term in the above inequality is $O\left(N^3 e^{-N}\right)$. Compared with the dominating term $O(N^22^{-N^{\beta''}})$, the other three terms $O(C_1^{-N^2})$, $O(N^2C_1^{-N^2})$ and $O(N^3 e^{-N})$ are all negligible when $N$ is large. This is due to the fact that the densities of both $Y$ and $X$ decrease exponentially to their square norms. For this reason, we can safely assume a boundary of the two variables $Y$ and $X$, which results in a maximum distortion $d_{\max}=4\kappa^2$, and then ignore the extra contribution of the variables outside the boundary. Therefore, in the following proof, we assume a maximum distortion $d_{\max}=4\kappa^2=O(N^2)$ between Gaussian distributed variables, and skip the marginal effect for the sake of brevity.

%Now we consider using the same encoder to quantize the Gaussian source $Y$. The resulted average distortion $D_{Q,Y}$ can be written as
%\begin{eqnarray}
%\begin{aligned}
%D_{Q,Y}&=\frac{1}{N}\sum_{u_{1:r}^{1:N},y^{1:N}}Q_{U_{1:r}^{1:N},Y^{1:N}}(u_{1:r}^{1:N},y^{1:N})d(y^{1:N},\mathcal{M}(u_{1:r}^{1:N}G_N))\\ \notag\
%&=\frac{1}{N}\sum_{u_{1:r}^{1:N},y^{1:N}} P_{Y^{1:N}}(y^{1:N})Q_{U_{1:r}^{1:N}|Y^{1:N}}(u_{1:r}^{1:N}|y^{1:N})\cdot d(y^{1:N},\mathcal{M}(u_{1:r}^{1:N}G_N)).
%\end{aligned}
%\end{eqnarray}

%Since the same encoder is used, for a same realization $y^{1:N}$, we have $Q_{U_{1:r}^{1:N}|Y^{1:N}}(u_{1:r}^{1:N}|y^{1:N})=Q_{U_{1:r}^{1:N}|Y'^{1:N}}(u_{1:r}^{1:N}|y^{1:N})$, and hence
%\begin{eqnarray}
%\begin{aligned}\notag\
%D_{Q,Y}-D_{Q,Y'}&=\frac{1}{N}\sum_{u_{1:r}^{1:N},y^{1:N}} (P_{Y^{1:N}}(y^{1:N})-P_{Y'^{1:N}}(y^{1:N}))\cdot Q_{U_{1:r}^{1:N}|Y^{1:N}}(u_{1:r}^{1:N}|y^{1:N})d(y^{1:N},\mathcal{M}(u_{1:r}^{1:N}G_N)) \\
%&\leq \frac{1}{N} N d_{\max}\sum_{y^{1:N}}|P_{Y^{1:N}}(y^{1:N})-P_{Y'^{1:N}}(y^{1:N})|+2\Delta \sum_{x\in \Lambda |x|>N} P_X(x),
%\end{aligned}
%\end{eqnarray}
%where we use the same argument as in \eqref{eq:2region1}-\eqref{eq:2region2}, i.e., divide the summations into two regions.

As a result, the achievable average distortion $D_{Q,Y}$ can be sub-exponentially close to $\Delta$ with $R>I(X;Y')\geq \frac{1}{2} \log\frac{\sigma_s^2}{\Delta}-\frac{5\epsilon_{\Lambda}(\tilde{\sigma}_\Delta)}{n}\log(e)$ ($n$ could be 1). When $N\to \infty$ and $\epsilon_{\Lambda}(\tilde{\sigma}_\Delta) \to 0$, we have $I(X;Y') \to \frac{1}{2} \log\frac{\sigma_s^2}{\Delta}$ and $R>\frac{1}{2} \log\frac{\sigma_s^2}{\Delta}$.

Now it is ready to explain the lattice structure. From the definition of $\mathcal{F}_\ell$ and \cite[Lemma 5]{LiuL15w}, it is easy to find that $\mathcal{F}_\ell \subseteq \mathcal{F}_{\ell-1}$ for $1<\ell\leq r$. When $u_\ell^{\mathcal{S}_\ell}$ is uniformly selected and $u_\ell^{\mathcal{F}_\ell}=0$ on each level, the constructed polar code at level $\ell-1$ is a subset of the polar code at level $\ell$. Therefore, the resulted multilevel code is actually a polar lattice and the MAP decision on the bits in $\mathcal{S}_\ell$ is a shaping operation according to $D_{\Lambda,\sigma_r}$. Moreover, since $D_{Q,Y}$ is an average distortion over all random choices of $u_\ell^{\mathcal{F}_\ell}$, there exists at least one specific choice of $u_\ell^{\mathcal{F}_\ell}$ on each level making the average distortion satisfying \eqref{eqn:distortionBound}. This is exactly a shift on the constructed polar lattice. Consequently, the shifted polar lattice achieves the rate-distortion bound of the Gaussian source.
\end{proof}

\section{Proof of Theorem \ref{Thm:WZ}}\label{appendix4}
\begin{IEEEproof}
We firstly show that the target distortion can be achieved. Recall $U_\ell^{1:N}=A_\ell^{1:N}G_N$ for each level $\ell$. Let $P_{U_{1:r}^{1:N}, \bar{X}^{1:N}}$ denote the joint distribution between $U_{1:r}^{1:N}$ and $\bar{X}_{1:r}^{1:N}$ when the encoder performs no compression on each level, i.e., the encoder applies encoding rule \eqref{eqn:lossyencoder1} for all indices $i \in [N]$ at level 1, encoding rule \eqref{eqn:lossyencoder4} for all $i \in [N]$ at level 2 and similar rules for higher levels, with the notation $X$ and $Y'$ being replaced by $A$ and $\bar{X}$, respectively. Let $Q_{U_{1:r}^{1:N}, \bar{X}^{1:N}}$ denote the joint distribution when only $U_\ell^{\mathcal{I}_\ell^Q}$ is recorded following the randomized rounding rule on each level. $U_\ell^{\mathcal{F}_\ell^Q}$ is a uniformly random sequence shared between the encoder and decoder, and $U_\ell^{\mathcal{S}_\ell^Q}$ is determined according to the MAP rule (see \eqref{eqn:lossyencoder2} and \eqref{eqn:lossyencoder5}).
As illustrated in \eqref{eqn:variantialbound},
\begin{eqnarray}
\mathbb{V}\left(P_{U_{1:r}^{1:N},\bar{X}^{1:N}},Q_{U_{1:r}^{1:N},\bar{X}^{1:N}}\right)=O\left(r \cdot 2^{-N^{\beta'}}\right).
\end{eqnarray}
Since $r=O(\log N)$, we can write $\mathbb{V}\left(P_{U_{1:r}^{1:N},\bar{X}^{1:N}},Q_{U_{1:r}^{1:N},\bar{X}^{1:N}}\right)=O\left(2^{-N^{\beta''}}\right)$ for $0<\beta''<\beta<1/2$. When quantization is performed for the source $X$, let $Q_{U_{1:r}^{1:N}, X^{1:N}}$ denote the resulted joint distribution. By Lemma \ref{lem:YY'distance} again,
\begin{eqnarray}
\begin{aligned}
&\sum_{u_{1:r}^{1:N},x^{1:N}}\left|Q_{U_{1:r}^{1:N},\bar{X}^{1:N}}(u_{1:r}^{1:N}, x^{1:N})-Q_{U_{1:r}^{1:N}, X^{1:N}}(u_{1:r}^{1:N}, x^{1:N})\right|\\
&=\sum_{x^{1:N}}\left|P_{\bar{X}^{1:N}}(x^{1:N})-P_{X^{1:N}}(x^{1:N})\right|\sum_{u_{1:r}^{1:N}}Q_{U_{1:r}^{1:N}|X^{1:N}}\left(u_{1:r}^{1:N}|x^{1:N}\right)\\\notag\
&=\sum_{x^{1:N}}\left|P_{\bar{X}^{1:N}}(x^{1:N})-P_{X^{1:N}}(x^{1:N})\right|\leq N\cdot 4 \epsilon_{\Lambda}(\tilde{\sigma}_q).
\end{aligned}
\end{eqnarray}
It has been shown in Proposition \ref{prop:morelevel} that $\epsilon_{\Lambda}(\tilde{\sigma}_q)=O\left(e^{-N}\right)$, we further have
\begin{eqnarray}\label{eqn:wzbound1}
\mathbb{V}\left(P_{U_{1:r}^{1:N},\bar{X}^{1:N}},Q_{U_{1:r}^{1:N},X^{1:N}}\right) &\leq& \mathbb{V}\left(P_{U_{1:r}^{1:N},\bar{X}^{1:N}},Q_{U_{1:r}^{1:N},\bar{X}^{1:N}}\right)+\mathbb{V}\left(Q_{U_{1:r}^{1:N},\bar{X}^{1:N}},Q_{U_{1:r}^{1:N}, X^{1:N}}\right) \\ \notag
&=&O\left(2^{-N^{\beta''}}\right)+O\left(Ne^{-N}\right)=O\left(2^{-N^{\beta''}}\right).
\end{eqnarray}

As mentioned in Section \ref{sec:WZ}, the encoder only sends $U_\ell^{d\mathcal{I}_\ell}$ to the decoder, which then utilizes the side information to recover $U_\ell^{\mathcal{I}_\ell^C}$. Here we assume that $U_\ell^{\mathcal{I}_\ell^C}$ can be correctly decoded and $U_\ell^{\mathcal{S}_\ell^Q}$ is recovered according to the MAP rule. In this case, the decoder enjoys the same joint distribution $Q_{U_{1:r}^{1:N},X^{1:N}}$ as the encoder does. Recall that $B=\frac{\alpha_q}{\alpha_c}Y$ and $\bar{B}=\frac{\alpha_q}{\alpha_c}\bar{Y}$. Let $Q_{U_{1:r}^{1:N},X^{1:N}, B^{1:N}}$ denote the resulted joint distribution of $U_{1:r}^{1:N}$, $X^{1:N}$, and $B^{1:N}$ when the encoder performs compression, i.e., compresses $X^{1:N}$ to $U_{\ell}^{\mathcal{I}_\ell^Q}$ on each level. Let $P_{U_{1:r}^{1:N},\bar{X}^{1:N}, \bar{B}^{1:N}}$ denote the resulted joint distribution of $U_{1:r}^{1:N}$, $\bar{X}^{1:N}$, and $\bar{B}^{1:N}$ when the encoder performs no compression for $\bar{X}^{1:N}$.
\begin{eqnarray}
\begin{aligned}
&2\mathbb{V}\left(P_{U_{1:r}^{1:N}, \bar{X}^{1:N}, \bar{B}^{1:N}}, Q_{U_{1:r}^{1:N}, X^{1:N}, B^{1:N}}\right)\\
&= \sum_{u_{1:r}^{1:N}, x^{1:N}, b^{1:N}}\left|P\left(u_{1:r}^{1:N}, x_{1:r}^{1:N}, b^{1:N}\right)-Q\left(u_{1:r}^{1:N}, x_{1:r}^{1:N}, b^{1:N}\right)\right| \\
&= \sum_{u_{1:r}^{1:N}, x^{1:N}, b^{1:N}}\left|P(u_{1:r}^{1:N}, x_{1:r}^{1:N})P(b^{1:N}|u_{1:r}^{1:N}, x_{1:r}^{1:N})-Q(u_{1:r}^{1:N}, x_{1:r}^{1:N})Q(b^{1:N}|u_{1:r}^{1:N}, x_{1:r}^{1:N})\right|.
\end{aligned}
\end{eqnarray}
According to Fig. \ref{fig:WZrvs}, $\alpha_qX' \rightarrow X \rightarrow B$ and $A \rightarrow \bar{X} \rightarrow \bar{B}$ are two Markov chains. We have
\begin{eqnarray}
P\left(b^{1:N}|u_{1:r}^{1:N}, x_{1:r}^{1:N}\right)&=&Q\left(b^{1:N}|u_{1:r}^{1:N}, x_{1:r}^{1:N}\right) \\
&=&\prod_{i=1}^N \frac{1}{\sqrt{2 \pi \sigma_{z'}^2}} \exp\left(-\frac{(b^i-x^i)^2}{2\sigma_{z'}^2}\right).
\end{eqnarray}
Therefore,
\begin{eqnarray}
\mathbb{V}\left(P_{U_{1:r}^{1:N}, \bar{X}^{1:N}, \bar{B}^{1:N}}, Q_{U_{1:r}^{1:N}, X^{1:N}, B^{1:N}}\right)=\mathbb{V}\left(P_{U_{1:r}^{1:N},\bar{X}^{1:N}},Q_{U_{1:r}^{1:N},X^{1:N}}\right)=O\left(2^{-N^{\beta''}}\right).
\end{eqnarray}
Recall that the reconstruction of $\bar{X}$ is given by $\check{X}=A+\gamma(\bar{B}-A)$. The average distortion $\Delta_{P}$ caused by $P_{U_{1:r}^{1:N}, \bar{X}^{1:N}, \bar{B}^{1:N}}$ can be expressed as
\begin{eqnarray}
\Delta_{P}=\frac{1}{N}\sum_{u_{1:r}^{1:N},x^{1:N}, b^{1:N}}P_{U_{1:r}^{1:N}, \bar{X}^{1:N}, \bar{B}^{1:N}}\left(u_{1:r}^{1:N}, x^{1:N}, b^{1:N}\right)d\left(x^{1:N},\check{x}^{1:N}\right),
\end{eqnarray}
where $\check{x}^{1:N}=\gamma b^{1:N}+ (1-\gamma)\mathcal{M}\left(u_{1:r}^{1:N}G_N\right)$, where $\mathcal{M}\left(u_{1:r}^{1:N}G_N\right)$ is a mapping from $u_{1:r}^{1:N}$ to $a^{1:N}$ according to the lattice Gaussian distribution. Clearly, given $u_{1:r}^{1:N}$, there is a one-to-one mapping between $b^{1:N}$ and $\check{x}^{1:N}$ when $r$ is sufficiently large. Thus, $\Delta_{P}$ can be written as
\begin{eqnarray}
\begin{aligned}
\Delta_{P}&=\frac{1}{N}\sum_{u_{1:r}^{1:N},\check{x}^{1:N}, x^{1:N}}P_{U_{1:r}^{1:N}, \check{X}^{1:N}, \bar{X}^{1:N}}\left(u_{1:r}^{1:N}, \check{x}^{1:N}, x^{1:N}\right)d\left(\check{x}^{1:N}, x^{1:N}\right)\\
&=\frac{1}{N}\sum_{\check{x}^{1:N}, x^{1:N}}P_{\check{X}^{1:N}, \bar{X}^{1:N}}\left(\check{x}^{1:N},x^{1:N}\right)d\left(\check{x}^{1:N}, x^{1:N}\right)\\ \notag\
&=\frac{1}{N}\cdot N \sum_{\check{x},x}P_{\check{X},\bar{X}}(\check{x},x)d(\check{x},x) \\ \notag\
&=\int_{\check{x}} f_{\check{X}}(\check{x})\int_{-\infty}^{+\infty}\frac{1}{\sqrt{2\pi \Delta}} \exp\left(-\frac{(x-\check{x})^2}{2\Delta}\right)(x-\check{x})^2 dxd\check{x}\\ \notag\
&=\Delta.
\end{aligned}
\end{eqnarray}
The expected distortion $\Delta_Q$ achieved by $Q_{U_{1:r}^{1:N}, X^{1:N}, B^{1:N}}$ satisfies
\begin{eqnarray}
\begin{aligned}\notag\
\Delta_Q-\Delta_{P}&=\frac{1}{N}\sum_{u_{1:r}^{1:N},x^{1:N},b^{1:N}}\left(P_{U_{1:r}^{1:N}, \bar{X}^{1:N}, \bar{B}^{1:N}}-Q_{U_{1:r}^{1:N}, X^{1:N}, B^{1:N}}\right)d\left(\check{x}^{1:N}, x^{1:N}\right)\\
&\leq \frac{1}{N} N d_{\max}\sum_{u^{1:N},x^{1:N},b^{1:N}}\left|P_{U_{1:r}^{1:N}, \bar{X}^{1:N}, \bar{B}^{1:N}}-Q_{U_{1:r}^{1:N}, X^{1:N}, B^{1:N}}\right|\\
&=O\left(N^22^{-N^{\beta''}}\right),
\end{aligned}
\end{eqnarray}
where we invoke the max distortion $d_{\max}=4\kappa^2$ within the interval $[-\kappa,\kappa)$ and ignore the non-dominating terms caused by parts outside the boundary using the same argument as in \eqref{eqn:DPYboundFinal}, \eqref{eqn:DQYPart1} and \eqref{eqn:DQYPart22}.

Now we show that the decoder is able to decode $U_\ell^{\mathcal{I}_\ell^C}$ with vanishing error probability.
\begin{eqnarray}\label{eqn:lessVari}
2\mathbb{V}(P_{U_{1:r}^{1:N},\bar{B}^{1:N}},Q_{U_{1:r}^{1:N},B^{1:N}})&=&\sum_{u_{1:r}^{1:N}, b^{1:N}}\left|P(u_{1:r}^{1:N}, b^{1:N})-Q(u_{1:r}^{1:N}, b^{1:N})\right| \notag \\
&=&\sum_{u_{1:r}^{1:N}, b^{1:N}}\left|\sum_{x^{1:N}}\bigg[P(u_{1:r}^{1:N}, x^{1:N}, b^{1:N})-Q(u_{1:r}^{1:N}, x^{1:N}, b^{1:N})\bigg]\right|\notag \\
&\leq& \sum_{u_{1:r}^{1:N}, b^{1:N}}\sum_{x^{1:N}}\left|P_{U_{1:r}^{1:N}, \bar{X}^{1:N}, \bar{B}^{1:N}}-Q_{U_{1:r}^{1:N}, X^{1:N}, B^{1:N}}\right| \notag \\
&=&O\left(2^{-N^{\beta''}}\right).
\end{eqnarray}
By the result of \cite[Theorem 5]{LiuL15w}, $P_{U_{1:r}^{1:N},\bar{B}^{1:N}}$ results in an expectation of error probability $E_P^{\ell}[P_e]$ on each level such that $E_P^{\ell}[P_e]=O\left(2^{-N^{\beta'}}\right)$. To see this, let $\mathcal{E}_i$ denote the set of pairs of $u_\ell^{1:N}$ and $b^{1:N}$ such that the SC decoding error occurs at the $i$th bit for level $\ell$, then the block decoding error event is given by $\mathcal{E}^\ell \equiv \bigcup_{i \in \mathcal{I}_\ell^{C}} \mathcal{E}_i$. Then the expectation of decoding error probability over all random mapping is expressed as
\begin{eqnarray}\label{eqn:errorP}
\begin{aligned}
E_P^{\ell}[P_e]&=\sum_{u_{1:\ell}^{1:N},b^{1:N}} P_{U_{1:\ell}^{1:N}, \bar{B}^{1:N}}\left(u_{1:\ell}^{1:N}, b^{1:N}\right) \mathds{1}\left[\left(u_{\ell}^{1:N},b^{1:N}\right)\in \mathcal{E}^\ell\right] \\
&\leq \sum_{i\in \mathcal{I}_\ell^{C} \cup \mathcal{S}_\ell^{Q}} \sum_{u_{1:\ell}^{1:N},b^{1:N}} P_{U_{1:\ell}^{1:N}, \bar{B}^{1:N}}\left(u_{1:\ell}^{1:N}, b^{1:N}\right) \mathds{1}\left[\left(u_{\ell}^{1:N},b^{1:N}\right)\in \mathcal{E}_i\right] \\
&\leq\sum_{i\in \mathcal{I}_\ell^{C} \cup \mathcal{S}_\ell^{Q}}\sum_{u_\ell^{1:i}, u_{1:\ell-1}^{1:N},b^{1:N}}P\left(u_\ell^{1:i-1}, u_{1:\ell-1}^{1:N} ,b^{1:N}\right)P\left(u_\ell^i|u_\ell^{1:i-1}, u_{1:\ell-1}^{1:N},b^{1:N}\right)\\
&\hspace{8em}\cdot\mathds{1}\left[P(u_\ell^i|u_\ell^{1:i-1}, u_{1:\ell-1}^{1:N},b^{1:N})\leq P(u_\ell^i\oplus1|u_\ell^{1:i-1}, u_{1:\ell-1}^{1:N},b^{1:N})\right]\\
&\leq\sum_{i\in \mathcal{I}_\ell^{C} \cup \mathcal{S}_\ell^{Q}}\sum_{u_\ell^{1:i}, u_{1:\ell-1}^{1:N},b^{1:N}}P\left(u_\ell^{1:i-1}, u_{1:\ell-1}^{1:N} ,b^{1:N}\right)P\left(u_\ell^i|u_\ell^{1:i-1}, u_{1:\ell-1}^{1:N},b^{1:N}\right)\\
&\hspace{8em}\sqrt{\frac{P\left(u_\ell^i\oplus1|u_\ell^{1:i-1}, u_{1:\ell-1}^{1:N},b^{1:N}\right)}{P\left(u_\ell^i|u_\ell^{1:i-1}, u_{1:\ell-1}^{1:N},b^{1:N}\right)}}\\
&\leq N \cdot Z\left(U_\ell^i|U_\ell^{1:i-1},A_{1:\ell-1}^{1:N}, \bar{B}^{1:N}\right)\\
&=O\left(2^{-N^{\beta'}}\right).
\end{aligned}
\end{eqnarray}

Then, by the union bound, we immediately obtain that the expectation of multistage decoding error probability for the polar lattice $E_P[P_e]=O\left(2^{-N^{\beta''}}\right)$ for $0<\beta''<\beta'<1/2$.
%Note that the expectation is taken over all choices of frozen bits $U_\ell^{\mathcal{F}_\ell^C}$ and shaping bits $U_\ell^{d\mathcal{S}_\ell}$.
Let $P_e^{WZ}$ denote the expectation of error probability caused by $Q_{U_{1:r}^{1:N},B^{1:N}}$, i.e., it is an average error probability over all choices of the frozen bits $U_\ell^{\mathcal{F}_\ell^C}$ and shaping bits $U_\ell^{d\mathcal{S}_\ell}$ for each level. Let $\mathcal{E}$ denote the set of the pairs $\left(u_{1:r}^{1:N},b^{1:N}\right)$ such that a lattice decoding error occurs. We have
\begin{eqnarray}
P_e^{WZ}-E_P[P_e] &=& \sum_{u_{1:\ell}^{1:N},b^{1:N}} \left(P(u_{1:\ell}^{1:N}, b^{1:N})-Q(u_{1:\ell}^{1:N}, b^{1:N})\right) \cdot \mathds{1}\left[(u_{1:r}^{1:N},b^{1:N})\in \mathcal{E}\right]
\\ &\leq& 2\mathbb{V}\left(P_{U_{1:r}^{1:N},\bar{B}^{1:N}},Q_{U_{1:r}^{1:N},B^{1:N}}\right)\\
&\leq& O\left(2^{-N^{\beta''}}\right).
\end{eqnarray}
%\color{red} Need discussion with Dr. Ling. \color{black}

With regard to the data rate, we have
\begin{eqnarray}
\sum_{\ell=1}^{r} \frac{\left|\mathcal{I}_\ell^Q\right|}{N} \to {\frac{1}{2}\log \bigg(\frac{\sigma_x^2\sigma_z^2-\sigma_y^2\Delta}{\sigma_z^2\Delta}\bigg)}^+,
\end{eqnarray}
and
\begin{eqnarray}
\sum_{\ell=1}^{r} \frac{\left|\mathcal{I}_\ell^C\right|}{N} \to {\frac{1}{2}\log \bigg(\frac{\sigma_x^2\sigma_z^2-\sigma_y^2\Delta}{\sigma_z^4}\bigg)}^-.
\end{eqnarray}
Finally,
\begin{eqnarray}
R=\sum_{\ell=1}^{r} \frac{\left|\mathcal{I}_\ell^Q\right|}{N}-\frac{\left|\mathcal{I}_\ell^C\right|}{N} \to \frac{1}{2}{\log\left(\frac{\sigma_z^2}{\Delta}\right)}^+.
\end{eqnarray}
\end{IEEEproof}

\section{Proof of Theorem \ref{Thm:GP}}\label{appendix5}
\begin{IEEEproof}
We firstly show the power constraint $P$ can be satisfied. Recall $U_\ell^{1:N}=A_\ell^{1:N}G_N$ for each level $\ell$. Similar to the previous proof, denote by $P_{U_{1:r}^{1:N}, \bar{T}^{1:N}}$ the joint distribution between $U_{1:r}^{1:N}$ and $\bar{T}_{1:r}^{1:N}$ when the encoder applies randomized rounding rule for all indices $i \in [N]$ at level $\ell$. Denote by $Q_{U_{1:r}^{1:N}, T^{1:N}}$ the joint distribution when $U_\ell^{\mathcal{I}_\ell^Q}$ and $U_\ell^{d\mathcal{S}_\ell}$ are encoded following the randomized rounding rule on each level. $U_\ell^{\mathcal{F}_\ell^C}$ is a uniformly random sequence shared between the encoder and decoder, $U_\ell^{d\mathcal{F}_\ell}$ is a uniform message sequence and $U_\ell^{\mathcal{S}_\ell^C}$ is determined according to the MAP rule.

Notice that $\epsilon_{\Lambda}(\tilde{\sigma}_q) \leq \epsilon_{\Lambda}(\tilde{\sigma}_c)=O\left(e^{-N}\right)$. Similar to the previous proof in \eqref{eqn:wzbound1}, we have
\begin{eqnarray}
\mathbb{V}(P_{U_{1:r}^{1:N},\bar{T}^{1:N}},Q_{U_{1:r}^{1:N},T^{1:N}})\leq O\left(2^{-N^{\beta''}}\right)+O\left(Ne^{-N}\right)=O\left(2^{-N^{\beta''}}\right).
\end{eqnarray}
Thus, the average transmit power realized by $Q_{U_{1:r}^{1:N}, T^{1:N}}$ can be arbitrarily close to that realized by $P_{U_{1:r}^{1:N}, \bar{T}^{1:N}}$, i.e.,
\begin{eqnarray}
&&\frac{1}{N}\sum_{u_{1:r}^{1:N},t^{1:N}} \left[Q_{U_{1:r}^{1:N}, T^{1:N}}(u_{1:r}^{1:N},t^{1:N})-P_{U_{1:r}^{1:N}, \bar{T}^{1:N}}(u_{1:r}^{1:N},t^{1:N})\right] d\left(\frac{1}{\alpha_q} \mathcal{M}(u_{1:r}^{1:N}G_N), t^{1:N}\right) \\
&&\leq \frac{2}{N} N d_{\max} \mathbb{V}\left(P_{U_{1:r}^{1:N},\bar{T}^{1:N}},Q_{U_{1:r}^{1:N},T^{1:N}}\right),\\
&&= O\left(N^22^{-N^{\beta''}}\right),
\end{eqnarray}
where $\mathcal{M}\left(u_{1:r}^{1:N}G_N\right)$ denotes a mapping from $u_{1:r}^{1:N}$ to $t^{1:N}$ according to the lattice Gaussian distribution. Similarly, we invoke the boundary $\kappa$ and ignore the non-dominating terms caused by parts outside the boundary using the same argument as in \eqref{eqn:DPYboundFinal}, \eqref{eqn:DQYPart1} and \eqref{eqn:DQYPart22}. Recall that for a constant $\sigma_a^2$ and $\sigma_s^2$, we set $\kappa=O(N)$ and $d_{\max}=4\kappa^2$, which is the maximum distortion in the region of $[-\kappa, \kappa)$.

Now we show that the average transmit power realized by $P_{U_{1:r}^{1:N}, \bar{T}^{1:N}}$ is arbitrarily close to $P$. When $r$ is sufficiently large, $P_{A_{1:r}} \to D_{\Lambda, \sigma_a^2}$, and $\bar{T}=A+N(0,\alpha_q P)$ as shown in Fig. \ref{fig:GPMMSE}. Then, the variable $\frac{1-\alpha_q}{\alpha_q} A+(A-\bar{T})=\frac{1-\alpha_q}{\alpha_q} A+N(0,\alpha_q P)$ corresponds to a variable resulted from adding a lattice Gaussian distributed variable to an independent Gaussian noise. Notice that $\frac{1-\alpha_q}{\alpha_q}$ only involves a scale on $D_{\Lambda,\sigma_a^2}$. When $\epsilon_{\Lambda}(\tilde{\sigma}_q) = O\left(e^{-N}\right)$, the flatness factor associated with the AWGN channel from $\frac{1-\alpha_q}{\alpha_q} A$ to $\frac{1-\alpha_q}{\alpha_q} A+N(0,\alpha_q P)$ is also upper bounded by $O\left(e^{-N}\right)$.

Check that $\left(\frac{1-\alpha_q}{\alpha_q}\right)^2\sigma_a^2=(1-\alpha_q)P$. Let $\dot{X}$ and $\ddot{X}$ denote $\frac{1-\alpha_q}{\alpha_q} A+N(0,\alpha_q P)$ and Gaussian random variable with distribution $N(0, P)$, respectively. By Lemma \ref{lem:YY'distance}, $\mathbb{V}(P_{\dot{X}}, P_{\ddot{X}}) \leq O\left(e^{-N}\right)$. Letting $x^{1:N}=\frac{1}{\alpha_q} \mathcal{M}\left(u_{1:r}^{1:N}G_N\right)-t^{1:N}$, we have
\begin{eqnarray}
&&\frac{1}{N}\sum_{u_{1:r}^{1:N},t^{1:N}} P_{U_{1:r}^{1:N}, \bar{T}^{1:N}}\left(u_{1:r}^{1:N},t^{1:N}\right) d\left(\frac{1}{\alpha_q} u_{1:r}^{1:N}G_N, t^{1:N}\right)\\
&&=\frac{1}{N}\sum_{x^{1:N}}P_{\dot{X}^{1:N}}d\left(x^{1:N},0\right)\\
&&=E_{\dot{X}}\left[x^2\right].
\end{eqnarray}
Since $\mathbb{V}(P_{\dot{X}}, P_{\ddot{X}}) = O\left(e^{-N}\right)$, we can show that $E_{\dot{X}}\left[x^2\right]-E_{\ddot{X}}\left[x^2\right] = O\left(N^2e^{-N}\right)$ by dividing $x$ into the two sets according to $|x|\leq \kappa$ and $|x|>\kappa$. Consequently,
\begin{eqnarray}
P_T&=&\frac{1}{N}\sum_{u_{1:r}^{1:N},t^{1:N}} Q_{U_{1:r}^{1:N}, T^{1:N}}\left(u_{1:r}^{1:N},t^{1:N}\right)d\left(\frac{1}{\alpha_q} \mathcal{M}(u_{1:r}^{1:N}G_N), t^{1:N}\right) \\
&\leq& \frac{1}{N}\sum_{u_{1:r}^{1:N},t^{1:N}} P_{U_{1:r}^{1:N}, \bar{T}^{1:N}}\left(u_{1:r}^{1:N},t^{1:N}\right)d\left(\frac{1}{\alpha_q} \mathcal{M}(u_{1:r}^{1:N}G_N), t^{1:N}\right)+ O\left(N^22^{-N^{\beta''}}\right)\\
&\leq& E_{\ddot{X}}\left[x^2\right]+O\left(N^2e^{-N}\right)+O\left(N^22^{-N^{\beta''}}\right)\\
&=&P+O\left(N^22^{-N^{\beta''}}\right).
\end{eqnarray}

Now we prove the reliability. Recall that $Y=S+X+Z$, where $X=S'-\rho S$ is independent of $S$. Scaling $Y$ by $\rho$ gives us
\begin{eqnarray}
\rho Y&=&\rho S'+ \rho(1-\rho) S+\rho Z \\
&=&\alpha_c S'+\rho(1-\rho) S-(\alpha_c-\rho)S'+\rho Z.
\end{eqnarray}
It can also be checked that
\begin{eqnarray}
\alpha_c-\alpha_q=\frac{P(P+\sigma_z^2)}{P\sigma_i^2+(P+\sigma_z^2)^2}=(1-\alpha_q)\rho,
\end{eqnarray}
leaving us $\alpha_c-\rho=\alpha_q(1-\rho)$. Scale $\rho Y$ by $\frac{\alpha_q}{\alpha_c}$. Then,
\begin{eqnarray}
\frac{\alpha_q}{\alpha_c}\rho Y &=&\alpha_q S'+ \frac{\alpha_q}{\alpha_c} (1-\rho) (\rho S-\alpha_q S')+ \frac{\alpha_q}{\alpha_c}\rho Z.
\end{eqnarray}
Note that both $\rho S-\alpha_q S'$ and $Z$ are independent of $S'$. Replacing $\alpha_q S'$ with $A$, we have
\begin{eqnarray}
\frac{\alpha_q}{\alpha_c}\rho \dot{Y} &=&A+ \frac{\alpha_q}{\alpha_c} (1-\rho) (\rho S-A)+ \frac{\alpha_q}{\alpha_c}\rho Z,
\end{eqnarray}
which corresponds to the reverse solution shown in Fig. \ref{fig:GPMMSE}. Let $\bar{Y}$ denote the channel output when $S$ is replaced by $\bar{S}$, i.e.,
\begin{eqnarray}
\frac{\alpha_q}{\alpha_c}\rho \bar{Y} &=&A+ \frac{\alpha_q}{\alpha_c} (1-\rho) (\rho \bar{S}-A)+ \frac{\alpha_q}{\alpha_c}\rho Z.
\end{eqnarray}
Recall $\bar{T}=\rho \bar{S}$, $T=\rho S$, and $\bar{B}=\frac{\alpha_q}{\alpha_c}\rho \bar{Y}$. Also let $\dot{B}=\frac{\alpha_q}{\alpha_c}\rho \dot{Y}$. According to the previous proof, we already have $\mathbb{V}(P_{U_{1:r}^{1:N},\bar{T}^{1:N}},Q_{U_{1:r}^{1:N},T^{1:N}}) = O(2^{-N^{\beta''}})$. Note that $Z$ is an independent Gaussian noise, it is not difficult to obtain that
\begin{eqnarray}
\mathbb{V}\left(P_{U_{1:r}^{1:N},\bar{T}^{1:N}, \bar{B}^{1:N}},Q_{U_{1:r}^{1:N},T^{1:N}, \dot{B}^{1:N}}\right)=\mathbb{V}\left(P_{U_{1:r}^{1:N},\bar{T}^{1:N}},Q_{U_{1:r}^{1:N},T^{1:N}}\right) = O\left(2^{-N^{\beta''}}\right),
\end{eqnarray}
since $P_{\bar{B}^{1:N}|U_{1:r}^{1:N},\bar{T}^{1:N}}=Q_{\dot{B}^{1:N}|U_{1:r}^{1:N},T^{1:N}}$. Similarly to \eqref{eqn:lessVari}, we obtain
\begin{eqnarray}
\mathbb{V}\left(P_{U_{1:r}^{1:N},\bar{B}^{1:N}},Q_{U_{1:r}^{1:N},\dot{B}^{1:N}}\right) \leq \mathbb{V}\left(P_{U_{1:r}^{1:N},\bar{T}^{1:N}, \bar{B}^{1:N}},Q_{U_{1:r}^{1:N},T^{1:N}, \dot{B}^{1:N}}\right) = O\left(2^{-N^{\beta''}}\right).
\end{eqnarray}
Note that $\bar{B}$ is a variable obtained by adding $A$ to a Gaussian noise, and $\dot{B}$ is the real signal received because the side information $S$ is Gaussian distributed. Similar to $\eqref{eqn:errorP}$, the expectation of error probability $E_P[P_e]$ caused by $P_{U_{1:r}^{1:N},\bar{B}^{1:N}}$ can be upper bounded as $E_P[P_e] \leq O(2^{-N^{\beta''}})$. Finally, the expectation of error probability $P_e^{GP}$ caused by $Q_{U_{1:r}^{1:N},\dot{B}^{1:N}}$ satisfies
\begin{eqnarray}
P_e^{GP} &\leq& E_P[P_e]+2\mathbb{V}\left(P_{U_{1:r}^{1:N},\bar{B}^{1:N}},Q_{U_{1:r}^{1:N},\dot{B}^{1:N}}\right)\\
&=& O\left(2^{-N^{\beta''}}\right).
\end{eqnarray}
\end{IEEEproof}

\section{Proof of Lemma \ref{lem:TwoGaussianRegion2} \label{sec:Proof-of-Lemma-TwoGaussianRegion2}}
\begin{IEEEproof}
The PDF of $\left(X,Y\right)$ can be represented from the PDF of $\left(\bar{X},\bar{Y}\right)$ as follows: {\allowdisplaybreaks
\begin{equation}
\begin{aligned} & f_{\bar{X},\bar{Y}}\left(x,y\right)\\
 & =\sum_{a\in\Lambda}f_{\bar{X'}}(a)f_{\bar{X},\bar{Y}|\bar{X'}}\left(x,y|a\right)\\
 & =\sum_{a\in\Lambda}f_{\bar{X'}}(a)f_{Z_{1}Z_{2}}\left(x-a,y-\sqrt{\frac{\delta_{2}}{\delta_{1}}}a\right)\\
 & =\frac{1}{f_{\sqrt{\delta_{1}}}\left(\Lambda\right)}\sum_{a\in\Lambda}\frac{1}{\sqrt{2\pi\delta_{1}}}\exp\left(-\frac{a^{2}}{2\delta_{1}}\right)\frac{1}{2\pi\sqrt{\Delta_{1}\Delta_{2}-\left(\rho-\sqrt{\delta_{1}\delta_{2}}\right)^{2}}}\\
 & \hspace{1em}\cdot\exp\left[-\frac{1}{2\left(1-\frac{\left(\rho-\sqrt{\delta_{1}\delta_{2}}\right)^{2}}{\Delta_{1}\Delta_{2}}\right)}\left(\frac{\left(x-a\right)^{2}}{\Delta_{1}}+\frac{\left(y-\sqrt{\frac{\delta_{2}}{\delta_{1}}}a\right)^{2}}{\Delta_{2}}-\frac{2\left(\rho-\sqrt{\delta_{1}\delta_{2}}\right)\left(x-a\right)\left(y-\sqrt{\frac{\delta_{2}}{\delta_{1}}}a\right)}{\Delta_{1}\Delta_{2}}\right)\right]\\
 & =\frac{1}{2\pi\sqrt{1-\rho^{2}}}\exp\left[-\frac{\left(x^{2}+y^{2}-2\rho xy\right)}{2\left(1-\rho^{2}\right)}\right]\frac{1}{f_{\sqrt{\delta_{1}}}\left(\Lambda\right)}\sum_{a\in\Lambda}\sqrt{\frac{1-\rho^{2}}{2\pi\delta_{1}\left(\Delta_{1}\Delta_{2}-\left(\rho-\sqrt{\delta_{1}\delta_{2}}\right)^{2}\right)}}\\
 & \hspace{1em}\cdot\exp\left[-\frac{1-\rho^{2}}{2\delta_{1}\left(\Delta_{1}\Delta_{2}-\left(\rho-\sqrt{\delta_{1}\delta_{2}}\right)^{2}\right)}\left(a-\frac{\left(\delta_{1}-\rho\sqrt{\delta_{1}\delta_{2}}\right)x+\left(\sqrt{\delta_{1}\delta_{2}}-\delta_{1}\rho\right)y}{1-\rho^{2}}\right)^{2}\right],
\end{aligned}
\label{eq:Region2GaussianJointProb}
\end{equation}
}where $\frac{1}{2\pi\sqrt{1-\rho^{2}}}\exp\bigg(-\frac{x^{2}+y^{2}-2\rho xy}{2(1-\rho^{2})}\bigg)=f_{X,Y}(x,y)$ is the PDF of two joint Gaussian RVs. By the definition of the flatness factor (\ref{eq:flatnessFactor}), we have
{\allowdisplaybreaks
\begin{equation}
\begin{aligned} & \left|V\left(\Lambda\right)\sum_{a\in\Lambda}\sqrt{\frac{1-\rho^{2}}{2\pi\delta_{1}\left(\Delta_{1}\Delta_{2}-\left(\rho-\sqrt{\delta_{1}\delta_{2}}\right)^{2}\right)}}\right.\\
 & \left.\hspace{1em}\cdot\exp\left[-\frac{1-\rho^{2}}{2\delta_{1}\left(\Delta_{1}\Delta_{2}-\left(\rho-\sqrt{\delta_{1}\delta_{2}}\right)^{2}\right)}\left(a-\frac{\left(\delta_{1}-\rho\sqrt{\delta_{1}\delta_{2}}\right)x+\left(\sqrt{\delta_{1}\delta_{2}}-\delta_{1}\rho\right)y}{1-\rho^{2}}\right)^{2}\right]-1\right|\\
 & \leq\epsilon_{\Lambda}\left(\sqrt{\frac{\delta_{1}\left(\Delta_{1}\Delta_{2}-\left(\rho-\sqrt{\delta_{1}\delta_{2}}\right)^{2}\right)}{1-\rho^{2}}}\right)=\epsilon.
\end{aligned}
\label{eq:Region2GaussianFlatFactor}
\end{equation}
}

Since $\epsilon_{\Lambda}\left(\sigma\right)$ is a monotonically decreasing function of $\sigma$ and the fact that
\[
\begin{aligned} & \Delta_{1}\Delta_{2}-\left(\rho-\sqrt{\delta_{1}\delta_{2}}\right)^{2}-1+\rho^{2}\\
 & =2\rho\sqrt{\delta_{1}\delta_{2}}-\delta_{1}-\delta_{2}\\
 & =-\left(\sqrt{\delta_{1}}-\sqrt{\delta_{2}}\right)^{2}-2\sqrt{\delta_{1}\delta_{2}}\left(1-\rho\right)\leq0,
\end{aligned}
\]
we have
\[
\begin{aligned}\frac{\Delta_{1}\Delta_{2}-\left(\rho-\sqrt{\delta_{1}\delta_{2}}\right)^{2}}{1-\rho^{2}} & \leq1.\end{aligned}
\]
Therefore, it implies $\epsilon_{\Lambda}\left(\sqrt{\delta_{1}}\right)\leq\epsilon$ and more specifically
\begin{equation}
\left|V\left(\Lambda\right)f_{\sqrt{\delta_{1}}}\left(\Lambda\right)-1\right|\leq\epsilon.\label{eq:Region2GaussianFlatFactor2}
\end{equation}

From the above results (\ref{eq:Region2GaussianJointProb}), (\ref{eq:Region2GaussianFlatFactor}) and (\ref{eq:Region2GaussianFlatFactor2}), we have
\[
f_{X,Y}\left(x,y\right)\left(1-2\epsilon\right)\leq f_{X,Y}\left(x,y\right)\frac{1-\epsilon}{1+\epsilon}\leq f_{\bar{X},\bar{Y}}\left(x,y\right),
\]
and
\[
f_{\bar{X},\bar{Y}}\left(x,y\right)\leq f_{X,Y}\left(x,y\right)\frac{1+\epsilon}{1-\epsilon}\leq f_{X,Y}\left(x,y\right)\left(1-4\epsilon\right),
\]
when $\epsilon<0.5$. Finally,
\begin{align*}
\int_{\mathbb{R}^{2}}\left|f_{\bar{X},\bar{Y}}(x,y)-f_{X,Y}(x,y)\right|dxdy\\
\leq4\epsilon\int_{\mathbb{R}^{2}}f_{X,Y}(x,y)dxdy=4\epsilon.
\end{align*}

Similarly, the Kullback-Leibler divergence between $f_{\bar{X},\bar{Y}}(x,y)$ and $f_{X,Y}(x,y)$ can be upper-bounded as
\begin{equation}
\begin{aligned}\mathbb{D}(f_{\bar{X},\bar{Y}}\|f_{X,Y}) & =\int_{\mathbb{R}^{2}}f_{\bar{X},\bar{Y}}(x,y)\log\frac{f_{\bar{X},\bar{Y}}(x,y)}{f_{X,Y}(x,y)}dxdy\\
 & \leq\int_{\mathbb{R}^{2}}f_{\bar{X},\bar{Y}}(x,y)\log(1+4\epsilon)dxdy\\
 & =\log(1+4\epsilon).
\end{aligned}
\label{eq:GaussianDistanceRegion2}
\end{equation}

For any $\sqrt{\frac{\delta_{1}\left(\Delta_{1}\Delta_{2}-\left(\rho-\sqrt{\delta_{1}\delta_{2}}\right)^{2}\right)}{1-\rho^{2}}}>0$, $\epsilon$ can be made arbitrarily small by scaling $\Lambda$. To show that $I\left(\bar{X},\bar{Y};\bar{X'}\right)$ can be arbitrarily close to $I\left(X,Y;X'\right)$, we rewrite $\mathbb{D}(f_{\bar{X},\bar{Y}}\|f_{X,Y})$ as
\[
\begin{aligned} & \mathbb{D}(f_{\bar{X},\bar{Y}}\|f_{X,Y})\\
 & =\int_{\mathbb{R}^{2}}f_{\bar{X},\bar{Y}}(x,y)\log\frac{f_{\bar{X},\bar{Y}}(x,y)}{f_{X,Y}(x,y)}dxdy\\
 & =-\int_{\mathbb{R}^{2}}f_{\bar{X},\bar{Y}}(x,y)\log f_{X,Y}(x,y)dxdy-h(\bar{X},\bar{Y})\\
 & =-\int_{\mathbb{R}^{2}}f_{\bar{X},\bar{Y}}(x,y)\log\left(\frac{1}{2\pi\sqrt{1-\rho^{2}}}\exp\bigg(-\frac{x^{2}+y^{2}-2\rho xy}{2(1-\rho^{2})}\bigg)\right)dxdy-h(\bar{X},\bar{Y})\\
 & =\log\big(2\pi\sqrt{1-\rho^{2}}\big)+\frac{\mathit{E}_{\bar{X},\bar{Y}}[x^{2}+y^{2}-2\rho xy]}{2(1-\rho^{2})}\log(e)-h(\bar{X},\bar{Y})\\
 & =\log\big(2\pi\sqrt{1-\rho^{2}}\big)+\log(e)-h(\bar{X},\bar{Y}),
\end{aligned}
\]
based on the fact that $\mathit{E}_{\bar{X},\bar{Y}}\left[x,y\right]=\rho$ and $\mathit{E}_{\bar{X},\bar{Y}}\left[x^{2}\right]=\mathit{E}_{\bar{X},\bar{Y}}\left[y^{2}\right]=1$. Trivially we have
\[
\begin{aligned}\mathbb{D}(f_{\bar{X},\bar{Y}}\|f_{X,Y}) & \geq\log\big(2\pi\sqrt{1-\rho^{2}}\big)+(1-\epsilon)\log(e)-h(\bar{X},\bar{Y})\\
 & =h(X,Y)-h(\bar{X},\bar{Y})-\epsilon\log(e).
\end{aligned}
\]

Using (\ref{eq:GaussianDistanceRegion2}), we obtain
\[
\begin{aligned}I(X,Y;X')-I(\bar{X},\bar{Y};\bar{X'}) & =h(X,Y)-h(\bar{X},\bar{Y})\\
 & \leq\log(1+4\epsilon)+\epsilon\log(e)\\
 & \leq5\epsilon\log(e).
\end{aligned}
\]
\end{IEEEproof}

\section{Proof of Theorem \ref{thm:Gaussian_Region2_Them} \label{sec:Proof-of-Theorem-Achivability}}
\begin{IEEEproof}
Firstly, for the sources $\left(\bar{X},\bar{Y}\right)$ and reconstruction RVs $\left(\bar{X'},\bar{Y'}\right)$, we consider the average performance of the multilevel polar codes with all possible choice of frozen sets $u_{\ell}^{\mathcal{F}_{\ell}}$ defined by (\ref{eqn:asymdefinition2}) on each level. If the encoding rule shown by (\ref{eqn:lossyencoder4}) is used for all $i\in[N]$ on each level, the resulted average distortions
of $\left(\bar{X},\bar{Y}\right)$ are given by
\[
\begin{aligned}\Delta_{P,\bar{X}} & =\frac{1}{N}\sum_{u_{1:r}^{1:N},\bar{x}^{1:N}}P_{U_{1:r}^{1:N},\bar{X}^{1:N}}\left(u_{1:r}^{1:N},\bar{x}^{1:N}\right)d\left(\bar{x}^{1:N},\mathcal{M}\left(u_{1:r}^{1:N}G_{N}\right)\right),\\
\Delta_{P,\bar{Y}} & =\frac{1}{N}\sum_{u_{1:r}^{1:N},\bar{y}^{1:N}}P_{U_{1:r}^{1:N},\bar{Y}^{1:N}}\left(u_{1:r}^{1:N},\bar{y}^{1:N}\right)d\left(\bar{y}^{1:N},\mathcal{M}\left(\sqrt{\frac{\delta_{2}}{\delta_{1}}}u_{1:r}^{1:N}G_{N}\right)\right),
\end{aligned}
\]
where $\mathcal{M}\left(u_{1:r}^{1:N}G_{N}\right)$ denotes a mapping from $u_{1:r}^{1:N}$ to $\bar{x'}{}^{1:N}$ given by \eqref{constructionD-finite-power}. Due to the linear relation $\bar{Y'}=\sqrt{\frac{\delta_{2}}{\delta_{1}}}\bar{X'}$ for region $\varepsilon_{2}$, we have $\bar{x'}{}_{\ell}^{1:N}=u_{\ell}^{1:N}G_{N}$ and $\bar{y'}{}_{\ell}^{1:N}=\sqrt{\frac{\delta_{2}}{\delta_{1}}}u_{\ell}^{1:N}G_{N}$ for each level. When $r\rightarrow\infty$, there exist an one-to-one mapping from $u_{1:r}^{1:N}$ to $\bar{x'}{}^{1:N}$ and $\bar{y'}{}^{1:N}$. Thus, we have {\allowdisplaybreaks
\[
\begin{aligned} \Delta_{P,\bar{X}}
 & =\frac{1}{N}\sum_{\bar{x'}^{1:N},\bar{x}^{1:N}}P_{\bar{X'}^{1:N},\bar{X}^{1:N}}\left(\bar{x'}^{1:N},\bar{x}^{1:N}\right)d^{N}\left(\bar{x'}^{1:N},\bar{x}^{1:N}\right)\\
 & =N\cdot\frac{1}{N}\sum_{\bar{x'},\bar{x}}P_{\bar{X'},\bar{X}}\left(\bar{x'},\bar{x}\right)d\left(\bar{x'},\bar{x}\right)\\
 & =\sum_{\bar{x'}\in\Lambda}P_{\bar{X'}}\left(\bar{x'}\right)\int_{-\infty}^{+\infty}\frac{1}{\sqrt{2\pi\Delta_{1}}}\exp\left(-\frac{\left(\bar{x}-\bar{x'}\right)^{2}}{2\Delta_{1}}\right)\left(\bar{x}-\bar{x'}\right)^{2}d\bar{x}\\
 & =\Delta_{1}.
\end{aligned}
\]
}
We can also apply this distortion to source $\bar{Y}$ and derive $\Delta_{P,\bar{Y}}=\Delta_{2}$. The results $\Delta_{P,\bar{X}}=\Delta_{1}$ and $\Delta_{P,\bar{Y}}=\Delta_{2}$ are reasonable, since the encoder does not conduct any compression. Next we replace $P_{U_{1:r}^{1:N},\bar{X}^{1:N}}\left(u_{1:r}^{1:N},\bar{x}^{1:N}\right)$ to $Q_{U_{1:r}^{1:N},\bar{X}^{1:N}}\left(u_{1:r}^{1:N},\bar{x}^{1:N}\right)$ and $P_{U_{1:r}^{1:N},\bar{Y}^{1:N}}\left(u_{1:r}^{1:N},\bar{y}^{1:N}\right)$ to $Q_{U_{1:r}^{1:N},\bar{Y}^{1:N}}\left(u_{1:r}^{1:N},\bar{y}^{1:N}\right)$,
so that the encoder compresses $\left(\bar{X}^{1:N},\bar{Y}^{1:N}\right)$ to $U_{\ell}^{\mathcal{I}_{\ell}}$ on each level according to (\ref{eqn:lossyencoder4}). The result average distortion $\Delta_{Q,\bar{X}}$ can be bounded
as {\allowdisplaybreaks
\[
\begin{aligned} \Delta_{Q,\bar{X}}
 & =\frac{1}{N}\sum_{u_{1:r}^{1:N},\bar{x}^{1:N}}Q_{U_{1:r}^{1:N},\bar{X}^{1:N}}\left(u_{1:r}^{1:N},\bar{x}^{1:N}\right)d\left(\bar{x}^{1:N},\mathcal{M}\left(u_{1:r}^{1:N}G_{N}\right)\right)\\
 & \leq\frac{1}{N}\sum_{u_{1:r}^{1:N},\bar{x}^{1:N}}P_{U_{1:r}^{1:N},\bar{X}^{1:N}}\left(u_{1:r}^{1:N},\bar{x}^{1:N}\right)d\left(\bar{x}^{1:N},\mathcal{M}\left(u_{1:r}^{1:N}G_{N}\right)\right)\\
 & \hspace{1em}+\frac{1}{N}\sum_{u_{1:r}^{1:N},\bar{x}^{1:N}}\left|P_{U_{1:r}^{1:N},\bar{X}^{1:N}}\left(u_{1:r}^{1:N},\bar{x}^{1:N}\right)-Q_{U_{1:r}^{1:N},\bar{X}^{1:N}}\left(u_{1:r}^{1:N},\bar{x}^{1:N}\right)\right|d\left(\bar{x}^{1:N},\mathcal{M}\left(u_{1:r}^{1:N}G_{N}\right)\right)\\
 & \leq\Delta_{P,\bar{X}}+\frac{1}{N}\cdot Nd_{\max,x}2\mathbb{V}\left(P_{U_{1:r}^{1:N},\bar{X}^{1:N}}\left(u_{1:r}^{1:N},\bar{x}^{1:N}\right),Q_{U_{1:r}^{1:N},\bar{X}^{1:N}}\left(u_{1:r}^{1:N},\bar{x}^{1:N}\right)\right)\\
 & =\Delta_{1}+O\left(N^22^{-N^{\beta''}}\right),
\end{aligned}
\]
}where we introduce the maximum distortion $d_{\max, x}=4\kappa^2$ between $\bar{x}$ and $\bar{x'}$ in the region of $[-\kappa, \kappa)$, and ignore the non-dominating terms caused by parts outside the boundary as we did in \eqref{eqn:DPYboundFinal}, \eqref{eqn:DQYPart1} and \eqref{eqn:DQYPart22}. The last equality follows from \eqref{eqn:variantialbound}.
Similarly, we also have $\Delta_{Q,\bar{Y}}=\Delta_{2}+O\left(N^22^{-N^{\beta''}}\right)$ for source $\bar{Y}$.

Now we quantize the Gaussian sources $\left(X,Y\right)$ by the same encoder. Again we take the source $X$ as example. The resulted distortion $\Delta_{Q,X}$ can be written as {\allowdisplaybreaks
\[
\begin{aligned} \Delta_{Q,X}
 & =\frac{1}{N}\sum_{u_{1:r}^{1:N},x^{1:N}}Q_{U_{1:r}^{1:N},X^{1:N}}\left(u_{1:r}^{1:N},x^{1:N}\right)d\left(x^{1:N},\mathcal{M}\left(u_{1:r}^{1:N}G_{N}\right)\right)\\
 & =\frac{1}{N}\sum_{u_{1:r}^{1:N},x^{1:N}}P_{X^{1:N}}\left(x^{1:N}\right)Q_{U_{1:r}^{1:N}\mid X^{1:N}}\left(u_{1:r}^{1:N}\mid x^{1:N}\right)d\left(x^{1:N},\mathcal{M}\left(u_{1:r}^{1:N}G_{N}\right)\right),
\end{aligned}
\]
}
Since the same encoder is used, we apply the same realizations $\left(x^{1:N},y^{1:N}\right)$ for both RV pairs $\left(X^{1:N},Y^{1:N}\right)$ and $\left(\bar{X}^{1:N},\bar{Y}^{1:N}\right)$. Then, the relation holds
\[
\begin{aligned}Q_{U_{1:r}^{1:N}\mid X^{1:N}}\left(u_{1:r}^{1:N}\mid x^{1:N}\right) & =Q_{U_{1:r}^{1:N}\mid\bar{X}^{1:N}}\left(u_{1:r}^{1:N}\mid x^{1:N}\right),\end{aligned}
\]
and hence
{\allowdisplaybreaks
\[
\begin{aligned} \Delta_{Q,X}-\Delta_{Q,\bar{X}}
 & =\frac{1}{N}\sum_{u_{1:r}^{1:N},x^{1:N}}\left(P_{X^{1:N}}\left(x^{1:N}\right)-P_{\bar{X}^{1:N}}\left(x^{1:N}\right)\right)\\
 & \hspace{2em}\cdot Q_{U_{1:r}^{1:N}\mid X^{1:N}}\left(u_{1:r}^{1:N}\mid x^{1:N}\right)\cdot d\left(x^{1:N},\mathcal{M}\left(u_{1:r}^{1:N}G_{N}\right)\right)\\
 & \leq\frac{1}{N}\sum_{x^{1:N}}\left|P_{X^{1:N}}\left(x^{1:N}\right)-P_{\bar{X}^{1:N}}\left(x^{1:N}\right)\right|Nd_{\max, x}.
\end{aligned}
\]

}

By the telescoping expansion,
{\allowdisplaybreaks
\[
\begin{aligned} & \sum_{x^{1:N}}\left|P_{X^{1:N}}\left(x^{1:N}\right)-P_{\bar{X}^{1:N}}\left(x^{1:N}\right)\right|\\
 & =\sum_{x^{1:N}}\sum_{i=1}^{N}\left|P_{X^{i}}\left(x^{i}\right)-P_{\bar{X}^{i}}\left(x^{i}\right)\right|P_{X^{1:i-1}}\left(x^{1:i-1}\right)P_{\bar{X}^{i+1:N}}\left(x^{i+1:N}\right)\\
 & =\sum_{i=1}^{N}\sum_{x^{i}}\left|P_{X^{i}}\left(x^{i}\right)-P_{\bar{X}^{i}}\left(x^{i}\right)\right|\\
 & =\sum_{i=1}^{N}\sum_{x^{i}}\left|\sum_{y^{i}}\left(P_{X^{i},Y^{i}}\left(x^{i},y^{i}\right)-P_{\bar{X}^{i},\bar{Y}^{i}}\left(x^{i},y^{i}\right)\right)\right|\\
 & \leq\sum_{i=1}^{N}\sum_{x^{i}}\sum_{y^{i}}\left|P_{X^{i},Y^{i}}\left(x^{i},y^{i}\right)-P_{\bar{X}^{i},\bar{Y}^{i}}\left(x^{i},y^{i}\right)\right|\\
 & \leq N\cdot4\epsilon_{\Lambda}\left(\sqrt{\frac{\delta_{1}\left(\Delta_{1}\Delta_{2}-\left(\rho-\sqrt{\delta_{1}\delta_{2}}\right)^{2}\right)}{1-\rho^{2}}}\right).
\end{aligned}
\]
}The last inequality results from Lemma \ref{lem:TwoGaussianRegion2}.

As a result,
\begin{equation}
\begin{aligned}\Delta_{Q,X} & \leq\Delta_{1}+O\left(N^22^{-N^{\beta''}}\right)+N\cdot4\epsilon_{\Lambda}\left(\sqrt{\frac{\delta_{1}\left(\Delta_{1}\Delta_{2}-\left(\rho-\sqrt{\delta_{1}\delta_{2}}\right)^{2}\right)}{1-\rho^{2}}}\right)d_{\max ,x}.\end{aligned}
\label{eq:Gaussian_Distortion_Conditions}
\end{equation}
Similarly, the distortion of $Y$ can be bounded as
\begin{equation}
\Delta_{Q,Y}\leq\Delta_{2}+O\left(N^22^{-N^{\beta''}}\right)+N\cdot4\epsilon_{\Lambda}\left(\sqrt{\frac{\delta_{1}\left(\Delta_{1}\Delta_{2}-\left(\rho-\sqrt{\delta_{1}\delta_{2}}\right)^{2}\right)}{1-\rho^{2}}}\right)d_{\max ,y},\label{eq:Gaussian_Distortion_Conditions_Y}
\end{equation}
where $d_{\max,y}=4\kappa^2=O(N^2)$ is the maximum distortion between $\bar{y}$ and $\bar{y'}$ in the region of $[-\kappa, \kappa)$.

By scaling $\Lambda$, we can make
\[
\epsilon_{\Lambda}\left(\sqrt{\frac{\delta_{1}\left(\Delta_{1}\Delta_{2}-\left(\rho-\sqrt{\delta_{1}\delta_{2}}\right)^{2}\right)}{1-\rho^{2}}}\right)\ll\frac{1}{4N\cdot4\kappa^2}.
\]
Therefore, $\Delta_{Q,X}$ and $\Delta_{Q,Y}$ can be arbitrarily close to $\Delta_{1}$ and $\Delta_{2}$, respectively, with the rate
\[
\begin{aligned}R & >I\left(\bar{X},\bar{Y};\bar{X'}\right)\\
 & \geq\frac{1}{2}\log\frac{1-\rho^{2}}{\Delta_{1}\Delta_{2}-\left(\rho-\sqrt{\left(1-\Delta_{1}\right)\left(1-\Delta_{2}\right)}\right)^{2}}-5\epsilon_{\Lambda}\left(\sqrt{\frac{\delta_{1}\left(\Delta_{1}\Delta_{2}-\left(\rho-\sqrt{\delta_{1}\delta_{2}}\right)^{2}\right)}{1-\rho^{2}}}\right)\log\left(e\right).
\end{aligned}
\]
When $\epsilon_{\Lambda}\left(\sqrt{\frac{\delta_{1}\left(\Delta_{1}\Delta_{2}-\left(\rho-\sqrt{\delta_{1}\delta_{2}}\right)^{2}\right)}{1-\rho^{2}}}\right)\rightarrow0$,
we have
\[
I\left(\bar{X},\bar{Y};\bar{X'}\right)\rightarrow\frac{1}{2}\log\frac{1-\rho^{2}}{\Delta_{1}\Delta_{2}-\left(\rho-\sqrt{\left(1-\Delta_{1}\right)\left(1-\Delta_{2}\right)}\right)^{2}}
\]
 and
\[
R>\frac{1}{2}\log\frac{1-\rho^{2}}{\Delta_{1}\Delta_{2}-\left(\rho-\sqrt{\left(1-\Delta_{1}\right)\left(1-\Delta_{2}\right)}\right)^{2}}.
\]

Since $\Delta_{Q,X}$ and $\Delta_{Q,Y}$ are average distortions over all random choices of $u_{\ell}^{\mathcal{F}_{\ell}}$, there exists at least one specific choice of $u_{\ell}^{\mathcal{F}_{\ell}}$ on each level making the average distortions satisfying (\ref{eq:Gaussian_Distortion_Conditions}) and (\ref{eq:Gaussian_Distortion_Conditions_Y}). This is a shift on the constructed polar lattice. As a result, the shifted polar lattice achieves the rate-distortion bound of the Gaussian sources.
\end{IEEEproof}
\section{Proof of Lemma \ref{lem:Lemma L Gaussian source } \label{sec:Proof-of-Lemma-LGaussian}}
\begin{IEEEproof}
For $L$ joint Gaussian RVs $\mathbf{X}_{L}=\{X_{1},X_{2},\ldots X_{L}\}$ with covariance matrix $K_{L}$ as given in (\ref{eq:covariance matrix KL}). The determinant of the covariance matrix is
\[
\left|K_{L}\right|=\left(1+(L-1)\rho\right)\left(1-\rho\right)^{L-1},
\]
and the inverse of $K_{L}$ is
\[
\begin{aligned} & K_{L}^{-1}=\frac{1}{-(L-1)\rho^{2}+(L-2)\rho+1}\begin{bmatrix}(L-2)\rho+1 & -\rho & \cdots & -\rho\\
-\rho & (L-2)\rho+1 & \cdots & -\rho\\
\vdots & \vdots & \ddots & \vdots\\
-\rho & -\rho & \cdots & (L-2)\rho+1
\end{bmatrix}.\end{aligned}
\]

Therefore the joint distribution of $L$ Gaussian sources with covariance matrix $K_{L}$ can be given by {\allowdisplaybreaks
\begin{align*}
\begin{aligned}  f_{\mathbf{X}_{L}}(\mathbf{x}_{L})
 & =\frac{1}{\sqrt{(2\pi)^{L}\left|K_{L}\right|}}\exp\left(-\frac{1}{2}\mathbf{x}_{L}^{\mathrm{T}}K_{L}^{-1}\mathbf{x}_{L}\right)\\
 & =\frac{1}{\sqrt{(2\pi)^{L}\left(1+(L-1)\rho\right)\left(1-\rho\right)^{L-1}}}\\
 & \hspace{1em}\cdot\exp\left[-\frac{1}{2\left(-(L-1)\rho^{2}+(L-2)\rho+1\right)}\left(\left(\left(L-2\right)\rho+1\right)\sum_{i=1}^{L}x_{i}^{2}-2\rho\hspace{-2em}\sum_{\begin{aligned}i,j=1;i<j\end{aligned}
}^{L}\hspace{-2em}x_{i}x_{j}\right)\right].
\end{aligned}
\end{align*}

}

Since the components of $\mathbf{X}_{L}$ are conditionally independent given $W$, we have {\allowdisplaybreaks[4]
\begin{equation}
\begin{aligned} & f_{\mathbf{\bar{X}}_{L}}(\mathbf{x}_{L})\\
 & =\sum_{a\in\Lambda}f_{\bar{X}_{1},\bar{X}_{2},\ldots\bar{X}_{L},\bar{W}}\left(x_{1},x_{2},\ldots x_{L},a\right)\\
 & =\sum_{a\in\Lambda}f_{\bar{W}}(a)f_{\bar{X}_{1}|\bar{W}}(x_{1}|a)f_{\bar{X}_{2}|\bar{W}}(x_{2}|a)\ldots f_{\bar{X}_{L}|\bar{W}}(x_{L}|a)\\
 & =\frac{1}{f_{\sqrt{\rho}}\left(\Lambda\right)}\sum_{a\in\Lambda}\frac{1}{\sqrt{2\pi\rho}}\exp\left(-\frac{a^{2}}{2\rho}\right)\frac{1}{\sqrt{2\pi(1-\rho)}}\exp\left(-\frac{(x_{1}-a)^{2}}{2(1-\rho)}\right)\cdots\frac{1}{\sqrt{2\pi(1-\rho)}}\exp\left(-\frac{(x_{L}-a)^{2}}{2(1-\rho)}\right)\\
 & =\frac{1}{\sqrt{(2\pi)^{L}\left(1+(L-1)\rho\right)\left(1-\rho\right)^{L-1}}}\\
 & \hspace{2em}\cdot\exp\left[-\frac{1}{2\left(-(L-1)\rho^{2}+(L-2)\rho+1\right)}\left(\left(\left(L-2\right)\rho+1\right)\sum_{i=1}^{L}x_{i}^{2}-2\rho\hspace{-2em}\sum_{\begin{aligned}i,j=1;i<j\end{aligned}
}^{L}\hspace{-2em}x_{i}x_{j}\right)\right]\\
 & \hspace{2em}\cdot\frac{1}{f_{\sqrt{\rho}}\left(\Lambda\right)}\sum_{a\in\Lambda}\frac{1}{\sqrt{\frac{2\pi\rho(1-\rho)}{1+(L-1)\rho}}}\exp\left[-\frac{1}{\frac{2\rho(1-\rho)}{1+(L-1)\rho}}\left(a-\frac{\rho}{1+(L-1)\rho}\left(x_{1}+\cdots+x_{L}\right)\right)^{2}\right]\\
 & =\frac{f_{\mathbf{X}_{L}}(\mathbf{x}_{L})}{f_{\sqrt{\rho}}\left(\Lambda\right)}\cdot\sum_{a\in\Lambda}\frac{1}{\sqrt{\frac{2\pi\rho(1-\rho)}{1+(L-1)\rho}}}\exp\left[-\frac{1}{\frac{2\rho(1-\rho)}{1+(L-1)\rho}}\left(a-\frac{\rho}{1+(L-1)\rho}\left(x_{1}+\cdots+x_{L}\right)\right)^{2}\right].
\end{aligned}
\label{eq:Gaussian joint L sources}
\end{equation}

}By the definition of the flatness factor (\ref{eq:flatnessFactor}), we have
\begin{equation}
\begin{aligned} & \left|V(\Lambda)\sum_{a\in\Lambda}\frac{1}{\sqrt{\frac{2\pi\rho(1-\rho)}{1+(L-1)\rho}}}\exp\left[-\frac{1}{\frac{2\rho(1-\rho)}{1+(L-1)\rho}}\left(a-\frac{\rho\left(x_{1}+\cdots+x_{L}\right)}{1+(L-1)\rho}\right)^{2}\right]-1\right|\\
 & \leq\epsilon_{\Lambda}\left(\sqrt{\frac{\rho(1-\rho)}{1+(L-1)\rho}}\right)=\epsilon.
\end{aligned}
\label{eq:FlatnessFactorL}
\end{equation}

Moreover, we have $\epsilon_{\Lambda}\left(\sqrt{\rho}\right)\leq\epsilon$ since $\epsilon_{\Lambda}\left(\sigma\right)$ is monotonically decreasing of $\sigma$. Hence
\begin{equation}
\left|V(\Lambda)f_{\sqrt{\rho}}(\Lambda)-1\right|\leq\epsilon_{\Lambda}\left(\sqrt{\rho}\right)\leq\epsilon.\label{eq:FlatnessFactorL2}
\end{equation}

Combining $\left(\ref{eq:Gaussian joint L sources}\right)$, $\left(\ref{eq:FlatnessFactorL}\right)$ and $\left(\ref{eq:FlatnessFactorL2}\right)$ gives us
\[
f_{\mathbf{X}_{L}}(\mathbf{x}_{L})\left(1-2\epsilon\right)\leq f_{\mathbf{X}_{L}}(\mathbf{x}_{L})\frac{1-\epsilon}{1+\epsilon}\leq f_{\mathbf{\bar{X}}_{L}}(\mathbf{x}_{L}),
\]
and
\[
f_{\mathbf{\bar{X}}_{L}}(\mathbf{x}_{L})\leq f_{\mathbf{X}_{L}}(\mathbf{x}_{L})\frac{1+\epsilon}{1-\epsilon}\leq f_{\mathbf{X}_{L}}(\mathbf{x}_{L})\left(1+4\epsilon\right),
\]
when $\epsilon<\frac{1}{2}$. Finally,
\[
\begin{aligned}\int_{\mathbb{R}^{L}}\left|f_{\mathbf{\bar{X}}_{L}}(\mathbf{x}_{L})-f_{\mathbf{X}_{L}}(\mathbf{x}_{L})\right|dx_{1}dx_{2}\ldots dx_{L}\leq4\epsilon\int_{\mathbb{R}^{L}}f_{\mathbf{X}_{L}}(\mathbf{x}_{L})dx_{1}dx_{2}\ldots dx_{L}=4\epsilon.\end{aligned}
\]

Also, the Kullback-Leibler divergence between $f_{\mathbf{\bar{X}}_{L}}(\mathbf{x}_{L})$ and $f_{\mathbf{X}_{L}}(\mathbf{x}_{L})$ can be upper-bounded as
\begin{equation}
\begin{aligned}\mathbb{D}\left(f_{\mathbf{\bar{X}}_{L}}(\mathbf{x}_{L})\|f_{\mathbf{X}_{L}}(\mathbf{x}_{L})\right) & =\int_{\mathbb{R}^{L}}f_{\mathbf{\bar{X}}_{L}}(\mathbf{x}_{L})\log\frac{f_{\mathbf{\bar{X}}_{L}}(\mathbf{x}_{L})}{f_{\mathbf{X}_{L}}(\mathbf{x}_{L})}dx_{1}dx_{2}\ldots dx_{L}\\
 & \leq\int_{\mathbb{R}^{L}}f_{\mathbf{\bar{X}}_{L}}(\mathbf{x}_{L})\log(1+4\epsilon)dx_{1}dx_{2}\ldots dx_{L}\\
 & =\log(1+4\epsilon).
\end{aligned}
\label{eq:LGaussianKLdistance}
\end{equation}

For any $\sqrt{\frac{\rho(1-\rho)}{1+(L-1)\rho}}>0$, $\epsilon$ can be made arbitrarily small by scaling $\Lambda$. Therefore, $\bar{W}$ can be regarded as the common message when $\epsilon\rightarrow0$. To show that $I\left(\mathbf{\bar{X}}_{L};\bar{W}\right)$ can be arbitrarily close to the CI, we rewrite $\mathbb{D}\left(f_{\mathbf{\bar{X}}_{L}}(\mathbf{x}_{L})\|f_{\mathbf{X}_{L}}(\mathbf{x}_{L})\right)$ as
{\allowdisplaybreaks
\[
\begin{aligned} & \mathbb{D}\left(f_{\mathbf{\bar{X}}_{L}}(\mathbf{x}_{L})\|f_{\mathbf{X}_{L}}(\mathbf{x}_{L})\right)\\
 & =\int_{\mathbb{R}^{L}}f_{\mathbf{\bar{X}}_{L}}(\mathbf{x}_{L})\log\frac{f_{\mathbf{\bar{X}}_{L}}(\mathbf{x}_{L})}{f_{\mathbf{X}_{L}}(\mathbf{x}_{L})}dx_{1}dx_{2}\ldots dx_{L}\\
 & =-\int_{\mathbb{R}^{L}}f_{\mathbf{\bar{X}}_{L}}(\mathbf{x}_{L})\log f_{\mathbf{X}_{L}}(\mathbf{x}_{L})dx_{1}dx_{2}\ldots dx_{L}-h(\mathbf{\bar{X}}_{L})\\
 & =-\int_{\mathbb{R}^{L}}f_{\mathbf{\bar{X}}_{L}}(\mathbf{x}_{L})\log\left[\frac{1}{\sqrt{(2\pi)^{L}\left(1+(L-1)\rho\right)\left(1-\rho\right)^{L-1}}}\right.\\
 & \hspace{1em}\cdot\left.\exp\left[-\frac{1}{2\left(-(L-1)\rho^{2}+(L-2)\rho+1\right)}\left(\left(\left(L-2\right)\rho+1\right)\sum_{i=1}^{L}x_{i}^{2}-2\rho\hspace{-2em}\sum_{\begin{aligned}i,j=1;i<j\end{aligned}
}^{L}\hspace{-2em}x_{i}x_{j}\right)\right]\right]dx_{1}\ldots dx_{L}-h(\mathbf{\bar{X}}_{L})\\
 & =\log\left(\sqrt{(2\pi)^{L}\left(1+(L-1)\rho\right)\left(1-\rho\right)^{L-1}}\right)-h(\mathbf{\bar{X}}_{L})\\
 & \hspace{1em}+\mathsf{\mathit{E}_{\mathbf{\bar{X}}_{L}}}\left[\left(\left(L-2\right)\rho+1\right)\sum_{i=1}^{L}x_{i}^{2}-2\rho\hspace{-2em}\sum_{\begin{aligned}i,j=1;i<j\end{aligned}
}^{L}\hspace{-2em}x_{i}x_{j}\right]\frac{\log\left(e\right)}{2\left(-(L-1)\rho^{2}+(L-2)\rho+1\right)}\\
 & =\log\left(\sqrt{(2\pi)^{L}\left(1+(L-1)\rho\right)\left(1-\rho\right)^{L-1}}\right)-h(\mathbf{\bar{X}}_{L})+\frac{L}{2}\frac{\mathsf{\mathit{E}_{\bar{W}}}\left[w^{2}\right]+(L-2)\rho+1}{(L-1)\rho+1}\log\left(e\right).
\end{aligned}
\]

}Note that $\mathsf{\mathit{E}}_{\bar{W}}\left[w^{2}\right]\geq\rho(1-2\epsilon)$ according to \cite[Lemma 5]{LingBel13} and \cite[Remark 3]{LingBel13}. Hence{\allowdisplaybreaks
\[
\begin{aligned} & \mathbb{D}\left(f_{\mathbf{\bar{X}}_{L}}(\mathbf{x}_{L})\|f_{\mathbf{X}_{L}}(\mathbf{x}_{L})\right)\\
 & \geq\log\left(\sqrt{(2\pi)^{L}\left(1+(L-1)\rho\right)\left(1-\rho\right)^{L-1}}\right)-h(\mathbf{\bar{X}}_{L})+\left(\frac{L}{2}-\frac{L\rho}{(L-1)\rho+1}\epsilon\right)\log\left(e\right)\\
 & \geq\log\left(\sqrt{(2\pi)^{L}\left(1+(L-1)\rho\right)\left(1-\rho\right)^{L-1}}\right)-h(\mathbf{\bar{X}}_{L})+\left(\frac{L}{2}-\epsilon\right)\log\left(e\right)\\
 & =h(\mathbf{X}_{L})-h(\mathbf{\bar{X}}_{L})-\epsilon\log\left(e\right).
\end{aligned}
\]

}By $\left(\ref{eq:LGaussianKLdistance}\right)$, we obtain
\[
\begin{aligned}I\left(\mathbf{X}_{L};W\right)-I\left(\mathbf{\bar{X}}_{L};\bar{W}\right) & =h(\mathbf{X}_{L})-h(\mathbf{\bar{X}}_{L})\\
 & \leq\log(1+4\epsilon)+\epsilon\log\left(e\right)\\
 & \leq5\epsilon\log\left(e\right).
\end{aligned}
\]
\end{IEEEproof}

\section{Proof of Theorem \ref{thm:Theorem L Gaussian source} \label{sec:Proof-of-Theorem-LGaussian}}
\begin{IEEEproof}
%Let $\bar{W}$ be labeled by bits $\bar{W}_{1:r}=\{\bar{W}_{1},\ldots,\bar{W}_{r}\}$ according to a binary partition chain $\Lambda/\Lambda_{1}/\cdots/\Lambda_{r-1}/\Lambda'$ ($\Lambda'$ also refers to $\Lambda_{r}$). $D_{\Lambda,\sqrt{\rho}}$ induces a distribution $P_{\bar{W}_{1:r}}$ whose limit corresponds to $D_{\Lambda,\sqrt{\rho}}$ as $r\rightarrow\infty$.
By the chain rule of mutual information
\[
I(\mathbf{\bar{X}}_{L};\bar{W}_{1:r})=\sum_{\ell=1}^{r}I(\mathbf{\bar{X}}_{L};\bar{W}_{\ell}|\bar{W}_{1:\ell-1}),
\]
we obtain $r$ binary-input test channels $V_{\ell}$ for $1\leq\ell\leq r$. Given the realization $w_{1:\ell}$ of $\bar{W}_{1:r}$, denote by $\mathcal{A}_{\ell}(w_{1:\ell})$ the coset of $\Lambda_{\ell}$ indexed by $w_{1:\ell-1}$ and $w_{\ell}$. According to \cite{multilevel}, the channel transition PDF of the $\ell$-th channel $V_{\ell}$ is
given by
{\allowdisplaybreaks
\[
\begin{aligned} & f_{\mathbf{\bar{X}}_{L}|\bar{W}_{\ell},\bar{W}_{1:\ell-1}}\left(\mathbf{x}_{L}|w_{\ell},w_{1:\ell-1}\right)\\
 & =\frac{1}{f_{\sqrt{\rho}}(\mathcal{A}_{\ell}(w_{1:\ell}))}\sum_{a\in\mathcal{A}_{\ell}(w_{1:\ell})}f_{\sqrt{\rho}}(a)f_{\mathbf{\bar{X}}_{L}|\bar{W}}(\mathbf{x}_{L}|a)\\
 & =\frac{1}{f_{\sqrt{\rho}}(\mathcal{A}_{\ell}(w_{1:\ell}))}\sum_{a\in\mathcal{A}_{\ell}(w_{1:\ell})}\frac{1}{\sqrt{2\pi\rho}}\exp\bigg(-\frac{a^{2}}{2\rho}\bigg)\frac{1}{\sqrt{2\pi(1-\rho)}}\exp\bigg(-\frac{(x_{1}-a)^{2}}{2(1-\rho)}\bigg)\cdots\frac{1}{\sqrt{2\pi(1-\rho)}}\exp\bigg(-\frac{(x_{L}-a)^{2}}{2(1-\rho)}\bigg)\\
 & =\frac{1}{\sqrt{(2\pi)^{L}\left(1+(L-1)\rho\right)\left(1-\rho\right)^{L-1}}}\\
 & \hspace{1em}\cdot\exp\left[-\frac{1}{2\left(-(L-1)\rho^{2}+(L-2)\rho+1\right)}\left(\left(\left(L-2\right)\rho+1\right)\sum_{i=1}^{L}x_{i}^{2}-2\rho\hspace{-2em}\sum_{\begin{aligned}i,j=1;i<j\end{aligned}
}^{L}\hspace{-2em}x_{i}x_{j}\right)\right]\\
 & \hspace{1em}\cdot\frac{1}{f_{\sqrt{\rho}}(\mathcal{A}_{\ell}(w_{1:\ell}))}\sum_{a\in\mathcal{A}_{\ell}(w_{1:\ell})}\frac{1}{\sqrt{\frac{2\pi\rho(1-\rho)}{1+(L-1)\rho}}}\exp\left[-\frac{1}{\frac{2\rho(1-\rho)}{1+(L-1)\rho}}\left(a-\frac{\rho}{1+(L-1)\rho}\left(x_{1}+\cdots+x_{L}\right)\right)^{2}\right],
\end{aligned}
\]

}

Let $\tilde{V}_{\ell}$ be a symmetrized channel with input $\tilde{W}_{\ell}$ (assume to be uniformly distributed) and output $\left(\mathbf{\bar{X}}_{L},\bar{W}_{1:\ell-1},\bar{W}_{\ell}\oplus\tilde{W}_{\ell}\right)$,
built from the asymmetric channel $V_{\ell}$. Then the joint PDF
of $V_{\ell}$ can be represented by the transition PDF of $\tilde{V}_{\ell}$, as shown in the following
equation.
{\allowdisplaybreaks
\begin{equation}
\begin{aligned} & f_{\tilde{V}_{\ell}}(\mathbf{x}_{L},w_{1:\ell-1},w_{\ell}\oplus\tilde{w}_{\ell}|\tilde{w}_{\ell})\\
 & =f_{\mathbf{\bar{X}}_{L},\bar{W}_{1:\ell}}(\mathbf{x}_{L},w_{1:\ell})\\
 & =\frac{1}{\sqrt{(2\pi)^{L}\left(1+(L-1)\rho\right)\left(1-\rho\right)^{L-1}}}\\
 & \hspace{1em}\cdot\exp\left[-\frac{1}{2\left(-(L-1)\rho^{2}+(L-2)\rho+1\right)}\left(\left(\left(L-2\right)\rho+1\right)\sum_{i=1}^{L}x_{i}^{2}-2\rho\hspace{-2em}\sum_{\begin{aligned}i,j=1;i<j\end{aligned}
}^{L}\hspace{-2em}x_{i}x_{j}\right)\right]\\
 & \hspace{1em}\cdot\frac{1}{f_{\sqrt{\rho}}(\Lambda)}\sum_{a\in\mathcal{A}_{\ell}(w_{1:\ell})}\frac{1}{\sqrt{\frac{2\pi\rho(1-\rho)}{1+(L-1)\rho}}}\exp\left[-\frac{1}{\frac{2\rho(1-\rho)}{1+(L-1)\rho}}\left(a-\frac{L\rho}{1+(L-1)\rho}\frac{\left(x_{1}+\cdots+x_{L}\right)}{L}\right)^{2}\right].
\end{aligned}
\label{eq:AppTheoremSymm}
\end{equation}

}

Comparing with the $\Lambda_{\ell-1}/\Lambda_{\ell}$ channel \cite{polarlatticeJ}, we observe that the likelihood ratio of the symmetrized channel (\ref{eq:GaussianSymmetrimizedF}) is equivalent to that of a $\Lambda_{\ell-1}/\Lambda_{\ell}$ channel with noise variance $\frac{\rho(1-\rho)}{1+(L-1)\rho}$. Moreover, the mean value and variance of the Gaussian RV $\frac{X_{1}+\cdots+X_{L}}{L}$ are respectively given as
\[
\mathsf{\mathit{E}}\left[\frac{X_{1}+\cdots+X_{L}}{L}\right]=\frac{1}{L}\left(\mathsf{\mathit{E}}\left[X_{1}\right]+\ldots+\mathsf{\mathit{E}}\left[X_{L}\right]\right)=0,
\]
and
\[
\begin{aligned}\sigma^{2}\left(\frac{X_{1}+\cdots+X_{L}}{L}\right) & \mathsf{=\mathit{E}}\left[\left(\frac{X_{1}+\cdots+X_{L}}{L}\right)^{2}\right]-\left(\mathsf{\mathit{E}}\left[\frac{X_{1}+\cdots+X_{L}}{L}\right]\right)^{2}\\
 & =\frac{1}{L^{2}}\left(\mathsf{\mathit{E}}\left[\sum_{i=1}^{L}X_{i}^{2}+2\rho\hspace{-2em}\sum_{\begin{aligned}i,j=1;i<j\end{aligned}
}^{L}\hspace{-2em}X_{i}X_{j}\right]\right)\\
 & =\frac{L+2\rho\begin{pmatrix}L\\
2
\end{pmatrix}}{L^{2}}=\frac{1+\left(L-1\right)\rho}{L},
\end{aligned}
\]
where $\begin{pmatrix}L\\
r
\end{pmatrix}\triangleq\frac{L!}{r!\left(L-r\right)!}$.

Consider the construction of a polar lattice to quantize $\frac{X_{1}+X_{2}+\ldots+X_{L}}{L}$ with reconstruction distribution $D_{\Lambda,\sqrt{\rho}}$. Denote the variance of the source and the reconstruction by $\sigma_{s}^{2}=\frac{1+\left(L-1\right)\rho}{L}$ and $\sigma_{r}^{2}=\rho$, respectively. Thus, the variance of the
noise equals $\sigma_{z}^{2}=\sigma_{s}^{2}-\sigma_{r}^{2}=\frac{1-\rho}{L}$. To apply MMSE , the coefficient $\alpha$ and the noise variance $\tilde{\sigma}_{z}^{2}$ are respectively given by
\[
\alpha=\frac{\sigma_{r}^{2}}{\sigma_{s}^{2}}=\frac{L\rho}{1+\left(L-1\right)\rho},
\]

\[
\tilde{\sigma}_{z}^{2}=\alpha\cdot\sigma_{z}^{2}=\frac{\rho(1-\rho)}{1+\left(L-1\right)\rho},
\]
which are the same to those in the summation section of $\left(\ref{eq:AppTheoremSymm}\right)$.
\end{IEEEproof}

\bibliographystyle{IEEEtran}
\bibliography{Myreff,MyreffGeneral_JW}
% that's all folks
\end{document}